\documentclass[12pt]{article}
\pdfoutput=1
\usepackage{putex}
\usepackage{amsmath,hyperref,amssymb,amsfonts,bm,amscd,slashed,ytableau}
\usepackage{cite}
\usepackage{graphicx}
\usepackage{multirow}
\usepackage{verbatim}
\usepackage{appendix}
\usepackage{fancybox}
\usepackage{array}
\usepackage{url}
\usepackage{float}
\usepackage{mathtools}
\usepackage{mathrsfs}
\usepackage{bbm}
\usepackage[bbgreekl]{mathbbol}

\DeclareSymbolFontAlphabet{\mathbbm}{bbold}
\DeclareSymbolFontAlphabet{\mathbb}{AMSb}

\DeclareMathAlphabet{\mathpzc}{OT1}{pzc}{m}{it}

\newcommand{\bea}{\begin{eqnarray}}
\newcommand{\eea}{\end{eqnarray}}
\newcommand{\be}{\begin{equation}}
\newcommand{\ee}{\end{equation}}

\def \beaa {\begin{equation}\begin{aligned}}
\def \eeaa {\end{aligned}\end{equation}}

\newcommand{\Z}{{\mathbb Z}}
\newcommand{\R}{{\mathbb R}}
\newcommand{\C}{{\mathbb C}}

\newcommand{\cZ}{{\mathcal{Z}}}
\newcommand{\cF}{{\mathcal{F}}}
\newcommand{\cN}{{\mathcal{N}}}

\newcommand{\dd}{{\rm d}}

\def\Tr{{\rm Tr \,}}

\def\m{\mu}

\def\tilde{\widetilde}
\def\hat{\widehat}
\def\bar{\overline}

\def\cA{{\mathcal A}}
\def\cB{{\mathcal B}}
\def\cC{{\mathcal C}}
\def\cD{{\mathcal D}}
\def\cE{{\mathcal E}}
\def\cF{{\mathcal F}}
\def\cG{{\mathcal G}}
\def\cH{{\mathcal H}}
\def\cI{{\mathcal I}}
\def\cJ{{\mathcal J}}

\def\cL{{\mathcal L}}
\def\cM{{\mathcal M}}
\def\cN{{\mathcal N}}
\def\cO{{\mathcal O}}
\def\cP{{\mathcal P}}
\def\cQ{{\mathcal Q}}
\def\cR{{\mathcal R}}

\def\cT{{\mathcal T}}

\def\cV{{\mathcal V}}
\def\cW{{\mathcal W}}

\def\cZ{{\mathcal Z}}
\def\bE{{\mathbb E}}
\def\bfT{{\mathbf T}}
\def\sB{{\mathscr B}}
\def\sD{{\mathscr D}}
\def\sN{{\mathscr N}}
\def\bL{{\mathbb L}}
\def\rQ{{\mathbbmtt{Q}\,}}
\def\rH{{\mathbbmtt{H}\,}}
\def\rG{{\mathbbmtt{G}\,}}

\renewcommand{\bar}{\overline}
\renewcommand{\hat}{\widehat}

\numberwithin{equation}{section}
\newcommand{\ii}{\mathrm{i}}

\begin{document}
%\preprint{}

\institution{SCGP}{Simons Center for Geometry and Physics,\cr Stony Brook University, Stony Brook, NY 11794-3636, USA}

\title{Interfaces and Quantum Algebras, I:\\ Stable Envelopes}
\authors{Mykola Dedushenko  and Nikita Nekrasov}
	
\abstract{The stable envelopes of Okounkov et al. realize some representations of quantum algebras associated to quivers, using geometry. 
We relate these geometric considerations to quantum field theory. The main ingredients are the supersymmetric interfaces in gauge theories with four supercharges, relation of supersymmetric vacua to generalized cohomology theories, and Berry connections. We mainly consider softly broken 
compactified three dimensional
${\cN} =4$ theories. The companion papers will discuss applications of this construction to symplectic duality, Bethe/gauge correspondence, generalizations to higher dimensional theories, and other topics.}  
\date{}
	
\maketitle

\tableofcontents

\section{Introduction}

The study of quantum field theory in recent decades has been enriched by the appreciation of the r{\^o}le of extended observables, associated to surfaces, domain walls, boundaries and interfaces. Traditional gauge theory has, naturally, electric line observables and codimension three magnetic observables, better known as Wilson loops, and 't Hooft loops, in four dimensions. There is a growing sense of the need to include all kinds of defects, boundaries and corners, in order to have a real understanding of what quantim field theory is, and what is it good for, cf. \cite{Braverman:2004vv, Gukov:2006jk, Lurie:2009keu, Drukker:2011}, and \cite{NikIAS:2004,LosevTalk,NeitzkeTalk}.

Quantum field theory is, in many respects, an infinite-dimensional version of quantum mechanics. In quantum mechanics
observables form a noncommutative algebra $\cA$, 
as expectation values of products of observables depends on the time ordering,
\[ \langle {\cO}_{1}(t_1) {\cO}_{2}(t_2) \rangle \neq \langle {\cO}_{2}(t_1) {\cO}_{1}(t_2) \rangle \, , \ {\cO}_{1,2} \in {\cA} \ . \]
The noncommutativity makes geometric considerations blurred. Recall that algebraically, 
topological spaces are identified with the commutative algebras (algebras of continuous functions, say)
\[ x \in X \leftrightarrow f\mapsto f(x) , \]
while geometry can be encoded in additional restrictions and structures. 
Occasionally, the noncommutative algebra $\cA$ of observables of some quantum mechanical system contains a large enough commutative subalgebra $\cC \subset \cA$. One possible source of such emergence of (``the target space'') geometry within the framework of the quantum mechanics is a topological, or vacuum, sector of a quantum field theory with a mass gap, compactified so as to look macroscopically as a one dimensional theory.\\

\centerline{\includegraphics[width=10cm]{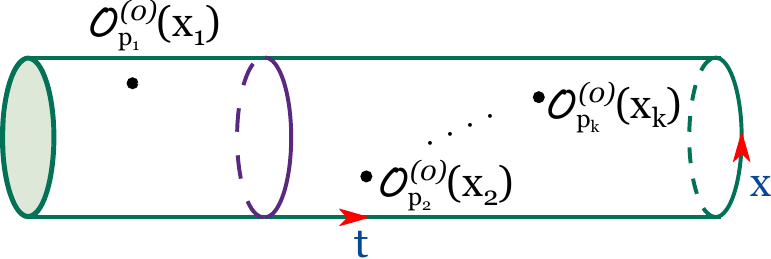}}

If the gapped theory can be endowed with the nilpotent symmetry ${\cQ}$, ${\cQ}^2 = 0$, such that the stress-tensor is 
a homotopy of the Hilbert space to the subspace of vacua, 
\[ T_{\mu\nu} = \{ {\cQ}, G_{\mu\nu} \} , \]
with some operator-valued tensor $G_{\mu\nu}$, then local operators ${\cO}^{(0)}_{i}(x)$, annihilated by ${\cQ}$
are independent of their spacetime position, up to $\cQ$-exact terms:
\be \dd {\cO}^{(0)} = \{ {\cQ} , {\cO}^{(1)} \} . \label{eq:descent}
\ee
This relation implies the commutativity of the corresponding quantum mechanical operators, as points in two and 
higher dimensions can be moved around each other. Meanwhile, interfaces represented by a purple circle in the picture above, do not commute and descend to noncommutative quantum mechanical operators. A large class of quantum integrable systems, whose integrability is explained by the use of the structure \eqref{eq:descent}, corresponds to supersymmetric gauge theories.

The connection between integrability and gauge theories has a long history. The projection method of Olshanetsky and Perelomov \cite{OP}, Kazhdan-Kostant-Sternberg construction of Calogero-Moser-Sutherland systems of particles can be viewed as the examples of one-dimensional gauge theories equivalent to quantum integrable systems. 
The discovery \cite{Witten:1988hf} of the connection between Jones polynomial and Chern-Simons theory in three dimensions
made possible an embedding \cite{Witten:1989rw, Witten:1989wf,Witten:1991pz} of a large class of soluble models of statistical physics into the realm of topological field theories. 

The paper \cite{Gorsky:1993pe} raised the question of a possibility of connecting the exact results, e.g. \cite{Novikov:1983uc, Witten:1992xu} in supersymmetric gauge theories to quantum integrable systems. 
The construction of
\cite{Gorsky:1993pe} related the Sutherland model to the two-dimensional Yang-Mills theory, which, thanks to \cite{Witten:1992xu} can be viewed as a subsector of a deformation of the two dimensional ${\cN}=(2,2)$ super Yang-Mills theory. The relativistic generalization of the Sutherland model embeds \cite{Gorsky:1993dq} to three dimensional Chern-Simons theory which, in turn, admits an interpretation as the vacuum subsector of a twisted supersymmetric theory \cite{NikThesis, Baulieu:1997nj,Beasley:2005vf,Nedelin:2016gwu}. 
A close cousin of these many-body models, Lieb-Liniger system describing the $N$-particle sector of a one-dimensional Bose gas \cite{LL:1963}, 
also known as the quantum non-linear Schr{\"o}dinger system was found \cite{Moore:1997dj} to be related, perhaps in a similar \cite{Gerasimov:2007ap} way, to a deformation of the ${\cN}=2^{*}$ theory in two dimensions, a close cousin of the two dimensional Yang-Mills theory. The analogous elliptic models required somewhat more exotic generalizations of gauge theories, involving non-Lorentz invariant deformations by Chern-Simons terms multiplied by holomorphic differentials (see p.5 of \cite{Gorsky:1994dj},
pp. 88-89 of \cite{NikThesis}), a hybrid version of the holomorphic Chern-Simons theory introduced in \cite{Witten:1992fb}, motivated, among other things, by \cite{Gerasimov:1989qj,Etingof:1992br}, and used, implicitly, in \cite{Arutyunov:1996vy}. 

In a somewhat parallel way a connection between the quantum integrability and topological sigma models was found in \cite{Givental:1993nc, GiventalEq}, see \cite{ASL} for a mathematical physicist's perspective.  The paper \cite{Moore:1997dj}, in the hindsight, gave an important example of such connection  (the models found in \cite{Givental:1993nc} had continuum spectrum, so the Bethe equations could not be detected), which was further expanded to the Bethe/gauge correspondence between the vacua of supersymmetric gauge theories with two dimensional ${\cN}=2$ super-Poincare invariance and quantum integrable systems amenable to Bethe ansatz in \cite{Nekrasov:2009uh, Nekrasov:2009ui, Nekrasov:2009rc}. In the approach of Faddeev's school \cite{Faddeev:1994nk, Faddeev:1996iy, Takhtajan:1979iv} the latter is a consequence of a (hidden) noncommutative algebraic structure of the system: the presence of a spectrum generating quantum  algebra whose commutative subalgebra is the set of quantum integrals of motion.

In the context of supersymmetric (gauge) theory, such commutative algebra is typically the algebra of local operators, commuting with some (equivariantly) nilpotent supercharge, more precisely its cohomology. More generally, these operators could be local in two dimensions where the translational part of the ${\cN}=2$ super-Poincare algebra acts, while extended in other dimensions. For example, a three dimensional theory compactified on a circle $S^1$ may have the line operators wrapped on $S^1$, forming such a subalgebra. 

The question of recovering the full quantum algebra, e.g. the Yangian or quantum loop algebra, has been asked in \cite{Nekrasov:2009rc}.  In \cite{NikRome:2009} it was proposed that the answer should involve some sort of supersymmetric interfaces, i.e. boundary conditions compatible with a fraction of supersymmetry, connecting two, possibly different, quantum field theories.

The proposal of \cite{NikThesis} was left unnoticed until the similar proposal was independently made in \cite{Costello:2013zra, Costello:2013sla}, and greatly developed in \cite{Costello:2017dso, Costello:2018gyb, Costello:2018txb}. In this approach, the main ingredient of
the algebraic Bethe ansatz \cite{Faddeev:1994nk} approach to integrability, an $R$-matrix depending on a spectral parameter, is derived
from the perturbative analysis of the four dimensional Chern-Simons theory, just like the finite dimensional quantum group
constant $R$-matrix can be derived, to some extent, from the three dimensional Chern-Simons theory \cite{Polyakov:1988my, Polyakov:1988md, Guadagnini:1989th,Guadagnini:1989am, Kontsevich:1993, Fock:1998nu, Morozov:2010kv}.

\ 

In a completely parallel development, partly inspired by the ideas of \cite{Nekrasov:2009uh}, but also by completely independent discoveries of R.~Bezrukavnikov, the geometric approach to the construction of $R$-matrices and the associated quantum algebras
was initiated in \cite{Maulik:2012wi}, followed by \cite{Aganagic:2016jmx}. The representations of quantum loop algebras and Yangians in equivariant $K$-theory and cohomology of Nakajima quiver varieties were constructed earlier \cite{Nakajima:2001, Varagnolo2000}. A generalization to a singular ``higher spin'' $A_1$ quiver variety in the rational case has appeared in \cite{Bykov:2019cst}. 

\ 

This is the part I of a series of three papers, which aim to bridge several proposals, relating quantum algebras and quantum field theory. Our main tool is the use of supersymmetric backgrounds in gauge theories with ${\cN}=4$ supersymmetry in three dimensions, compactified on a two-torus, with 
soft breaking of supersymmetry by background gauge fields. Sometimes we make the backgrounds nearly singular, thereby creating
interface operators. This is somewhat similar in spirit to the definition of local fermionic operators through singular gauge transformations applied to background vector fields, gauging some global symmetry \cite{Witten:1987tv, Losev:1995cr}. In our work we  mostly activate the scalar superpartners, e.g. masses, of these background vector fields. Of course, backgrounds with varying masses are well-studied
in the context of applications of quantum field theory to condensed matter physics, cf. \cite{Witten:2015aoa}, but here we dress them with the supersymmetric background allowing, in principle, to perform exact evaluations of certain correlation functions. More generally, backgrounds with spacetime-dependent parameters (and the resulting monodromies) have been previously used in the literature to realize various algebraic and geometric structures, e.g., see \cite{Cecotti:1991me,Gukov:2006jk,Cecotti:2010qn,Cecotti:2011iy,Gadde:2013wq,Gukov:2014gja}.

In this paper we are going to relate the stable envelope construction of \cite{Maulik:2012wi, Aganagic:2016jmx} to supersymmetric gauge theory interfaces. In part II we will study cigar partition functions, how they are acted on by the duality interface built from the stable envelopes, and use the $\Omega$-deformed theories (cigar backgrounds) to relate correlators
of operators extended in different number of dimensions. In part III we will study the R-matrices, establish the Bethe/gauge correspondence and the connection to the four dimensional Chern-Simons theory.

\subsection{Overview}
Let us provide a brief overview of this paper. Our work is founded on two main ideas, which are simple enough to be described in one paragraph. The first idea is that the supersymmetric ground states of a supersymmetric quantum field theory compactified to $0+1$ dimension, in the presence of flavor symmetry backgrounds, are described by an equivariant cohomology theory of the moduli space of vacua (its Higgs branch in simple cases). Precisely which cohomology theory it is depends on spacetime dimensionality and on the choice of supercharge. The second idea is that there exist supersymmetric Janus interfaces interpolating between the large real masses in one half-space and zero real masses in another. Treating the normal direction to the interface as time, such interfaces are certain BPS operators  acting in the Hilbert space of the theory, whose restrictions to the vacuum sector give maps in the corresponding cohomology theories. If $X$ is the Higgs branch of the theory with zero masses, the massive theory has $X^{\mathbf{A}}$ for its Higgs branch, i.e., the fixed point locus of the flavor symmetry torus $\mathbf{A}$ corresponding to the masses we switched on. Thus one gets maps going in both directions between the equivariant cohomology theories of $X$ and $X^{\mathbf{A}}$. The main claim of the current paper is that this is the physical realization of the stable envelopes of \cite{Maulik:2012wi,Aganagic:2016jmx}. Despite the simplicity of these ideas, making them even relatively precise involves understanding a lot of technical details, which is partly responsible for the length of this paper.

After reviewing the background material, such as the theories with eight supercharges in three, two, and one spacetime dimensions, in the Section \ref{sec:prel}, we scrutinize the first idea in the Section \ref{sec:VacAndCoh}. In this work, we deal  with de Rham cohomology, K-theory, and elliptic cohomology in parallel. This corresposponds to studying the vacua of a 1d theory on $\R$, 2d theory on $\R\times S^1$, or 3d theory on $\R\times \mathbb{E}_\tau$ (where $\mathbb{E}_\tau$ is an elliptic curve of complex structure $\tau$), with the supercharge $\cQ$ we introduce in Section \ref{sec:choice_of_Q}. Somewhat similarly, in studies of Bethe/Gauge correspondence, one works with the cohomology of a different supercharge $\cQ_A$. Our setup in 1d and 2d can be seen as dimensional reduction of the Bethe/Gauge setup, replacing the quantum cohomology/K-theory by their classical analogs. The 3d story is more subtle. 

At any rate, our setup is rich enough to see the constructions of stable envelopes and the corresponding quantum spectrum-generating algebras. We lift this setting to the full Bethe/Gauge correspondence and quantum cohomology/K-theory in the future work \cite{DN3}.

In Section \ref{sec:Januses} we construct the second main ingredient: the supersymmetric Janus interfaces. One can do this both for real masses and real Fayet-Iliopoulus (FI) parameters. The mass Janus plays central role in this paper, while the FI Janus, though not used here, will be featured more in part II. Both Janus interfaces have the property, which we refer to as the universality, that modulo $\cQ$-commutators they do not depend on the shape of the mass or FI profiles, they only depend on their asymptotic values. Furthermore, the dependence on the latter, when such asymptotic values are really large, is only captured by the chambers. Namely, the mass Janus depends on the chamber $\mathfrak{C}$ in the space of real masses, along which they are sent to infinity. The FI Janus similarly depends on the chamber $\mathfrak{C}'$ in the space of real FI parameters. When we choose different chambers $\mathfrak{C}_1$ and $\mathfrak{C}_2$ at $y=+\infty$ and $y=-\infty$ (throughout this paper, $y$ denotes the Euclidean time), this engineers the chamber R-matrices of \cite{Maulik:2012wi,Aganagic:2016jmx}, which will be the subject of part III. Here, we choose chamber $\mathfrak{C}$ on one side of the interface only, while the real masses vanish on the other. There are other interfaces, including those connecting Higgs and Coulomb branches, which we shall study later. 

In Section \ref{sec:SQM_section} we proceed to analyze the mass Janus in one-dimensional theories using the standard tools of the supersymmetric quantum mechanics \cite{Witten:1982im}. We show that the flows of the complexified flavor torus $\mathbf{A}_\C$ on $X$, which are involved in the construction of stable envelopes, are gradient flows for the Morse function typical to theories with four supercharges. This Morse function depends on the real masses $m$, which can be intuitively thought of as the external ``force'' fueling the flow. In case of theories with eight supercharges, the critical points of the Morse function all have the zero index. The corresponding flows do not lift the approximate vacua \cite{Witten:1982im}. However, if we let masses (and $f$) change over time, the gradient flows do play important role. 
In order to make the mass  (and, more generally, the Morse function $f$) change with time in a supersymmetric fashion, one adds a term $\sim\frac{\partial f}{\partial y}$ to the action. 

Then, something interesting happens.  The supersymmetric ground states, whose wave functions are, in the WKB approximation, peaked at the isolated classical massive vacua/critical points of $f$, become, actually, supported on the full repelling manifolds of those critical points/descending manifolds/Morse cells (cf. \cite{Frenkel:2006fy,Frenkel:2007ux,Frenkel:2008vz}). The latter are the unions of the trajectories (i.e. unparametrized images) of all the gradient flows starting at a given critical point of $f$. The subtle part of this definition concerns the contribution of the so-called broken flows. Another subtlety comes from the non-compact nature of  the repelling/attracting manifolds. We make the problem well defined by turning on the equivariant parameters. Indeed, one finds  non-trivial transition ampltiudes between the isolated vacua, transitions, induced by the changing mass.

In Section \ref{sec:StabEnv} we study the relation to stable envelopes in more detail. In particular, we argue that it is easier to compute them in the language of gauge theory (often referred to as the gauged linear sigma-model, or GLSM), rather than in the language of a non-linear sigma model. To this end, we show that in the limit of infinite large masses (in the chamber $\mathfrak{C}$), some degrees of freedom in the original GLSM $\cT$ freeze. What remains is called $\cT^{\mathfrak{C}}$, which breaks into a direct sum of quantum field theories $\oplus_p \cT^{\mathfrak{C}}_p$, labeled by the connected components $p\subset X^{\mathbf{A}}$ (i.e., isolated massive vacua when $X^{\mathbf{A}}$ is discrete). Each $\cT^{\mathfrak{C}}_p$ is itself a GLSM, whose Lagrangian description is canonically obtained from $\cT$. The $\cT^{\mathfrak{C}}$ is the theory whose Higgs branch is $X^{\mathbf{A}}$, and $\cT^{\mathfrak{C}}_p$ has the Higgs branch $p\subset X^{\mathbf{A}}$. The Janus interface construction leads to an interface between $\cT$ and each $\cT^{\mathfrak{C}}_p$, which also admits a simple Lagrangian description. We then proceed to compute matrix elements of such interfaces between vacua of $\cT$ and the single vacuum of $\cT^{\mathfrak{C}}_p$. This is done by replacing vacua with boundary conditions and computing the resulting interval partition functions on $(y_-,y_+)\times \mathbb{E}_\tau$. We then describe several examples and compare them to the known results in the literature, when available.

{\bf Acknowledgements} We are grateful to M.~Aganagic, C.~Closset, K.~Costello, E.~Frenkel, D.~Gaiotto, S.~Gukov, N.~Haouzi, S.~Jeong, Z.~Komargodski, N.~Lee, A.~Losev, G.~Moore, A.~Okounkov, A.~Smirnov, E.~Witten for discussions. MD is also grateful to A.~Okounkov for patience during the online course \cite{OkLect} on Enumerative Geometry and Geometric Representation Theory.

{\bf Note added:} in the process of completing this project we became aware of a related ongoing work of Mathew Bullimore and Daniel Zhang \cite{Bullimore:2021rnr}, and we are grateful to them for agreeing to coordinate the release.

\section{Preliminaries}\label{sec:prel}
\subsection{Gauge theories with eight and four supercharges}\label{sec:gauge_review}
We start by briefly reviewing the necessary facts about supersymmetric theories with eight supercharges \cite{Seiberg:1996bs,Seiberg:1996nz,Intriligator:1996ex,Hanany:1996ie,deBoer:1996ck,Kapustin:1999ha} in three, two, and one spacetime dimensions, commonly referred to as 3d $\cN=4$, 2d $\cN=(4,4)$ and 1d $\cN=8$ theories. We often view them as 3d $\cN=2$, 2d $\cN=(2,2)$, and 1d $\cN=4$ theories, respectively, with the other four supercharges possibly broken by (twisted) masses and/or background flat connections for a special symmetry denoted by $U(1)_\hbar$. It is a flavor symmetry of the theory with four supercharges, which is a part of the group of $R$-symmetry from the viewpoint of the theory with eight supercharges. Thus it commutes with four out of eight supercharges.

The structure of theories in question is rather uniform across the dimensions, so we start here by reviewing the three-dimensional theories, from which the 2d and 1d cases follow by the dimensional reduction. They are built from a 3d $\cN=4$ $\mathfrak{g}$-valued vector multiplet $\cV$ for some Lie group $G$, and a hypermultiplet $\cH$ valued in a quaternionic representation $\mathbf{R}$ of $G$. We only consider the theories of cotangent type, for which $\mathbf{R}=\cR \oplus \bar{\cR}$, where $\cR$ is a complex representation of $G$. From the 3d $\cN=2$ point of view, the vector multiplet $\cV$ decomposes into an $\cN=2$ vector $V$, 
and an adjoint-valued chiral multiplet $\Phi$; the hypermultiplet $\cH$ decomposes into an $\cR$-valued chiral multiplet $Q$, and a $\bar{\cR}$-valued chiral multiplet $\tilde{Q}$. The superpotential is:
\begin{equation}
W = \tilde{Q}\Phi Q.
\end{equation}
Global bosonic symmetries of the system include a flavor symmetry group $G_H$ (``H'' stands for Higgs), a Coulomb symmetry group $G_C$, and the $R$-symmetry group $SU(2)_H\times SU(2)_C$. The flavor group is defined as $G_H=N_{USp(\mathbf{R})}(G)/G$, where $N_{USp(\mathbf{R})}(G)$ is the normalizer of $G$ in the group of hyperkahler isometries of free $\mathbf{R}$-valued hypermultiplets. In all the theories that we study, $G_H$ acts on $\cR$ in some complex representation. Thus, throughout this paper, when we write
\begin{equation}
w\in\cR,
\end{equation}
we refer to the weights of the $G\times G_H$ action. When we need to emphasize the distinction between gauge and flavor weights, we write
\begin{equation}
(w,f)\in\cR,
\end{equation}
with $w$ being the $G$-weight and $f$ -- the $G_H$-weight.

Only the maximal torus of $G_C$ is visible in the UV: in 3d, each abelian gauge field $A$ gives rise to the current $J=*\dd A$ generating the ``topological'' $U(1)$,  such topological symmetries form the maximal torus of $G_C$. Thus, the rank of $G_C$ equals the dimension of the center $Z(G)$ of $G$. The $SU(2)_H$ and $SU(2)_C$ R-symmetries rotate complex structures of the Higgs and Coulomb branches respectively (they are hyper-K{\"a}hler). In the 3d $\cN=2$ language, there is only a $U(1)_R$ R-symmetry, which can be conveniently chosen as a diagonal subgroup of the product of the maximal tori:
\begin{equation}
U(1)_R = {\rm Diag } \left[U(1)_H\times U(1)_C\right] \subset SU(2)_H\times SU(2)_C,
\end{equation}
while the anti-diagonal is what we denote as $U(1)_\hbar$:
\begin{equation}
U(1)_\hbar = {\rm ADiag }\left[U(1)_H\times U(1)_C\right].
\end{equation}
To be more specific, if $R_H$ and $R_C$ are the Cartan generators, such that $Q$ has $R_H=\frac12$ and $\Phi$ has $R_C=1$, we define the $U(1)_R$ generator to be $R_H+R_C$, and the $U(1)_\hbar$ generator is defined as $R_H-R_C$.

The chiral multiplets $(Q,\tilde{Q})$ transform in some representation of the flavor group $G_H$, while none of the elementary fields transform under $G_C$. The only objects charged under $G_C$ are the disorder-type monopole operators \cite{Borokhov:2002cg,Borokhov:2003yu,Gaiotto:2008ak,Cremonesi:2013lqa}. We summarize charges under global symmetries in the following table:
\begin{table}[h]
	\centering
	\begin{tabular}{|c|c|c|c|c|}
		\hline
		3d $\cN=2$ multiplet & $G\times G_H$ & $G_C$ & U$(1)_\hbar$ & U$(1)_R$\\
		\hline
		Vector $V$ & \textbf{Adj} $\times \mathbf{1}$ & 0 & 0 & 0\\
		Chiral $\Phi$&\textbf{Adj} $\times \mathbf{1}$ & 0 & $-1$ & 1\\
		Chiral $Q$  & $\cR$ & 0 & $\frac12$ & $\frac12$\\
		Chiral $\tilde{Q}$& $\bar{\cR}$ & 0 & $\frac12$ & $\frac12$\\ 
		\hline
	\end{tabular}
	\caption{\label{tab:sym}Symmetry content.}
\end{table}

The vector multiplet $V$ contains $(A_\mu,\sigma, \lambda, \bar\lambda, D)$ -- a gauge field, a real scalar, a Dirac spinor, and a real auxiliary field. The chiral $\Phi$ contains $(\phi, \lambda_\phi,\bar\lambda_\phi, D_\C)$ -- a complex scalar, a Dirac spinor, and a complex auxiliary field; the chirals $Q$ and $\tilde{Q}$ have analogous components. The flat space SUSY transformations and actions are summarized in the Appendix \ref{app:conv}. We will use the following notations for the maximal tori and Cartan subalgebras:
\begin{align}
G_H:\quad &\text{Maximal torus } \mathbf{A},\quad \text{Cartan aubalgebra } \mathfrak{a}={\rm Lie}(\mathbf{A}),\cr
G_C:\quad &\text{Maximal torus } \mathbf{A}',\quad \text{Cartan aubalgebra } \mathfrak{a}'={\rm Lie}(\mathbf{A}'),\cr
G:\quad &\text{Maximal torus } \mathbf{H},\quad \text{Cartan aubalgebra } \mathfrak{h}={\rm Lie}(\mathbf{H}).\cr
\end{align}
Additionally, the torus of $\cN=2$ flavor symmetry group is denoted as
\begin{equation}
\bfT=\mathbf{A}\times U(1)_\hbar,\quad \mathfrak{t}={\rm Lie}(\bfT).
\end{equation}

\paragraph{Masses, FI, and CS couplings.} The standard way to generate masses is to give vevs to background vector multiplets. In 3d $\cN=4$ theories, this results in an $SU(2)_C$ triplet of masses $\vec{m}$ (valued in the Cartan of $G_H$) and an $SU(2)_H$ triplet of Fayet-Iliopoulos (FI) parameters $\vec{\zeta}$ (valued in the Cartan of $G_C$). From the $\cN=2$ point of view, the masses break into a real mass $M_\R$ and a complex mass $M_\C$: the real mass is a diagonal vev of $\sigma$ in the background vector multiplet gauging $G_H$, and the complex mass is realized via a superpotential term
\begin{equation}
W_M = \tilde{Q}M_\C Q,
\end{equation}
where $M_\C$ now is a complex mass matrix acting in the flavor group representation $\cF$. 

Likewise, from the $\cN=2$ point of view, FI parameters break into a real parameter $\zeta_\R$, and a complex one $\zeta_\C$. The real parameter analogously comes from the background vector multiplet for the topological symmetry, and is usually written in flat space as a Lagrangian coupling
\begin{equation}
\cL_{\rm FI}=i\text{Tr }(\zeta_\R D),
\end{equation}
where ``Tr'' only picks up components in the center of the gauge group. The complex FI term is another superpotential coupling:
\begin{equation}
W_{\rm FI} = i\text{Tr }(\zeta_\C\Phi).
\end{equation}
Notice that both complex masses $M_\C$ and complex FI parameters $\zeta_\C$ are charged under $U(1)_\hbar$. Since we are going to use this symmetry, we never turn on the complex parameters:
\begin{equation}
M_\C = \zeta_\C=0.
\end{equation}

As we said before, none of the fields in the Lagrangian are charged under the Coulomb branch symmetry $G_C$. Therefore, the ordinary couplings to the $G_C$ background vector multiplet, i.e. through the covariant derivatives of charged fields, are absent. Nonetheless, the $G_C$-vectors couple to the dynamical gauge multiplets via mixed $\cN=2$ Chern-Simons (CS), or BF, term. This term is of course responsible for the real FI parameter. Denote fields in the vector multiplet gauging the topological symmetry by $(A_\mu^{\rm top}, \sigma^{\rm top}, \lambda^{\rm top}, \bar\lambda^{\rm top}, D^{\rm top})$. Recall that the topological symmetry current is $J=*F$, where $F$ is an abelian field strength. The usual gauging procedure involves the current coupling $A_\mu^{\rm top} J^\mu =\frac12 A_\mu^{\rm top} \varepsilon^{\mu\nu\rho}F_{\nu\rho}$, thus producing the mixed CS term, whose $\cN=2$ SUSY completion, really a truncation of the full $\cN=4$ BF coupling \cite{Brooks:1994nn,Kapustin:1999ha}, is
\begin{align}
\label{SBF}
S_{\rm BF} = \frac{i}{2\pi}\int A^{\rm top}\wedge F + \frac{i}{2\pi}\int \dd^3 x\, \left( \sigma^{\rm top} D + D^{\rm top}\sigma -\frac12 \bar\lambda^{\rm top}\lambda -\frac12 \bar\lambda \lambda^{\rm top} \right).
\end{align}
We see from this action that giving a vev to $\sigma^{\rm top}$ generates a real FI term for the dynamical vector multiplet, as stated earlier. The coupling to $A^{\rm top}$ will also play an important role later in this paper. The bare 3d action does not include any other CS terms. While they are certainly possible \cite{Gaiotto:2008sd}, they go beyond the class of theories considered here.

\paragraph{Reduction to 2d and 1d.} The lower dimensional versions of the above theories can be obtained by reduction. In general, the reduction of a 3d $\cN=2$ theory to two dimension is very subtle, giving different 2d theories depending on how one scales parameters \cite{Aharony:2017adm}. It was also argued in \cite{Aharony:2017adm} that there exists a special scaling resulting in the ``same'' theory in 2d, i.e. the one admitting a UV completion by a purely 2d gauge theory, whose content is determined by a classical dimensional reduction. In principle, this is the only type of scaling we need, since we are only interested in purely 2d and purely 1d gauge theories constructed from the same gauge group and matter content as in 3d. 

Furthermore, we study 3d $\cN=4$ theories and their reductions to 2d and 1d. Before turning on the $U(1)_\hbar$ background, such theories preserve all eight supercharges, and do not allow for arbitrary superpotentials or twisted superpotentials in 2d and 1d. For such theories, the existence of scaling that preserves the structure of the gauge theory is even more clear, and we always assume it. One could only worry that once we turn on the $U(1)_\hbar$ background that breaks four supercharges, a non-trivial twisted superpotential is generated. Such a twisted superpotential can be made arbitrarily small by tuning the $U(1)_\hbar$ deformation parameter to sufficiently small values, and so its effect on the 2d and 1d physics is under control. Furthermore, we will see that there exists a proper scaling of the $U(1)_\hbar$ parameter (that will be introduced shortly), which allows to obtain the purely 2d (and purely 1d) answer out of the 3d computation.

Thus we can study the structure of 2d and 1d theories relying on the classical dimensional reduction. The general structure of multiplets is preserved, the only major difference being that the vector multiplet gains one more real adjoint-valued scalar $\sigma_1$ in 2d, and yet another one, denoted $\sigma_2$, after passing to 1d. The BF coupling \eqref{SBF} becomes an FI-Theta term in the twisted superpotential in 2d,
\begin{equation}
\tilde{W}_t=t\Sigma,
\end{equation}
where $\Sigma$ is the twisted chiral multiplet of the 2d vector multiplet, and $t=\frac{\Theta}{2\pi} + i\zeta$, where the theta-angle comes from the holonomy of $A^{\rm top}$ along the circle. If we further classically reduce to 1d, the theta-angle term disappears in the strict 1d limit, and only the real FI term survives. This can be summarized as follows:
\begin{equation}
\boxed{\text{3D: } S_{\rm BF}}\quad \longrightarrow\quad \boxed{\text{2D: } \tilde{W}_t}\quad \longrightarrow\quad \boxed{\text{1D: } \cL_{\rm FI}}
\end{equation}

In our applications, we will be studying three-dimensional theories on the space 
\begin{equation}
\R\times \bE_\tau, 
\end{equation}
where $\bE_\tau$ is an elliptic curve with the complex structure $\tau$ and Ramond spin structure along both circles (to preserve SUSY). We identify $\bE_\tau = \C^\times / \tilde{q}^\Z$, where $\tilde{q}=e^{2\pi i \tau}$. The A and B cycles are chosen so that for a constant differential $\dd {\rm z}$, we have
\begin{equation}
\int_{\rm A} \dd {\rm z}=1,\quad \int_{\rm B} \dd {\rm z} = \tau.
\end{equation}
We always define the 2d limit as $\tau\to0$, i.e., by shrinking the B cycle. In this limit, $\tilde{q}\to1$, or alternatively $q=e^{-\frac{2\pi i}{\tau}}\to0$.

For a non-degenerate elliptic curve, we turn on flat connections in the maximal tori of the global symmetry groups $G_H\times U(1)_\hbar$ and $G_C$. The flat connection $A^{\rm f}$ for the maximal torus $\bfT=\mathbf{A}\times U(1)_\hbar$ of the flavor group $G_H\times U(1)_\hbar$ is characterized by
\begin{equation}
\tilde{a} = \oint_B A^{\rm f} -\tau \oint_A A^{\rm f}.
\end{equation}
This is a doubly-periodic variable valued in $\mathfrak{t}_\C \mod Q^\vee\oplus \tau Q^\vee$, where $Q^\vee$ denotes the co-root lattice of $\bfT$, and $\mathfrak{t}_\C\equiv\mathfrak{t}\otimes\C$.

In the 2d limit $\tau\to 0$, the elliptic curve $\bE_\tau$ degenerates to a circle. We would like to take the limit in such a way that the low energy description coincides with that of the corresponding 2d gauge theory \cite{Aharony:2017adm}. In particular, the kinetic term normalization implies that $\int_B A\propto \tau\sigma_1$, $\int_B A^{\rm f}\propto\tau m_1$, where $\sigma_1$ is the new vector multiplet scalar, and $m_1$ is a mass. It is convenient to introduce a variable $a=\frac{\tilde{a}}{\tau}\in \mathfrak{t}_\C \mod Q^\vee\oplus \frac1{\tau} Q^\vee$ that remains finite in this limit, so
\begin{equation}
a =\frac1\tau \oint_B A^{\rm f} -\oint_A A^{\rm f}.
\end{equation}
Such rescaled variables are associated with the S-transformed elliptic curve $\bE_{-\frac1{\tau}}$, whose complex structure is $-\frac1{\tau}$. We also identify it as
\begin{equation}
\bE_{-\frac1{\tau}}=\C^\times/q^\Z,\quad \text{where } q=e^{-\frac{2\pi i}{\tau}}.
\end{equation}
The corresponding exponentiated elliptic variable is
\begin{equation}
X= e^{2\pi i a}.
\end{equation}
We prefer to work with $X$ (or $a$) and $q$ variables, as they are more convenient for addressing the 2d limit, which now becomes $q\to0$.

The variables $X$ will be interpreted as elliptic \emph{equivariant parameters} for the action of $\bfT$, but for now they are simly $\bfT$-valued flat connections on the elliptic curve. Because $\bfT=\mathbf{A}\times U(1)_\hbar$, we will split $X=(x,\hbar)$, and work only with such variables in what follows. They can be seen to parameterize the abelian variety of equivariant parameters:
\begin{equation}
\cE_\bfT =\left( \bE_{-\frac1{\tau}}\right)^{{\rm rk}(\bfT)}.
\end{equation}

In the 2d limit $\tau\to 0$, we can drop the second direct summand in the lattice $Q^\vee\oplus \frac1{\tau} Q^\vee$, and the variable $a$, or equivalently $X=(\hbar,x)$, becomes effectively cylindrical
\begin{equation}
(\hbar,x) \in (\C^\times)^{{\rm rk}(\bfT)}\quad \text{in the 2d limit.}
\end{equation}
Further reduction to 1d corresponds to shrinking the remaining circle, which also requires an analogous rescaling of variables. With a slight abuse of notations, the $\C^\times$-valued equivariant parameters $(x,\hbar)$ in this case should be replaced by $(e^{\epsilon x}, e^{\epsilon\hbar})$, where $\epsilon\to0$ is the 1d limit, and the exponents are interpreted as the $\C$-values equivariant parameters in 1d:
\begin{equation}
(\hbar,x) \in \C^{{\rm rk}(\bfT)} \quad \text{in the 1d limit.}
\end{equation}

In complete analogy, we also turn on a flat connection $A^{\rm top}$ along $\bE_\tau$ for the topological symmetry, i.e., the maximal torus $\mathbf{A}'$ of $G_C$. It is characterized by
\begin{equation}
\tilde\xi = \oint_B A^{\rm top} - \tau\oint_A A^{\rm top},
\end{equation}
which is a doubly-periodic variable in $\mathfrak{a}_\C'$, with periods $Q^\vee_C \oplus \tau Q^\vee_C$, where $Q^\vee_C$ is a co-root lattice of $\mathbf{A}'$. Again, the rescaled variable associated to the curve $\bE_{-\frac1{\tau}}$ is 
\begin{equation}
\xi = \frac{\tilde{\xi}}{\tau} \in \mathfrak{a}_\C' \mod Q^\vee_C \oplus \frac1{\tau} Q_C^\vee,
\end{equation}
and the exponentiated elliptic parameter is denoted 
\begin{equation}
z=e^{2\pi i \xi},
\end{equation}
which parameterizes $\cE_{\mathbf{A}'}=\left(\bE_{-\frac1{\tau}} \right)^{{\rm rk}(\mathbf{A}')}$. These $z$ will be identified with the elliptic \emph{K\"ahler parameters} later. As we can see, holonomies $z$ for the topological symmetries are analogous to flavor holonomies $X$ in most ways.

However, there is a distinction between the $X$ and $z$ holonomies in how we take the 2d and 1d limits. Recall that for flavor symmetries, $\int_B A^{\rm f}$ had to scale as $\tau$ in the limit $\tau\to0$, while $\int_A A^{\rm f}$ was kept constant (recall that in our conventions, in the 2d limit  B is the vanishing cycle). For the topological symmetry $\mathbf{A}'$, it is $\int_B A^{\rm top}$ that is kept constant, while the dependence on $\int_A A^{\rm top}$ disappears, and we can simply put it to zero. Let us elucidate this point.

Unlike the flavor gauge field, which explicitly enters covariant derivatives, the topological gauge field enters the action only through the BF term \eqref{SBF}, namely $\frac{i}{2\pi}\int A^{\rm top}\wedge F$. In the 2d limit, without any unusual scaling, the BF term 
becomes the $\Theta$-angle coupling for the abelian gauge field, with $\Theta=\int_B A^{\rm top}$. 
Thus, $\int_B A^{\rm top}$ is kept constant in the 2d limit. Naively, we get an additional term in 2d originating from $\int_{\R\times B}(\int_A A^{\rm top})F$. However, as $\int_B A \propto \tau\sigma_1$, we get the coupling
$\tau \int_{{\R} \times A} A^{\rm top} \wedge d{\sigma}_{1}$. Such a term is negligible in the $\tau\to 0$ limit.

If we simply put $\int_A A^{\rm top}=c$, then $\xi = \frac1{\tau}\int_B A^{\rm top} - \int_A A^{\rm top} = \frac{\Theta}{\tau} - c$, and the K\"ahler parameter becomes
\begin{equation}
z = e^{2\pi i\xi} = q^{-\frac{\Theta}{2\pi}}e^{-2\pi i c}.
\end{equation}
The expressions we will be getting involve ratios of the form $\frac{\vartheta(xz)}{\vartheta(z)}$, and in the $q\to0$ limit with $z=q^{-\Theta} e^{-2\pi i c}$, the $c$-dependence drops out. Therefore, we just put $c=0$, and choose the definition of the 2d limit as tending $q\to 0$, with the substitution
\begin{equation}
z = q^{-\frac{\Theta}{2\pi}},
\end{equation}
as well as scaling of the equivariant parameters explained earlier. This is exactly the kind of scaling that is used in the mathematical literature \cite{Okounkov:2016sya,Smirnov:2018drm}, where the Theta-angle is denoted by $s=\Theta$ and called a slope.

If we further take the 1d limit by shrinking the size of the remaining circle to zero, we again see a distinction from the case of flat connections for flavor symmetries. The parameter $z$, which starts its life in 3d as a flat connection for the topological symmetry, and reduces to a theta-angle in 2d, completely disappears in 1d. The 1d limit of the theta-term
\begin{equation}
\frac{\Theta}{2\pi}\int F
\end{equation}
is $\frac{\Theta}{2\pi}\int_\R \partial_y\sigma_2$, where $\sigma_2$ is the new real scalar that the 1d vector multiplet gains from the dimensional reduction. Again, proper normalization of its kinetic term eliminates this theta-like interaction in the 1d limit.

Finally, let us note that in the discussion above, there was no urgent need to consider both the $\bE_\tau$ and $\bE_{-\frac1{\tau}}$ elliptic curves: it is enough to formulate everything in terms of the $\bE_\tau$, and study the $\tau\to 0$ limit. However, this corresponds to the hard $\tilde{q}\to1$ limit, which is simplified by passing to the curve $\bE_{-\frac1{\tau}}$, leading to a much better behaved $q\to 0$ limit. This also makes contact with the conventions in the mathematical literature \cite{Aganagic:2016jmx,Okounkov:2016sya,Smirnov:2018drm}.

\paragraph{Boundary conditions.} 3d $\cN=4$ theories admit rich classes of half-BPS boundary conditions preserving either $\cN=(2,2)$ or $\cN=(0,4)$ supersymmetry at the boundary \cite{Gadde:2013wq,Okazaki:2013kaa,Chung:2016pgt,Bullimore:2016nji,Dimofte:2017tpi}. In the presence of a softly-breaking $U(1)_\hbar$ background, when only the $\cN=2$ subalgebra is left unbroken in the bulk, both classes of boundary conditions preserve $\cN=(0,2)$. Three-dimensional $\cN=2$ theories also admit $\cN=(1,1)$ boundary conditions, which break R-symmetry at the boundary. Depending on how one decomposes 3d $\cN=4$ multiplets into 3d $\cN=2$ multiplets, a given $\cN=(2,2)$ boundary condition can be understood both as an $\cN=(1,1)$ and $\cN=(0,2)$ boundary condition (the two interpretations related by the R-symmetry rotation in the bulk). However, the $\cN=(1,1)$ language makes certain symmetries (such as $U(1)_R$ and $U(1)_\hbar$) non-manifest at the boundary. As those symmetries are quite important to us, we have to work with the $\cN=(0,2)$ boundary conditions.

To describe $(0,2)$ boundary conditions, it is quite convenient, like in \cite{Dimofte:2017tpi}, to reformulate 3d $\cN=2$ multiplets as 2d $\cN=(0,2)$ multiplets valued in the infinite-dimensional target of the type ${\rm Maps}(\R,\dots)$. Here ellipsis represents the target of the original 3d multiplet, which is a gauge algebra for the vector multiplet, and its representation in the case of matter multiplets. In the 2d description, each 3d $\cN=2$ multiplet decomposes into a pair of 2d $\cN=(0,2)$ multiplets, and we concisely summarize this data in the following table:\\
\begin{table}[h]
	\centering
	\begin{tabular}{|c|c|c|c|}
		\hline
		3d $\cN=2$ multiplet & 1st 2d multiplet & 2nd 2d multiplet & special interactions  \\
		\hline
		Vector $V$ & vector $V_{\rm 2d}$ & adjoint chiral S & FI for V = FI for $V_{\rm 2d}$ \\ 
		\hline
		Chiral $\Phi$ & 2d chiral $\Phi$ & Fermi $\Psi_{\Phi}$ & 
		superpotential $J=\Psi_{\Phi}(\partial_y-S)\Phi$\\
		& & & superpotential $E=\frac{\partial W}{\partial\Phi}$\\
		\hline
	\end{tabular}
	\caption{2d $\cN=(0,2)$ description of 3d $\cN=2$ multiplets.}
	\label{tab:3d2d}
\end{table}\\
We used the standard notations for superpotentials: $W$ for the 3d $\cN=2$ and $J, E$ for the two superpotentials of the 2d $\cN=(0,2)$ theory \cite{Witten:1993yc}.

The description of boundary conditions is especially convenient in this language: for each 3d $\cN=2$ multiplet, one simply has to choose one of the two constituent $(0,2)$ multiplets, and make it vanish entirely at the boundary. Often a generalization is possible, where certain fields in this multiplet are instead set to some fixed non-zero values at the boundary. Such boundary deformations should be treated in the same sense as couplings in the Lagrangians: they may be given certain values in the UV, and then be subject to a nontrivial RG-flow on the way to the IR. 

The basic $\cN=(0,2)$ boundary conditions are constructed as follows:
\begin{align}
(0,2) \text{ Dirichlet for }V &\equiv \{V_{\rm 2d}\big|=0\},\\
(0,2) \text{ Neumann for }V &\equiv \{S\big|=0\},\\
(0,2) \text{ Dirichlet for }\Phi &\equiv \{\Phi\big|=0\},\\
(0,2) \text{ Neumann for }\Phi &\equiv \{\Psi_\Phi\big|=0\}.
\end{align}
The Dirichlet boundary condition for $V$ admits a generalization where the gauge field is fixed to be some non-trivial flat connection at the boundary. Similarly, the Dirichlet boundary condition for a chiral multiplet $\Phi$ admits a natural generalization, where the complex scalar is fixed to some covariantly constant value at the boundary. One can also contemplate the generalization of the Neumann boundary condition for $V$, in which the real scalar $\sigma$ in the vector multiplet, instead of zero, is given a covariantly constant value at the boundary. In non-abelian case, such a boundary vev ``Higgses'' the gauge group along the boundary.

The $\cN=(2,2)$ boundary conditions for a 3d $\cN=4$ vector multiplet $\cV=(V,\Phi)$ are constructed as follows:
\begin{align}\label{Dir22V}
(2,2) \text{ Dirichlet for } \cV &\equiv \{(0,2)\text{ Dirichlet for } V_{\rm 2d} \text{ and $(0,2)$ Dirichlet for } \Phi\},\\
(2,2) \text{ Neumann for } \cV &\equiv \{(0,2)\text{ Neumann for } V_{\rm 2d} \text{ and $(0,2)$ Neumann for } \Phi\}.
\end{align}
The $\cN=(2,2)$ boundary conditions for a 3d $\cN=4$ hypermultiplet $\cH$ are defined with the help of the complex symplectic geometry of the vector space $\sl V$ where $\cH$ takes values (there are also non-linear generalizations). Representing $\cH$ as a pair of chirals $(Q,\tilde{Q})$, the corresponding holomorphic symplectic form is 
\begin{equation}
\dd Q\wedge \dd \tilde{Q}
\end{equation}
with the understood natural pairing of vector spaces $\cR$ and $\bar\cR$, s.t. ${\sl V} = \cR\oplus\bar{\cR}$. Choose a Lagrangian splitting $L\oplus L^\perp$ of this complex symplectic space, for example $L = {\cR}$, $L^{\perp} = {\bar\cR}$, and denote the corresponding chiral multiplets by $(Q_L, \tilde{Q}_L)$. There are two basic $(2,2)$ boundary conditions associated with that splitting \cite{Bullimore:2016nji}:
\begin{align}
\cB_{Q_L} &\equiv \{(0,2) \text{ Neumann for } Q_L \text{ and $(0,2)$ Dirichlet for } \tilde{Q}_L\},\\
\cB_{\tilde{Q}_L} &\equiv \{(0,2) \text{ Dirichlet for } Q_L \text{ and $(0,2)$ Neumann for } \tilde{Q}_L\}.
\end{align}
These boundary conditions can also be generalized by giving non-trivial boundary vevs to scalars.

In this paper we study 3d theories on $\bE_\tau\times I$, where $I$ is an interval, subject to 
boundary conditions of the type described above. Although we only need $\cN=(2,2)$ boundary conditions and their $\cN=(0,2)$ versions, we also describe $\cN=(0,4)$ boundary conditions now, for the sake of completeness. The regular $\cN=(0,4)$ boundary conditions are constructed as follows:
\begin{align}
(0,4) \text{ Dirichlet for }\cV &\equiv \{(0,2) \text{ Dirichlet for } V_{\rm 2d} \text{ and $(0,2)$ Neumann for }\Phi\},\\
(0,4) \text{ Neumann for }\cV &\equiv \{(0,2) \text{ Neumann for } V_{\rm 2d} \text{ and $(0,2)$ Dirichlet for }\Phi\},\\
(0,4) \text{ Dirichlet for }\cH &\equiv \{(0,2) \text{ Dirichlet for both } Q \text{ and } \tilde{Q}\},\\
(0,4) \text{ Neumann for }\cH &\equiv \{(0,2) \text{ Neumann for both } Q \text{ and } \tilde{Q}\}.
\end{align}
The $\cN=(0,4)$ Dirichlet boundary conditions for the vector multiplet $\cV$ admit a generalization by the Nahm pole.

\paragraph{Moduli of vacua.} Three-dimensional supersymmetric theories have non-trivial moduli spaces of vacua that are not lifted quantum mechanically. They are stratified by the amount of unbroken gauge symmetry, which can be a maximal torus of the gauge group $G$ or less. The stratum that preserves the maximal torus of $G$ is called the Coulomb branch, and the stratum that breaks the gauge group entirely (up to, perhaps, a finite subgroup) is the Higgs branch. In between, there are mixed branches that preserve some portion of the maximal torus of $G$.

The Higgs branch can be described in the classical theory, as it is known to receive no quantum corrections. In gauge theory, the Higgs branch is the hyperK\"ahler quotient \cite{Hitchin:1986ea},, i.e. the space of $G$-equivalence classes of the space of
solutions to the real and complex moment map equations, ${\mu}_{\R} = 0, {\mu}_{\C} =0$:
\begin{equation}
\begin{aligned}
& \mu_\R  = Q Q^\dagger - \tilde{Q}^\dagger\tilde{Q}  - \zeta_\R, \\
&  \mu_\C  = Q\tilde{Q} - \zeta_\C.
\end{aligned}
\end{equation}
Thus, the Higgs branch is $\cM_H = \{\mu_\R=0,\ \mu_\C=0\}/G$. At $\zeta_\R=\zeta_\C=0$, it is singular; the real FI parameter $\zeta_\R$ resolves, the complex FI parameter $\zeta_\C$ deforms this singularity (this is of course relative to a specific complex structure of $\cM_H$). Solving the real moment map equation and dividing by $G$ can be traded for  quotient by a complexified gauge group $G_{\C}$ of the stable locus in the space of fields. On the smooth locus $\cM_H$
is also the K\"ahler quotient of the critical set of the superpotential $W$: 
\begin{equation}
\begin{aligned}
& {\cM}_{H} = \left( m^{-1}(0) \cap Crit W \right)/ G  \, , \\
& m = {\mu}_{\R} + [{\Phi}, {\Phi}^{\dagger} ] \, , \quad W = {\tr} {\Phi} {\mu}_{\C} 
\end{aligned}
\label{eq:n2form}
\end{equation}
where $\Phi$ is the complex adjoint scalar which belongs to the vector multiplet in the $\cN = 4$ formalism. The description \eqref{eq:n2form} is natural in the $\cN=2$ language (theory with four supercharges). 

In what follows we will keep $\zeta_\C=0$. 
The above description of the Higgs branch also applies to 2d $\cN=(4,4)$ and 1d $\cN=8$ theories (and to higher-dimensional theories with eight supercharges, but we do not consider them here). However, the physical meaning of the ``space of vacua'' is quite different in 2d and 1d. In three and higher dimensions, fixing the vevs of fields at infinity
forces the theory to sit in a specific vacuum corresponding to a point of the moduli space. In one and two dimensions it is not so. Instead, the ``moduli space of vacua'' only makes sense as a target space for an effective low energy nonlinear sigma model (NLSM). The wave functions of the true vacua spread over this space.

If we study a 3d theory on $\bE_\tau\times \R$ (or 2d theory on $S^1\times \R$), we can think of $\R$ as time, and the theory is one-dimensional macroscopically. Then the vacua behave as in 1d: instead of fixing a vev, we talk about the space of ground states. If at low energies the theory is described by an NLSM into some space $X$, then the space of vacua is an analog of harmonic forms on $X$ (in a proper sense that will be discussed later).

The ``branches of the moduli space of vacua'' now have the meaning of phases. The theory flows to an NLSM with the effective target space being a certain branch of the moduli space. Specifically which branch is selected is determined by what values the parameters (such as the FI parameters and masses) take. Changing the parameters may induce a change of branch, i.e., a phase transition. We will work with the theories that have the property that once the real FI parameters $\zeta_\R$ are generic enough, the theory is in the Higgs phase. Additionally turning on real masses $m_\R$ for a certain flavor symmetry from $\mathbf{A}\subset G_H$ reduces us to a submanifold of the Higgs branch, which is characterized as a fixed locus of that flavor symmetry. 

If we turn off the FI parameters $\zeta_\R$ while turning on the generic real masses $m_\R$, the theory is pushed onto the Coulomb branch (or Coulomb phase). When masses are non-generic or even zero (more generally, if both $\zeta_\R$ and $m_\R$ are non-generic), the theory can explore various mixed branches. The Coulomb branch of vacua is quite interesting on its own in that its classical description receives significant quantum corrections. Classically, it is parameterized by the triplet of commuting adjoint-valued scalars (from the vector multiplet), and the dual photon of the maximal torus of $G$. This allows to identify the classical Coulomb branch as $\mathfrak{h}_\C \times \mathbf{H}^\vee_\C / \cW = T^* \mathbf{H}^\vee_\C/\cW$, where $\cW$ is the Weyl group, and $\mathbf{H}^\vee_\C$ is the complexified dual torus of the maximal torus of $G$. Quantum-mechanical Coulomb branch is birational to this, but otherwise is quite different due to the quantum correction to its metric \cite{Seiberg:1996bs,Seiberg:1996nz,Intriligator:1996ex}. Mathematically, its construction is rather different from the Higgs branch. In the $m_\R=0$ case, the Coulomb branch is an affine variety defined as ${\rm Spec}(R_C)$, where $R_C$ is the Coulomb branch chiral ring. Turning on $m_\R$ corresponds to resolving singularities of this variety. The most non-trivial part in constructing the Coulomb branch is to identify the ring $R_C$, and there exists a few approaches to this problem, both in math \cite{Nakajima:2015txa,Braverman:2016wma,Braverman:2016pwk} and in physics \cite{Bullimore:2015lsa,Dedushenko:2017avn,Dedushenko:2018icp}.  
This of course only describes the complex structure the Coulomb branch. Of course, the Coulomb branch is lifted once the $U(1)_{\hbar}$
mass is turned on, producing an effective twisted superpotential, which can be computed exactly at one-loop \cite{DAdda:1978vbw,
	Witten:1993yc, Nekrasov:2009uh}.

Notice that in general, both Higgs and Coulomb branches are singular, and there might be not enough FI and real mass parameters in the theory to resolve one or another. In special classes of theories, however, they both admit smooth symplectic resolutions.

When the theory is in a Higgs phase in two dimensions, its low energy dynamics is that of a non-linear sigma model. The instantons of the latter
are the holomorphic maps of the worldsheet into the effective target space, the Higgs branch. It is well-known that the
point-like instanton singularities of the moduli space of holomorphic maps are resolved, in the gauge linear sigma model 
formulation, by the vortex-like solutions of the BPS equations, ``freckled'' instantons, where the instanton charge can be absorbed by a gauge flux \cite{Witten:1993yc, Witten:1993xi, Losev:1999nt}. Mathematically, these generalizations of  holomorphic maps
are called the ``quasi-maps'' \cite{CIOCANFONTANINE201417}. Their count \cite{Aganagic:2017gsx}, i.e. the computation of the partition function of the $\Omega$-deformed theory, 
is an alternative way of looking at the effective twisted superpotential, and the associated Bethe equations. We shall return to this problem in the part II of this series. 

\subsection{Quiver theories and quiver varieties}
An interesting class of 3d $\cN=4$ theories  are quiver gauge theories (for unitary gauge groups), which we now review. Their properties, especially in the context of Bethe/Gauge correspondence, are intimately connected to the Kac-Moody algebras encoded by the corresponding quivers.

Let $Q$ be a finite oriented graph with $\cI$ a set of vertices and $E$ a set of edges. 
For $e \in E$ let $s(e), t(e) \in \cI$ denote the source and the target of the edge $e$, respectively. For $k,l\in\cI$, let us denote the number of oriented edges joining $k$ and $l$  by $Q_{kl} = {\#} s^{-1}(k) \cap t^{-1} (l)$.

Take two collections of complex finite dimensional vector spaces labelled by the vertices: $(V_k)_{k\in\cI}$ and $(W_k)_{k\in\cI}$. The gauge group of the quiver theory is taken to be
\begin{equation}
G=\prod_{k\in \cI} U(V_k),
\label{eq:qgg}
\end{equation}
and the complex representation $\cR$ is chosen as
\begin{equation}
\cR = \bigoplus_{k\in\cI} {\rm Hom}(W_k, V_k) \oplus \bigoplus_{e \in E} {\rm Hom}(V_{s(e)}, V_{t(e)})\otimes {\C}^{Q_{s(e),t(e)}}.
\end{equation}
The quaternionic representation is then $T^*\cR = \cR\oplus \bar{\cR}$. We see that such theories have $w_k$ hypermultiplets in the fundamental representation of $U(V_k)$ for each $k$, also $Q_{kk}$ -- in the adjoint representation, and $Q_{kl}$ hypers -- in the bi-fundamental of $U(V_k)\times U(V_l)$ for each pair $k\neq l$ of vertices connected by an edge. The flavor symmetry group $G_H$ is a normalizer of $G$ in $USp(\cR)$ modulo $G$, 
\be
G_{H} = \left( \prod_{k,l \in \cI} \, U(Q_{kl}) \ \times \ \prod_{k \in \cI}\,  U(W_{k}) \right)/ \prod_{k \in \cI} U(1)
\label{eq:qfg}
\ee
and it contains factors like
\begin{equation}
G_H \supset \prod_{k\in\cI} SU(W_k),\quad \prod_{k\in\cI}U(Q_{kk}),\quad \prod_{k< l} SU(Q_{kl}).
\end{equation}
In the case of quivers of ADE type, the flavor symmetry $G_H$ is precisely given by $\prod_{k\in\cI} U(W_k) / U(1)^{\cI}$. More generally, e.g. for affine ADE quivers, $G_H = \prod_{k\in\cI} U(W_k) \times \prod_{e\in E} U(1)/ U(1)^{\cI}$. The topological symmetry, i.e., the maximal torus $\mathbf{A}'$ of the Coulomb branch symmetry group $G_C$ is
\begin{equation}
\mathbf{A}' = \prod_{k\in\cI} U(1),
\end{equation}
and may be enhanced to some non-abelian $G_C$ in the IR.

\subsection{Brane construction}

The gauge theories described above arise on the worldvolumes of D-branes in various string constructions \cite{Douglas:1996sw,
	Aganagic:2015cta,Aganagic:2017smx}. 
More specifically, the three dimensional ${\cN}=4$ supersymmetric quiver gauge theory, with the underlying graph being the affine Dynkin diagram of the $A,D,E$ type, with color 
is realized on the worldvolume of a stack of fractional $D2$ branes probing an ADE singularity ${\C}^{2}/{\Gamma}$ with 
${\Gamma}$ a finite subgroup of $SU(2)$,  McKay dual to the $A, D,E$ simple Lie algebra. 
The group $\Gamma$ defines the quiver, whose vertices are the irreducible representations ${\cR}_{i}$ of $\Gamma$, $\cI \cong {\Gamma}^{\vee}$. The vertices $i$ and $j$ are connected by $a_{ij}$ edges, where $a_{ij}$ is the multiplicity
\be
{\C}^{2} \otimes {\cR}_{i} = \bigoplus_{j} {\C}^{a_{ij}} \otimes {\cR}_{j}.
\ee
We consider the IIA string background with the spacetime of the form 
\be
M = Y \times N \times S,
\ee
with $Y \cong {\R}^{1,2}$ the three dimensional Minkowski space or its compactified versions ${\R}^{1,1} \times S^1$
or ${\R} \times T^2$, $N \cong {\R}^{3}$, and $S = {\C}^{2}/{\Gamma}$ or its resolution ${\tilde S}$ of singularities.

The $D2$ branes are spanning the submanifold $Y \times 0 \times 0 \subset M$ of the ten dimensional spacetime. 
The gauge group $G$ is given by \eqref{eq:qgg} with $v_k$ being the number of fractional $D2$ branes corresponding to the 
irrep $\cR_k$ of $\Gamma$. In other words, the Chan-Paton space living on the stack of $D2$ branes is a representation of $\Gamma$
\be
{\hat V} = \bigoplus_{k \in \cI}\, {\C}^{v_{k}} \otimes {\cR}_{k}.
\label{eq:cpg}
\ee
In addition to the \emph{color} $D2$
branes we have the \emph{flavor} $D6$ branes, wrapping $Y \times 0 \times S$. Since $S$ at infinity looks like the lens space $L = S^3/{\Gamma}$, the gauge group $U(W)$ of the $D6$ brane theory is broken down to a subgroup by the choice of asymptotic flat connection ${\rho}: {\Gamma} = {\pi}_{1}(L) \to U(W)$. The choice of the flat connection is the decomposition
\be
{\hat W} = \bigoplus_{k \in \cI} {\C}^{w_{k}} \otimes {\cR}_{k}
\label{eq:wsp}
\ee
of the Chan-Paton space of the $D6$ brane theory. The gauge group is, therefore, broken down to 
\be
G_{f} = \prod_{k \in \cI} U(w_k),
\ee
which is perceived by the degrees of freedom on the $D2$ branes as the flavour symmetry. 
There might be additional global symmetries, as well as $R$-symmetries, of geometric origin. 

Indeed, the spin cover $SU(2)$ of the rotational symmetry of $N$ acts as the $R$-symmetry of the quiver theory. 
At the orbifold point, ${\C}^{2}/{\Gamma}$ has another $SU(2)$ isometry for general $\Gamma$, and $SU(2) \times U(1)$
for the $A$-type $\Gamma = {\Z}_{r+1}$. 

Resolution of singularities ${\tilde S}$ is a hyperk\"ahler manifold. Passing from $S$ to ${\tilde S}$ is accompanied by turning on the Fayet-Illiopoulos terms on the $D2$ worldvolume. The geometric resolution of singularities can be measured by the periods 
\be
{\vec\zeta}_{i} = \int_{C_{i}} {\vec\omega}
\ee
of the triplet of symplectic forms integrated over the compact two-cycles $C_i \approx S^2$, which correspond to the vertices
$i \in {\cI} \backslash \{ 0 \}$ of a finite ADE graph. The string background is characterized not only by the geometric parameters ${\vec\zeta}_i$, but also by the periods 
\be
{\theta}_{i} = \int_{C_{i}} B
\ee
of the NSNS $B$-field, as well as the parameters ${\vec\zeta}_{0}$ corresponding to the trivial representation ${\cR}_{0}$. 
It appears to correspond to the unnormalized mode of the $B$-field 
\be
B \sim {\vec\zeta}_{0} \cdot {\vec\omega}.
\ee
The normalizable modes of the RR $3$-form $C \sim {\vec A}^{\rm top} \wedge {\vec \omega}$ produce the photons of the topological symmetry ${\bf A}^{\prime} \subset G_{C}$. String theory realizes $G_{C}$ as the enhancement of the gauge symmetry in six dimensions, which occurs in the limit of orbifold ${\tilde S}$ with vanishing $B$-field periods.  

In what follows we shall restrict the choices of ${\tilde S}$ to those having a $U(1)$ isometry preserving one of the
K\"ahler forms ${\omega}_{\R}$ and rotating the other two into each other, in other words, phase-rotating ${\omega}_{\C}$. 
In the $A$-type case we have an additional $U(1)$-isometry of ${\tilde S}$, preserving all three symplectic forms. It corresponds to the nontrivial $H_1$ of the quiver.

The supersymmetric interfaces we study in this paper correspond to more general supersymmetric profiles of the stacks of $D2$
and $D6$ branes. 

$D$-branes wrapping a special Lagrangian submanifold in a Calabi-Yau threefold lead to a supersymmetric theory. Suppose $S$ is a function on $Y$. It defines a Lagrangian submanifold $L$ in $Y \times N$ by:
\be
x_{\mu} = \frac{\partial S}{\partial y_\mu}\, , \ {\mu} = 1, 2, 3,
\label{eq:Sgf}
\ee
where $\vec x \in N$, $\vec y \in Y$. Then $D(p+3)$-branes wrapping $L$ carry supersymmetric gauge theory in $p+1$ dimensions, which can be thought of as hybrid topological-supersymmetric theory in $p+4$ spacetime dimensions. The ``internal'' three dimensional theory is topologically twisted so that the adjoint scalars $X_{\mu}$, ${\mu} = 1, 2, 3$ describing the normal bundle to the brane worldvolume inside $Y \times N$ are one-forms on $L$, combining with the gauge fields $A_{\mu}$ into a complexfied gauge field $A + {\ii}X$. One can actually view the $(p+4)$-dimensional theory on $L \times {\R}^{p,1}$ as the $(p+1)$-dimensional supersymmetric theory with an infinite dimensional gauge group ${\cG} = {\rm Maps}(L, G)$. The choice of $S$ in \eqref{eq:Sgf} is akin to a D-term condition for the group of volume preserving diffeomorphisms of $L$.

Starting with the affine $A,D,E$ quiver theory, we can Higgs one of the nodes, in particular $0$, to produce the theory corresponding to a finite $A,D,E$ Dynkin diagram. The fate of $D$-branes engineering this theory can be guessed
by using various string dualities. The fractional $D2$ branes become the ordinary $D4$ branes wrapping the compact cycles $C_i \subset {\tilde S}$. The flavor $D6$ branes become the $D4$ branes, wrapping non-compact $2$-cycles $\check{C}^{i}$,
as in \cite{Aganagic:2015cta} obeying
\be
\check{C}^{i} \cap C_{j} = {\delta}_{j}^{i}\, , \ C_{i} \cap C_{j} = a_{ij} - 2 {\delta}_{ij}.
\ee
It is possible to describe these cycles and the corresponding couplings rather explicitly in the $A_{r}$-case. 
The $A_{r}$-type ALE space ${\tilde S}_{r}$ is a Gibbons-Hawking manifold with the metric
\be
ds^2_{r} = V^{-1} \left( d{\tau} + {\theta} \right)^2. + V d{\vec r}^2,
\ee
with 
\be
V({\vec r}) = \sum_{i=0}^{r}  \frac{1}{| {\vec r} - {\vec r}_{i} |}\, , \ dV = \star_{{\R}^3} d{\theta}.
\ee
The map ${\vec r}: {\tilde S}_{r} \to {\R}^{3}$ is the hyperk\"ahler moment map, associated with the $U(1)$ HK isometry, generated by the vector field ${\partial}_{\tau}$. 

The compact two-cycles $C_i$, $i = 1, \ldots, r$, are the preimages under this map of the straight intervals connecting the poles ${\vec r}_{i-1}$ and ${\vec r}_{i}$,
\be
{\vec r} (C_{i}) = {\ell}_{i} = \left\{ {\, (1-t) \vec r}_{i-1} + t{\vec r}_{i} \, | \, 0 \leq t \leq 1 \, \right\}\ ,
\ee
while the non-compact $2$-cycles $\check{C}^{i}$ are the preimages of the semi-axis 
\be
L_i = \left\{ \, {\vec r}_{i} + ({\vec r}_{i} - {\vec r}_{0}) t \, | \,  t \geq 0 \, \right\} \ .
\ee
The normalizable $2$-forms $B_i \in H^{2}_{L^2}({\tilde S}_{r})$ are given by:
\be
{\omega}_{i} = da_i\, , \ a_i = V^{-1} V_{i} \left( d{\tau} + {\theta} \right) - {\theta}_i,
\ee
where
\be
V_{i} = \frac{1}{| {\vec r} - {\vec r}_{i} |}\, , \ dV_i = \star_{{\R}^3} d{\theta}_i,
\ee
and the Dirac string for $\theta_i$ is chosen, e.g., along $L_i$. The $B$-field is given by
\be
B = \sum_{i=1}^{r} b_{i} {\omega}_{i}
\ee

In wrapping $D4$-brane on $C_i$ we generate the effective $2+1$-dimensional theory with the gauge coupling
\be
\frac{1}{g_{i}^{2}} = \sqrt{|{\vec r}_{i-1} - {\vec r}_{i}|^2 + b_{i}^2}
\ee

\ 

String theory gives some perspective on the relations between the quiver theories in various dimensions.
Compactifying $Y$ and performing $T$-duality along the compact directions moves us down in spacetime dimension of gauge theory, 
see the Figure \ref{fig:strings} below. 
\begin{figure}[h]
	\centering
	\includegraphics[width=15cm]{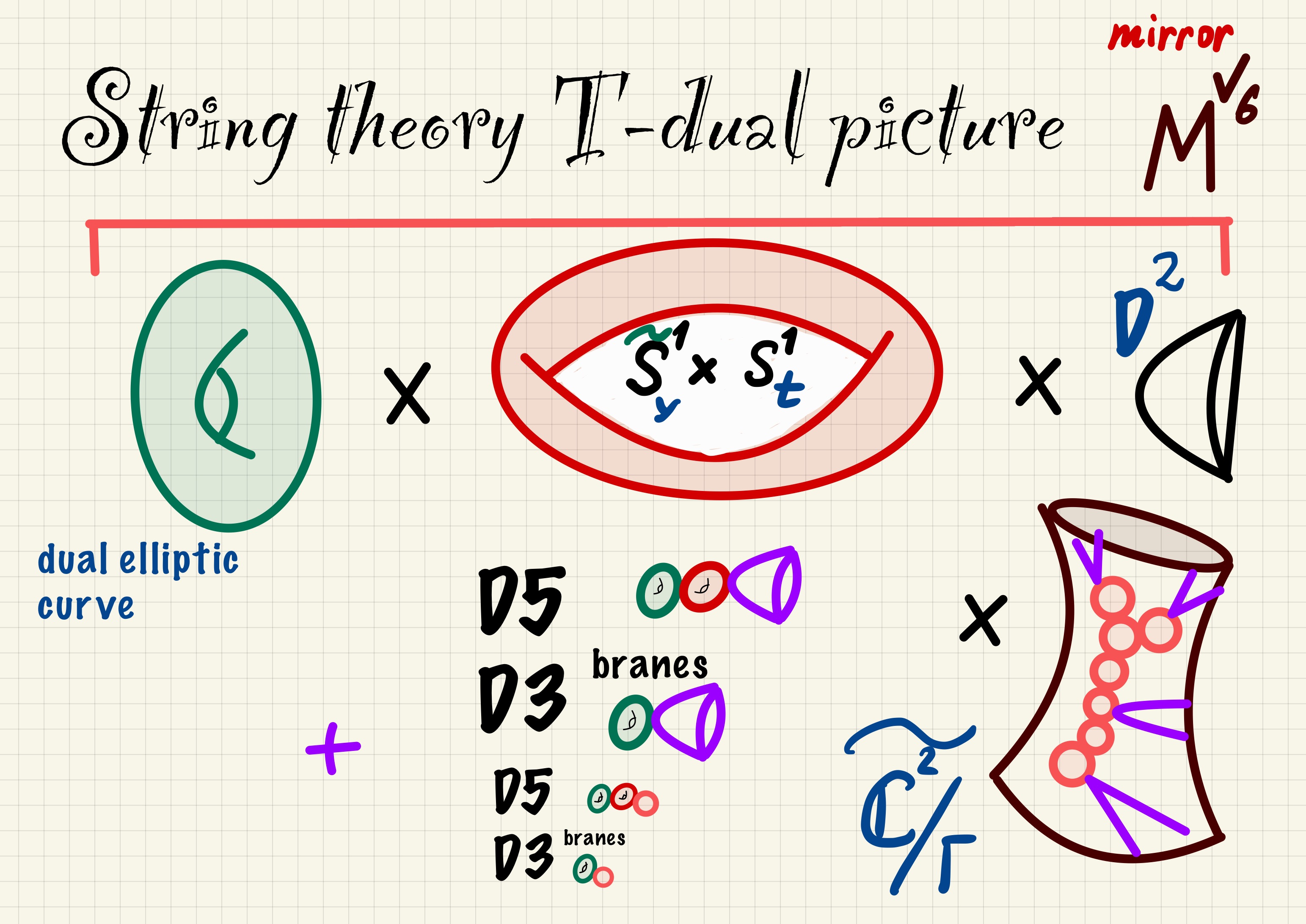}
	\caption{Artistic vision of the dual Type IIB setup.}\label{fig:strings}
\end{figure}\\
%\centerline{\includegraphics[width=15cm]{StringPicture.jpg}}

\section{Vacua and Q-cohomology}\label{sec:VacAndCoh}
In this section we describe the choice of supercharge in the 3d $\cN=2$ algebra, the connection between the Q-cohomology and the space of vacua, and how it is related to the generalized cohomology theories of the target space (in particular, the Higgs branch). There are three interesting supercharges, which can be seen as dimensional uplifts (from 2d ${\cN}=(2,2)$) of the A model supercharge $\cQ_A$, the B model supercharge $\cQ_B$, and the $\Omega$-deformed A or B model supercharge $\cQ$. The latter plays central role in this paper, but the A-type supercharge will also makes appearance in the follow-up work. Also note that while in general Omega-background deforms the action, the deformation vanishes if we study our theory on the cylinder, with the $U(1)$ isometry being just the cylinder rotation. In this case the action is the flat space one, and the ``Omega-deformation'' supercharge is simply a special linear combination of the Poincare supercharges. This is precisely our context.

\subsection{The choice of supercharges}\label{sec:choice_of_Q}
Consider again our 3d theory on $T^2 \times\R$ with Euclidean metric, along with its 2d and 1d versions.
The tower of reductions ${\rm 3d} \to {\rm 2d} \to {\rm 1d}$ accompanies our choice of the vector fields $V_{A}$ and $V_{B}$ on $T^2$, generating the action of $U(1) \times U(1)$ by isometries. The basis of $A$ and $B$ $1$-cycles in $H_{1}(T^{2}, {\Z})$ is represented by the closed orbits of $V_A$ and $V_{B}$, respectively. The 2d theory 
on $S^{1}_{A} \times {\R}$ is obtained by the dimensional reduction along $V_B$, so that $S_{A}^{1} = T^{2}/{\exp} \, {\R} V_{B}$,  while further reduction along $V_A$ leads to the definition of the 1d theory
on $\R$. The complex structure of $T^2$ is determined by the parameter ${\tau}$, so that $V_{B} - {\tau} V_{A}$ is an antiholomorphic vector field (it annihilates the holomorphic coordinate $z$), so that $T^2 \approx {\bE}_{\tau} = {\C}/\left( {\Z}+{\tau}{\Z} \right)$ . 

Choose a basis $\left( Q_{+}, Q_{-}, \bar{Q}_{+}, \bar{Q}_{-}\right)$ of 3d $\cN=2$ supercharges, such that a pair $(Q_+, \bar{Q}_+)$ anticommutes to $V_{B} - {\tau} V_{A}$. We say that this basis is \emph{adapted} to the complex structure of $\bE_\tau$. In terms of the adapted basis, a $(0,2)$ boundary condition along $\bE_\tau$ (at some fixed $y\in\R$) preserves $(Q_+,\bar{Q}_+)$. If we reduce to two dimensions, we are left with $S^1_A\times\R$. In this context, a complex structure on $S^1_A\times\R$, canonically determined by its Euclidean metric and a choice of orientation, becomes important. Thus we might also choose a different basis of 3d $\cN=2$ supercharges, $(q_+,q_-,\bar{q}_+,\bar{q}_-)$ adapted to the complex structure on $S^1_A\times \R$, meaning  $(q_+,\bar{q}_+)$ anticommute to the antiholomorphic vector field along $S^1_A\times\R$. Its relation to the basis $(Q_\pm,\bar{Q}_\pm)$ is a simple $\frac{\pi}{2}$ rotation:
\begin{align}
q_\pm &= \frac1{\sqrt{2}} (Q_+\pm Q_-),\cr
\bar{q}_\pm &=\frac1{\sqrt{2}}(\bar{Q}_+ \pm \bar{Q}_-).
\end{align} 
This basis conveniently matches the standard 2d conventions on $S^1_A\times\R$, in terms of which the definitions of the A and B model supercharges \cite{Witten:1988ze,Witten:1991zz} are easily lifted to 3d:
\begin{align}
\cQ_A &= \bar{q}_+ + q_-,\cr
\cQ_B &= \bar{q}_+ + \bar{q}_-.
\end{align}
In terms of $(Q_\pm,\bar{Q}_\pm)$, the supercharge $\cQ_A$ looks rather generic, while the B supercharge is fairly simple,
\begin{equation}
\cQ_B=\sqrt{2}\,\bar{Q}_+.
\end{equation}
This one is part of the 2d $\cN=(0,2)$ subalgebra that preserves $\bE_\tau \times \{y\}\subset \bE_\tau\times\R$, and can be identified with the supercharge of the holomorphic-topological twist in 3d, as considered in \cite{Costello:2020ndc}. 
It was considered earlier in \cite{NikThesis} in the studies of partially twisted ${\cN}=1$ four dimensional theories 
on ${\bE}_{\tau} \times {\Sigma}$. 

The 3d A-type supercharge is also known in the literature in the context of 3d A-twist \cite{Benini:2015noa,Benini:2016hjo,Closset:2016arn,Closset:2017zgf}, and was also considered earlier in \cite{Baulieu:1997nj} in the framework of the general $d, d+1, d+2$ towers of mixed topological/holomorphic theories.

The Euclidean path integral does not care about which coordinate is called time. However, once we pick a direction and Wick rotate it to the Minkowski signature, we gain extra structure. In fact, the physical unitary QFT is defined in Minkowski signature with
a Hilbert space (carrying a unitary inner product) assigned to a space-like slice. With respect to its unitary inner product, there is a notion of conjugation, so that the supercharges obey the following relation
\begin{equation}
B^{-1}Q^\dagger = \bar{Q},
\end{equation}
where we suppressed spinor indices, and $B$ is a matrix encoding how gamma matrices behave under the complex conjugation. The choice of $B$ depends on which direction is regarded as timelike, the standard relation being $B=iC\gamma^0$ (here $C$ is a charge conjugation matrix, see \cite{Freedman:2012zz} for more details).

In our case, if we Wick rotate the $y$ coordinate into time $t$, then the space of
states on the spatial slice $\bE_\tau$ becomes a Hilbert space
which we denote by
\begin{equation}
\cH[\bE_\tau].
\end{equation}
In our conventions, denoting the corresponding conjugation operation by $(\dots)^\dagger$, this choice results in the following relation among the supercharges:
\begin{equation}
Q_{\pm}^\dagger = \bar{Q}_{\mp}.
\end{equation} 
This can be easily understood: the $\pm$ notation here refers to the chirality along $\bE_\tau$, and since it has Euclidean signature, the chirality changes sign under the complex conjugation. Notice that with ${\bE}_{\tau}$ interpreted as a spatial slice, the boundary conditions along $\bE_\tau$ coincide with the \emph{initial conditions} in the path integral.

Another choice is to declare $V_A$ as the vector field generating the Euclidean time translations. 
By cutting $T^2$ open along $S^1_B$ and Wick rotating into Minkowski signature, we get a Hilbert space associated to the spatial cylinder $S^1_B\times \R$. The conjugation following from this choice will be denoted as $(\dots)^{\tilde\dagger}$. We find the following relation:
\be
\label{altConj}
Q^{\tilde\dagger}_{\pm} = \bar{Q}_{\pm}\ .
\ee
The $\pm$ notation here still refers to the 2d chirality along the plane corresponding to the $\bE_\tau$ directions. However, now it has Minkowski signature, so the chirality does not change under the conjugation, confirming \eqref{altConj}. With this choice of time direction, the boundary conditions along $\bE_\tau$ have the more conventional meaning of actual \emph{boundary conditions} on a timelike slice, as opposed to initial conditions. 

The latter interpretation is more familiar in the literature \cite{Hori:2000ck,Kapustin:2001ij}. In the 2d context, one usually defines A and B branes as those preserving the corresponding supercharges:
\begin{align}
\text{A-branes preserve: }\quad &\cQ_A \text{ and } (\cQ_A)^{\tilde\dagger}=q_+ + \bar{q}_-,\cr
\text{B-branes preserve: }\quad &\cQ_B \text{ and } (\cQ_B)^{\tilde\dagger}=q_+ + q_-.\cr
\end{align}
In particular, we see that in 3d notations, $\left( {\cQ}_{B}\, , \, (\cQ_B)^{\tilde\dagger} \right)$ correspond to $(\bar{Q}_+, Q_+)$. Hence, the $B$-branes lift to the $\cN=(0,2)$ boundary conditions in 3d ${\cN}=2$. It is easy to check that $\left( {\cQ}_{A}, ( {\cQ}_{A})^{\tilde\dagger} \right)$ correspond to the ${\cN}=(1,1)$ subalgebra, i.e., $A$-branes lift to the $\cN=(1,1)$ boundary conditions in 3d $\cN=2$.

Yet another interesting supercharge, simply denoted  by $\cQ$, is preserved by both $A$ and $B$ branes:
\begin{equation}
\cQ = \frac1{\sqrt 2}(\cQ_A+(\cQ_A)^{\tilde\dagger}) = \frac1{\sqrt 2}(\cQ_B+(\cQ_B)^{\tilde\dagger}).
\end{equation}
In 3d notation, this supercharge is identified with
\begin{equation}
\cQ=Q_+ + \bar{Q}_+.
\end{equation}
It indeed belongs to both the $\cN=(0,2)$ and $\cN=(1,1)$ subalgebras.  In three dimensions, when acting on fields, it squares to the covariant antiholomorphic derivative in the $\bE_\tau$ direction:
\begin{equation}
\cQ^2 = 2i D_{\bar z} \ . 
\end{equation}
Note that this supercharge depends on the choice of complex structure of $E_\tau$.
Since $\cQ$ squares to a spacetime symmetry we shall call it an $\Omega$-deformed supercharge. After passing to two dimensions, $D_{\bar z}$ becomes the sum of an infinitesimal isometry of $S^1_A\times \R$ plus a central charge. Indeed, such supercharge is used to define the $\Omega$-deformed theory in 2d.

To summarize, we have three interesting supercharges in 3d $\cN=2$: the lift of an $A$-model supercharge $\cQ_A$, the lift of a $B$-model supercharge $\cQ_B$, and the lift of the $\Omega$-deformed supercharge $\cQ$. We mostly work with $\cQ$, but also occasionally use $\cQ_A$. Although $\cQ$ is consistent with both A and B branes, we will only use the B branes, as the A branes break the $U(1)_\hbar$.

\subsection{Properties of $\cQ$, $\cQ_A$ and their cohomology}

We are going to act on the Hilbert space $\cH[\bE_\tau]$ with operators, dressed by the Euclidean time evolution operators, i.e. compressors, as opposed to the unitaries.  

Pick a complex supercharge $Q$. If it is nilpotent, $Q^2=0$, we may study its cohomology in $\cH[\bE_\tau]$, or, more generally, if $Q^2=\cZ\neq 0$,  the cohomology of $Q$ on ${\rm ker}{\cZ}$,  as is done mathematically in the context of equivariant cohomology. 
By the usual Hodge theory argument, each $Q$-cohomology class has a harmonic representative, i.e. an element of $\ker \{Q,Q^\dagger\}$, where $\dagger$ uses the Hilbert space structure on $\cH[\bE_\tau]$. We may want to find a harmonic representative uniformly across spacetime dimensions, namely also in the Hilbert space $\cH[S^1]$ of the 2d theory, and $\cH[\cdot]$, or simply $\cH$, of the 1d theory, with the corresponding supercharges $Q$ descending from the one in 3d. What is the interpretation of $Q$-cohomology? Let us first recall how this works in the most basic case of the $\cN=2$ quantum mechanics. There are only two supercharges, which can be taken as conjugates of each other:\footnote{More precisely, this is $\cN=(1,1)$ quantum mechanics. In the $\cN=(0,2)$ quantum mechanics, a central charge may appear.}
\begin{equation}
\{Q,Q^\dagger\}\sim H,
\end{equation}
where $H$ is the Hamiltonian. Thus the cohomology is identified with the space of ground states of $H$ \cite{Witten:1982im}. If the theory is formulated as an NLSM into some manifold, then it is identified with the usual de Rham cohomology of that manifold when $\cZ=0$, and with the equivariant de Rham cohomology when $\cZ$ is a nontrivial isometry of the target 
%\footnote{Note that the identification is exact, since we consider a complex target (in a generic real case, there are instanton corrections \cite{Witten:1982im}).}
.

{}We wish to apply this to our two supercharges, $\cQ$ and $\cQ_A$. First of all, we check that
\begin{eqnarray}
\{\cQ,\cQ^\dagger\} = \{\cQ_A,(\cQ_A)^\dagger\}=2H, 
\end{eqnarray}
where $H$ is the Hamiltonian acting on $\cH[\bE_\tau]$, i.e., the generator of time translations along $\R$. Thus, the cohomology with respect to either of the two supercharges is identified with the space of supersymmetric vacua in $\cH[\bE_\tau]$. We can express this fact as the isomorphism of vector spaces
\begin{equation}
H_{\cQ}(\cH[\bE_\tau])\cong H_{\cQ_A}(\cH[\bE_\tau])\cong \mathscr{V}_0,
\end{equation}
where $\mathscr{V}_0$ is the space of ground states. However, as we will see, there are additional structures on $H_{\cQ}(\dots)$ and $H_{\cQ_A}(\dots)$, which are not quite the same. Yet, they turn out to be completely equivalent upon the reduction to 1d.

As already mentioned earlier, the $\cQ$ supercharge, when acting on various fields, squares to the anti-holomorphic covariant derivative along $\bE_\tau$:
\begin{equation}
\label{equiQ}
\cQ^2 = 2iD_{\bar z}.
\end{equation}
Thus $\cQ$-cohomology should be understood in the equivariant sense, meaning taking the $\cQ$-cohomology
on the space of states, covariantly holomorphic with respect to both global and local symmetries\footnote{Physically, setting ${\cZ}$ to zero is  forced on us by the BPS inequality, since the eigenvalues of $H$ are bounded below by some multiple of the norm $\Vert {\cZ} \Vert$.}.  Let us now introduce a background flat flavor connection $A^{\rm f}$ on $\bE_\tau$.  It is gauge equivalent to a connection represented by a constant $1$-form valued in the maximal torus of the flavor group. Since it appears in the covariant derivative, it shifts the above equation by $-A^{\rm f}_{\bar z}$. This $A^{\rm f}_{\bar z}$ should be thought of as the flavor symmetry equivariant parameter. Notice that, being a flat connection on $\bE_\tau$, this parameter is an elliptic variable, as was already discussed in Section \ref{sec:gauge_review}. There, we characterized it by
\begin{equation}
a = \frac{1}{\tau} \oint_B A^{\rm f} -\oint_A A^{\rm f}.
\end{equation}
The exponentiated variable was denoted $X$, and then separated into $x$, corresponding to the $\cN=4$ flavor group $G_H$, and $\hbar$, corresponding to $U(1)_\hbar$. So indeed, these variables play the role of equivariant parameters. 

As a vector space, the $\cQ$-cohomology is the space of vacua, but working over the family of backgrounds parametrized by $A^{\rm f}$ corresponds to the mathematical equivariant elliptic cohomology.  Passing to 2d or 1d does not affect this isomorphism, but it affects the periodicity of equivariant parameters; the equation \eqref{equiQ} gets modified upon reduction to two and one dimensions. In 2d it becomes:
\begin{equation}
\label{Q2d}
\cQ^2 = -(\sigma_1+D_\varphi),
\end{equation}
where $\varphi$ parameterizes the remaining $S^1_A$ circle, and like in Section \ref{sec:gauge_review}, $\sigma_1$ is an extra real scalar in 2d. In 1d we earn another scalar $\sigma_2$, and the equation \eqref{equiQ} becomes\footnote{Not to be confused with Pauli matrices}:
\begin{equation}
\label{1dcQ}
\cQ^2 = -(\sigma_1 + i\sigma_2).
\end{equation}
When we turn on the background for flavor symmetries, these $\sigma_1$ and $\sigma_2$ get shifted by the masses $m_1$ and $m_2$. In the two-dimensional case \eqref{Q2d}, we have a flavor flat connection $A^{\rm f}_\varphi$ instead of $m_2$. Altogether, we find a $(\C^\times)^{{\rm rk}(\mathbf{T})}$-valued equivariant parameter $\exp\left(m_1 + i\oint_{A} A^f\right)$ in two dimensions, and a $\C^{{\rm rk}(\mathbf{T})}$-valued equivariant parameter $m_1 + im_2$ in one dimension. They originate from the $(\bE_{-1/\tau})^{{\rm rk}(\bfT)}$-valued equivariant parameters $(x,\hbar)$ in 3d. The reader might compare it to the discussion of Section \ref{sec:gauge_review}.

As for the $\cQ_A$ supercharge, its equivariant parameter is slightly different. In three dimensions, we find:
\begin{equation}
\label{QAsqrd}
\cQ_A^2=-i(\sigma_3+D_\alpha),
\end{equation}
where $\sigma_3$ is the real scalar from 3d $\cN=2$ vector multiplet, $D_\alpha$ is a covariant derivative along the $B$-cycle (which shrinks in the 2d limit), and we slightly abuse the notation (again) by simply writing $\sigma_3$, which really means the action by $\sigma_3$ in the appropriate representation. Turning on the flavor background corresponds to shifting $\sigma_3$ by the real mass $m_3$, and turning on a flavor holonomy $\oint_B A^{\rm f}$. As a result, we obtain a $(\C^\times)^{{\rm rk}(\mathbf{T})}$-valued equivariant parameter $\exp\left(m_3 + i\oint_B A^{\rm f}\right)$, but now in the context of 3d theory, unlike for the $\cQ$ supercharge.

Passing to two dimensions, we get:
\begin{equation}
\label{2dAmod}
(\cQ_{A})^2 = -i(\sigma_3+i\sigma_1),
\end{equation}
which is a familiar property of the A-model supercharge. The corresponding equivariant parameter for flavor symmetry $m_3+{\ii}m_1$ is $\C^{{\rm rk}(\mathbf{T})}$-valued.

Descending further to 1d does not change the Eq. \eqref{2dAmod}, as it already contains affine equivariant parameters. However, it becomes similar to the equation \eqref{1dcQ}, which is also written in the 1d context: we see that the $R$-symmetry rotation in the $(\sigma_2,\sigma_3)$ plane of the $(\sigma_1,\sigma_2, \sigma_3)$ space relates the two supercharges. Therefore, the distinction between $\cQ$ and $\cQ_A$ disappears in 1d, as claimed before. 

For completeness, let us briefly mention what observables can be found in the cohomology of the space of local operators for each of the three supercharges in 3d. For $\cQ_B$, there are local observables that form the VOA \cite{Costello:2020ndc}, as well as possibly extended operators (Wilson and vortex lines, surface operators). The cohomology of $\cQ_A$ contains line operators wrapping the $S^1_B$, as follows from \eqref{QAsqrd}. The basic such line is a Wilson loop (in the representation $R$):
\begin{equation}
\label{QALoop}
\Tr_R \ {\rm Pexp} \oint_{S^{1}_{B}}  \left( \sigma\, \dd\alpha + i A \right),
\end{equation}
but more complicated lines (in general defined by coupling to the worldline quantum mechanics \cite{Assel:2015oxa,Dey:2021jbf}) also exist. One can also consider $\cQ_A$-invariant surfaces (interfaces) defined by coupling to some 2d $\cN=(1,1)$ theory. Finally, because $\cQ^2 = 2iD_{\bar z}$, observables in the $\cQ$-cohomology must be invariant under the $\bar{z}$-translations, which means they can only be surfaces (interfaces) wrapping the $\bE_\tau$. A large class of such observables is defined by coupling some 2d $\cN=(0,2)$ degrees of freedom on the interface to the bulk. As we will see very soon, other ways to build $\cQ$-closed interfaces also exist, and we will explore relations between different constructions in the future work.

\subsection{Vacua and generalized cohomology}\label{sec:gen_coho}

Let us clarify the geometric meaning of the $\cQ$ and $\cQ_A$ cohomology. We start with the simplest case -- that of the 1d theory. There, $\cQ$ and $\cQ_A$ are equivalent (related by an R-symmetry rotation), while the corresponding equivariant parameters are ${\C}$-valued. This is an ${\cN}=8$ quantum mechanics, broken to ${\cN}=4$ by the $U(1)_{\hbar}$ twisted mass. Its low-energy description is given by the quantum mechanics on the Higgs branch $X$, sometimes called a 1d non-linear sigma model (NLSM). Thinking of it as of the $\cN=2$ theory, its Hilbert space is identified with the de Rham complex $\Omega^\bullet(X)$. Thinking of the family of theories parametrized by the twisted masses and/or background flat connections for flavor symmetries, the space of states can be viewed as
$\Omega^\bullet(X)\otimes S^\bullet\mathfrak{t}$, more natural space in the equivariant setting. Were $X$ compact, its $\cQ$-cohomology would be identified with the de Rham cohomology $H^{\bullet}(X)$  \cite{Witten:1982im}. In our story $X$ is never compact, yet, thanks to the equivariant setting, the $\cQ$-cohomology 
is identified with the equivariant de Rham cohomology $H^\bullet_\bfT (X)$, assuming the fixed point set 
$X^{\bf T}$ is compact (it is compact for generic flavor equivariant parameters). 

Moving up in dimension, we already know $\cQ$ and $\cQ_A$ are not equivalent in 2d. In particular, $\cQ_A$ is the ordinary $A$-model supercharge. Its equivariant parameter is still $\C$-valued, and its cohomology is interpreted mathematically as quantum equivariant cohomology of $X$, which is simply $H^\bullet_\bfT(X)$ with the deformed ring structure. The $\cQ_A$-cohomology of local operators is the twisted chiral ring, or the quantum cohomology ring of $X$. The $\cQ_A$-cohomology of states is just the space of Ramond vacua on the circle. We can generate the $\cQ_A$-cohomology classes in $\cH[S^1]$ by starting with the boundary states $|\cB\rangle$ for A-branes $\cB$, then evolving them in Euclidean time $T>0$, thus regularizing the otherwise unphysical states $|\cB\rangle$ that initially do not belong to $\cH[S^1]$ (see Appendix \ref{app:bdry_states} for a discussion). If one generates boundary conditions by coupling to the Chan-Paton vector bundles, one obtains a map from the equivariant K-theory into the $\cQ_A$-cohomology of $\cH[S^1]$, i.e., ${\rm ch}: K_\bfT(X) \to H_\bfT(X)$.

The $\cQ$ supercharge, however, has $\C^\times$-valued equivariant parameters in 2d. It is thus more natural to think of its cohomology as $K_\bfT(X)$. There is a map from $K_\bfT(X)$, thought of as classifying boundary conditions in 2d, to the $\cQ$-cohomology of states. Both this map and ${\rm ch}$ above involve quotient by the torsion, as the space of vacua is simply the $\C$-vector space. When the fixed points $X^\bfT$ of the $\bfT$-action on $X$ are isolated, one trivially sees that the space of vacua is isomorphic to $K_\bfT(X)\otimes\C$. Indeed, on the one hand, the fixed points are in one-to-one correspondence with the massive vacua (that exist when we turn on generic large equivariant parameters). On the other hand, the equivariant localization ensures that fixed points provide a basis in $K_\bfT(X)\otimes\C$. Thus the isolated massive vacua can be thought of as the basis in $K_\bfT(X)\otimes\C$. This argument also applies to $H_\bfT(X)$, the only apparent distinction between the two cases being the $\C$ or $\C^\times$ valuedness of the equivariant parameter. $H_\bfT(X)$ and $K_\bfT(X)$ also differ in the structure of pushforward, which is especially transparent in the ``sigma-model'' view on K-theory, where we interpret $K_\bfT(X)$ as the $S^1$-equivariant cohomology of the loop space, $H_{\bfT\times S^1}(LX)$. Here the circle group acts by rotating the loops, its equivariant parameter naturally identified with $\frac{1}{R}$, the KK mass. Then pushforward of a K-theory class with respect to a map $X\to Y$ corresponds to the pushforward of the cohomology class via the map $LX \to LY$. The latter means integrating over the loop space $L ($ a fiber of $X\to Y)$, which produces the relative ${\hat A}$-genus.

The explanation via the fixed points does not quite work when the fixed locus $X^\bfT$ is positive-dimensional. However, we assume that the space of vacua is still $K_\bfT(X)\otimes\C$ in such cases, at least when $X^\bfT$ is compact. This assumption follows from the sigma-model viewpoint sketched earlier. Also, shrinking the circle corresponds to the 1d limit, in which we know rigorously that the space of vacua is $H_\bfT(X)$. The K-theory, intuitively, degenerates to the cohomology in the limit.

Now move to three dimensions. In fact, we could start there and argue that the rest follows by reduction. In 3d, the supercharge $\cQ_A$ has $\C^\times$-valued equivariant parameters, and its operator cohomology contains line operators, such as Wilson loops \eqref{QALoop}. Their product gives rise to the quantum equivariant K-theory of $X$, which is the same as $K_\bfT(X)$ with the deformed product. It appears that this suggestion was first made in \cite{Kapustin:2013hpk}. It was studied in \cite{Jockers:2018sfl} (see also \cite{Jockers:2021omw}), where the main idea was to compute the cigar index $C\times_\varepsilon S^1$ in 3d, write the difference operators that insert $\cQ_A$-closed Wilson loops (like \eqref{QALoop}) at the tip of the cigar, and identify this whole structure with what is found in the quantum (permutation-symmetric) K-theory of \cite{Givental:2017fng}. Observables are inserted at the tip of the cigar due to the twist by $\varepsilon$, the analog of $\Omega$-deformation \cite{Nekrasov:2002qd}. Furthermore, the A-twisted 3d theory is the setting for the K-theoretic count of quasimaps in \cite{Okounkov:2015spn}.

We could also look at the $\cQ_A$-cohomology of states. By analogy with the 2d case, we can generate it by the appropriate set of boundary states (regularized by the Euclidean evolution), coming from the A-branes on $\mathbb{T}^2$, which is the boundary of $C\times_\varepsilon S^1$. Again by analogy with the 2d case, such boundary conditions can be thought of as generated by couplings to the Chan-Paton bundles on $LX$, the free loop space of $X$. The boundary degrees of freedom living on $\mathbb{T}^2$ are like the quantum mechanics into $LX$. Thus one expects to find that the boundary conditions are classified by
\begin{equation}
\label{Kth3d}
K_{\bfT\times S^1}(LX)\otimes_{\Z[q^{\pm1}]} \C(q) \cong K_\bfT (X) \otimes \C(q),
\end{equation}
the equivariant K-theory of $LX$, additionally equivariant with respect to the loop rotations $S^1$. By localization, it is related to the right hand side in \eqref{Kth3d}: everything is expected to localize on constant loops, and an additional factor of $\C(q)$ contains the equivariant parameter for loop rotations, which is, basically, the exponential of $\varepsilon$ (up to a constant factor). This supports identifying the $\cQ_A$-cohomology, or the space of vacua, with the equivariant K-theory. Additionally, when $X^\bfT$ is a discrete set of points, the argument about isolated massive vacua provides extra support for this claim. 

It has been long suspected that the $S^1$-equivariant K-theory on the loop space is related to the elliptic cohomology \cite{Witten:1986bf,Witten:1987cg}. See for example \cite{kitchloo_morava_2007,luecke2019completed}, where the left hand side of \eqref{Kth3d} (with $\C(q)$ replaced by $\Z[q^{-1}][[q]]$) is used to define the completed version of $S^1$-equivariant K-theory on $LX$, which is then connected to the elliptic cohomology (with the Tate curve taken as the elliptic curve in their case). Presumably, such understanding is more general than presented here (see e.g. \cite{grojnowski_2007} and \cite{Berwick-Evans:2018ryn}). In the case of the $\cQ_A$ supercharge in three dimensions, we nevertheless prefer the interpretation in terms of $K_\bfT(X)$, which is heuristically dictated by the fact that the equivariant parameters are $\C^\times$-valued. One could however argue that the boundary conditions in 3d are naturally labeled by the elliptic cohomology, in the sense of $K_{S^1}(LX)$, and \cite{Jockers:2018sfl} coin the term elliptic or ``E-branes'' for this very reason.

Finally, for the supercharge $\cQ$ in three dimensions, the equivariant parameters are elliptic, associated to a fixed elliptic curve $E_{-1/\tau}$ in the notations of Section \ref{sec:gauge_review}. In this case a natural interpretation for the $\cQ$-cohomology is in terms of the equivariant elliptic cohomology ${\rm Ell}_\bfT(X)$. We follow the approach of \cite{Aganagic:2016jmx} (see also references therein), in which ${\rm Ell}_\bfT(X)$ is a scheme. Since the $\cQ$-cohomology (of states) is just a vector space, we have to explain in what sense it is related to this scheme. As it turns out, this vector space is related to fibers of a certain line bundle over ${\rm Ell}_\bfT(X)$. In the rest of this subsection, we will discuss this question in greater detail.

\paragraph{Spectral manifolds and cohomology.} Consider the simple case where the generic equivariant parameters correspond to a symmetry subgroup leaving only the isolated fixed points in $X$. Ordinary equivariant cohomology $H_\bfT(X)$, considered as a ring, defines an affine scheme ${\rm Spec}\, H_\bfT(X)$, with fixed points in $X$ corresponding to irreducible components in ${\rm Spec}\, H_\bfT(X)$. Passing to $\C^\times$-valued variables via the exponentiation map, one obtains ${\rm Spec}\, K_\bfT(X)$, also an affine scheme. In the elliptic case, ${\rm Ell}_\bfT (X)$ is defined as a scheme itself, which is seen as a reduction of ${\rm Spec}\, K_\bfT(X)$ mod $q$. It is not affine, however, so there is no ring whose ${\rm Spec}$ it would be. It is still true that its irreducible components correspond to the points in $X$ fixed by $\bfT$.

What is the meaning of the schemes we mention above? Suppose we are interested in $H_\bfT(X)$ in the context of a 2d theory, i.e., in the $\cQ_A$ cohomology. The ring relations encoded in the effective twisted superpotential $\tilde{W}(\Sigma)$ of the 2d theory are known to be
\begin{equation}
\exp\left( \frac{\partial \tilde{W}}{\partial \Sigma}\right)=1.
\end{equation}
These are solved, roughly, by $\Sigma^{(i)} = F_i(x,t)$, $i=1..n$, expressing complex Cartan-valued scalars in the 2d $\cN=(2,2)$ gauge multiplets in terms of equivariant and K\"ahler parameters, $x$ and $t$. The number $n$ of solutions is precisely the number of massive vacua. If we want to consider ``ordinary'' (not quantum) cohomology, then we simply send $t\to +i\infty$, i.e., take the large-volume limit of $X$. The scheme ${\rm Spec}\, H_\bfT(X)$ can be seen as the graph of these solutions. The 3d case (compactified on a finite circle), with the same supercharge $\cQ_A$, is not very different. Now the same equation $\exp(\partial\tilde{W}/\partial\Sigma)=1$ encodes relations in the appropriate quantum K-theory of $X$, and $\Sigma$ are the $\C^\times$-valued variables containing real scalars and gauge holonomies along $S^1$. Again, taking the FI parameters to infinity, one gets back the classical relations. These relations determine ${\rm Spec}\, K_\bfT(X)$, which is an $n$-component scheme. Schematically, this looks as follows:
\begin{figure}[h]
	\centering
	\includegraphics[scale=0.7]{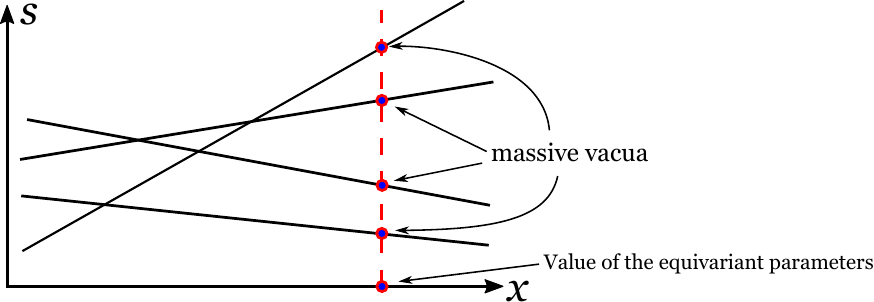}
	\caption{The horizontal axis here represents background equivariant parameters. The vertical axis represents gauge parameters of the similar nature, e.g., gauge holonomies on $\bE_\tau$ in the elliptic case. The collection of lines represents either ${\rm Spec}\, H_\bfT(X)$, or ${\rm Spec}\, K_T(X)$, or ${\rm Ell}_\bfT(X)$. Each line corresponds to a massive vacuumn, i.e., a fixed point in $X^\bfT$.}\label{fig:spec}
\end{figure}\\
The collection of lines in this picture represents a scheme over the base parameterized by the equivariant parameters for the torus $\bfT$ of global symmetries.

We want to have a similar understanding of the $\cQ$-cohomology. In 1d, as we know, it is isomorphic to the $\cQ_A$ case, and we simply get the ordinary equivariant cohomology $H_\bfT(X)$. Indeed, any possible instanton corrections represented by holomorphic curves in 2d, do not exist in 1d. The equations that determine the spectral manifold are the same as in the large-volume limit in 2d. They describe the classical ${\rm Spec}\, H_T(X)$ illustrated in Figure \ref{fig:spec}. Physically, these equations simply determine classical values of the vector multiplet scalars in various vacua. In other words, ${\rm Spec}\, H_\bfT(X)$ describes the classical vacua in 1d.

\textbf{Example:} consider $X=T^* \C P^{n-1}$. The flavor equivariant parameters are $x_i$, $i=1, \ldots , n$, such that $\sum_i x_i=0$. Additionally, we have an equivariant parameter $\hbar$ for $U(1)_\hbar$ that rotates the fiber. Denoting the gauge parameter by $s$, we get that ${\rm Spec}\, H_\bfT(X)$ is described by
\begin{equation}
\prod_{i=1}^n (s + x_i)=0.
\end{equation}
Physically, it is more natural to write this as
\begin{equation}
\prod_{i=1}^n \left(s + x_i + \frac{\hbar}{2}\right)=0,
\end{equation}
which differs by a shift of $s$. The reason is that in this case $s$ has a meaning of the complex scalar in the 1d $\cN=4$ dynamical vector multiplet (which also has a real scalar). The corresponding theory has hypermultiplets $(I_i, J_i)$, $i=1..n$, and $U(1)_\hbar$ rotates both $I_i$ and $J_i$ with charge $\frac12$. The $i$-th fixed point in the base of $T^*\C P^{n-1}$ is characterized by the vev $I_j = \sqrt{\zeta}\delta_{ij}$, and the total (gauge + flavor + $U(1)_\hbar$) weight of the field getting a vev must vanish. The total weight of $I_i$ is $(s+x_i+\frac{\hbar}{2})$, which is indeed what we find in the above product. We see that the equation describes $n$ irreducible components, $s=-x_i-\hbar/2$, $i = 1, \ldots , n$, which correspond to the $n$ fixed points.

Moving up to 2d, equivariant parameters become multiplicative. The classical description can be obtained from the 1d answer given above simply by rewriting equations in the multiplicative way. For example, in the $T^* \C P^{n-1}$ case, ${\rm Spec}\, K_\bfT(X)$ is simply
\begin{equation}
\prod_{i=1}^n (s x_i \hbar^{\frac12}-1)=0.
\end{equation}
This again describes the classical vacua: equations like $s x_i \hbar^{1/2}=1$ are the classical conditions saying that $s$ cancels the effect of $x_i \hbar^{1/2}$ in the $i$-th vacuum. In two dimensions on $\R\times S^1$, these parameters are $\C^\times$-valued, combining the holonomy around the circle and the imaginary part of the vector multiplet scalar (the one descending from the gauge field in 3d). Again, like on the Figure \ref{fig:spec}, for a chosen point in the base, i.e., for given equivariant parameters, there are $n$ points above it corresponding to the massive vacua $s=x_i^{-1}\hbar^{-1/2}$. This also describes massive vacua quantum mechanically, because the action is $\cQ$-exact, including the superpotentials, so the $\cQ$-cohomology can be analyzed in the UV. The $i^{\rm th}$ vacuum is given by a wave functional peaked around the $i^{\rm th}$ irreducible component $s=x_{i}^{-1}\hbar^{-1/2}$. We can say that the corresponding vector in $K_\bfT(X)$ is the class of the structure sheaf of this irreducible component.

Moving up to 3d, the story repeats itself, except that now we should take all variables mod $q$, reflecting that $\bE_{-1/\tau}$ is $\C^\times/q$. Now the base in Figure \ref{fig:spec} is identified with $\cE_\bfT$, the space of elliptic equivariant parameters. We still have $n$ points in the fiber over a point in the base, corresponding to $n$ isolated massive vacua. As we vary the base point, they assemble into the scheme ${\rm Ell}_\bfT(X)$. Each irreducible component of it is just a copy of $\cE_\bfT$, and this whole picture encodes the classical description of vacua. In the quantum 3d theory, however, due to non-perturbative effects, it is natural and necessary to also include the K\"ahler parameters $z\in\cE_{\mathbf{A}'}$ that couple to the topological symmetry. Therefore, one considers
\begin{equation}
{\rm E}_\bfT(X) = {\rm Ell}_\bfT(X)\times \cE_{\mathbf{A}'}
\end{equation}
as a scheme over $\cE_\bfT\times \cE_{\mathbf{A}'}$. Because neither ${\rm Ell}_\bfT(X)$ nor ${\rm E}_\bfT(X)$ are affine, there is no usual ``cohomology ring'', whose spectrum would be ${\rm Ell}_\bfT(X)$ or ${\rm E}_\bfT(X)$. Rather, one looks at sheaves or bundles on these schemes.

In fact, there is a natural way to obtains such bundles in quantum theory, where each massive ground state is associated with a one-dimensional vector space $\C$. This works similarly in all three cases: elliptic, K-theoretic, and cohomological. The vector space of supersymmetric vacua has a fixed dimension $n$, and changing equivariant (and K\"ahler in the 3d case) parameters, we obtain a rank-$n$ bundle over the parameter space. In the Figure \ref{fig:spec}, this bundle is over the base. In the cohomological or K-theoretic cases, the bundle can be identified with the pushforward of the structure sheaf upstairs (i.e., on the scheme ${\rm Spec}\, H_\bfT(X)$ or ${\rm Spec}\, K_\bfT(X)$) to the base. The resulting fiber is just $H_\bfT(X)$ or $K_\bfT(X)$, which we know to be the space of vacua. The elliptic case is similar: The bundle of vacua on $\cE_\bfT\times \cE_{\mathbf{A}'}$ can be understood as the pushforward of a certain line bundle $\cL$ from ${\rm E}_\bfT(X)$. This line bundle, when pulled back to an irreducible component $j$ of ${\rm E}_\bfT(X)$, becomes the massive vacuum number $j$ fibered over the space of parameters $\cE_\bfT\times \cE_{\mathbf{A}'}$.

\subsection{The bundle of vacua in the elliptic case}\label{sec:ell_vacua}
Let us understand the bundle of vacua over $\cE_\bfT\times \cE_{\mathbf{A}'}$ slightly better, relying on \cite{Dedushenko:2022pem}. For fixed equivariant parameters, the fiber of this bundle is the subspace of ground states in the Hilbert space $\cH[\bE_\tau]$. Suppose $|\Psi\rangle$ is an abstract state corresponding to a vacuum. We can formally characterize it by its wave functional. To do that, choose a polarization of the classical phase space $\cP[\bE_\tau]$ associated to the spatial slice $\bE_\tau$. The wave functional for a state $|\Psi\rangle$ ``written in the polarization'' can be understood as an overlap $\langle \cB|{\Psi}\rangle\equiv \Psi[\cB]$ with the boundary state engineered by the boundary condition $\cB$ that fixes values of the fields, or their normal derivatives, according to the polarization. As we will see, we can characterize ground states by evaluating their wave functionals against the special SUSY boundary conditions.% The resulting expressions are sections of a certain bundle over $\cE_\bfT\times \cE_{\mathbf{A}'}$.

Both $|\Psi\rangle$ and $\langle\cB|$ carry topological data, meaning they are sections of some bundles over $\cE_\bfT\times \cE_{\mathbf{A}'}$. Hence one could say that $|\Psi\rangle$ is a section of an abstract bundle of vacua, while $\langle \cB|\Psi\rangle$ is a section relative to the polarization. The latter is what we need for our applications, so both pieces of topological data are necessary.

\paragraph{Ground state $|\Psi\rangle$.} To determine the topological data in $|\Psi\rangle$, we need to know what happens as we go around the cycles of $\cE_\bfT\times \cE_{\mathbf{A}'}$, i.e., under the large $\bfT\times \mathbf{A}'$-gauge transformations which shift the $1$-forms representing the background flat connections by periods of the elliptic curve $\mathbb{E}_\tau$:
\begin{equation}
\label{large_gauge}
x_i\mapsto q x_i,\quad \text{or}\quad \hbar\mapsto q\hbar,\quad \text{or}\quad z_k\mapsto q z_k.
\end{equation}
Here  $(x_i,\hbar)$ are the monodromy parameters of flavor flat connections on $\mathbb{E}_\tau$, which are elliptic equivariant variables and also coordinates on $\cE_\bfT$. Likewise, $z_k$ are topological flat connections on $\mathbb{E}_\tau$ (K\"ahler variables), which are also coordinates on $\cE_{\mathbf{A}'}$.

Assume that there are $n$ isolated massive vacua $|\Psi_1\rangle,\dots, |\Psi_n\rangle$.  Upon the transport \eqref{large_gauge}, the complex lines ${\C}|{\Psi}_{j}\rangle$ are preserved. The basis state $|{\Psi}_{j}\rangle$ is multiplied by a phase:
\begin{equation}
|\Psi_j\rangle \mapsto e^{i\theta_j} |\Psi_j\rangle \ .
\end{equation}
However, the unitary transformation above would not preserve the holomorphy in $x_i$'s. There is a choice of counterterms enforcing holomorphy, which would make $\theta_j$'s complex. We would like to determine them. 

Start with $|\Psi_{j}\rangle$ at time $t=0$, and let $x, z, \hbar$ change adiabatically, such that $|\Psi_j(t))\rangle$ is the ground state at each instance of time.
In the process, the Berry phase may be generated. 
At some large time $t=\beta$, we assume 
\begin{equation}
\begin{aligned}
& x_i(\beta)=q^{n_{i}} x_i(0)\ , \\ 
& z_a(\beta)=q^{m_{a}} z_a(0)\ , \\ 
& {\hbar} ({\beta}) = q^{n_{\hbar}} {\hbar}(0) \\
\end{aligned}
\label{eq:trxzh}
\end{equation}
with integer $n$'s, $m$'s and $n_\hbar$. Let us denote the transported state $|{\Psi}_{j}\rangle$ by $|{\Psi}_{j}^{q}\rangle$. We close the Euclidean time to $S^1_\beta$ and compute the $\mathbb{E}_\tau\times S^1_\beta$ partition function in the large-$\beta$ limit. The leading contribution in this limit comes from the isolated ground states, and is given by:
\begin{equation}
Z_{\mathbb{E}_\tau\times S^1_\beta} \, = \, \sum_{j=1}^{n} \langle \Psi_j^{q} |\Psi_j \rangle + O(e^{-\beta \times \#}).
\end{equation}
From the point of view of effective field theory, each gapped vacuum contributes to the $\mathbb{E}_\tau\times S^1_\beta$ partition function via its effective Chern-Simons (CS) term (generated by massive degrees of freedom). Indeed, only topological terms survive in the large-$\beta$ limit. Hence we obtain
\begin{equation}
e^{i\theta_j}=\langle \Psi_j^{q}|\Psi_j \rangle = e^{-S_{\rm CS}^{(j)}},
\end{equation}
where we also assume vacua to be normalized. So the effective CS term in the $j$-th vacuum characterizes the topology of $|\Psi_j\rangle$. More precisely, it was shown in \cite{Dedushenko:2022pem} that the effective CS term determines the Berry connection for $|\Psi_j\rangle$, such that the Chern class of the vacuum bundle is equal to the effective CS level. To read off the Berry connection, all we need to do is evaluate $S_{\rm CS}^{(j)}$ on the background field configuration that consists of flat connections $z_a, \hbar, x_{i}$ on $\mathbb{E}_\tau$ that change slowly from these values to $q^{m_a} z_a, q^{n_\hbar} \hbar, q^{n_i} x_i$ as we go around $S^1_\beta$, generating  a nontrivial flux.

The specific formula for Berry connection coincides with the standard family index formula \cite{BGS}.
In terms of holonomy of the global symmetry flat connection along the A and B cycles of $\bE_{\tau}$, denoted $a_A^i$ and $a_B^i$, and the effective CS levels $k_{ij}$, $i,j = 1, \ldots, f$,  the Berry connection local one-form reads
\begin{equation}
B = \frac{k_{rs}}{4\pi} \left( {\rm log} \left( a_A^r \right)  \dd {\rm log} \left( a_B^s \right)  - {\rm log} \left( a_B^r \right)  \dd {\rm log} \left( a_A^s \right) \right) .
\end{equation}
In agreement with the family index theory, the curvature of $B$ is a pushforward of the 
the anomaly polynomial four-form $\frac{1}{4\pi} k_{rs} F^r \wedge F^s$. 

The $(0,1)$-part of $B$ determines a holomorphic structure (has curvature of type $(1,1)$), and the holomorphic sections were described in \cite{Dedushenko:2022pem} as
\begin{equation}
e^{\frac{i}{4\pi(\bar\tau-\tau)} k_{rs} \log(a^r) \log(a^s \bar{a}^s)}\Theta(a)\, , \quad 
\end{equation}
where 
\[ a_{r} = a_{B}^{r} \left( a_{A}^{r} \right)^{\tau} \,  , \]
are holomorphic coordinates on $\cE_\bfT\times \cE_{\mathbf{A}'}$. The transformation properties of $\Theta$ are
\begin{equation}
\label{holo_theta}
{\Theta}( q a_r) \, = \, q^{-\frac{k_{rr}}{2}} \prod_{s} a_{s}^{- k_{rs}} \ {\Theta}(a),
\end{equation}
where only the $r$'th variable is shifted.  One can easily construct such $\Theta(a)$'s from the familiar theta functions. Holomorphic sections naturally appear in 3d as supersymmetric partition functions in the presence of a boundary $\bE_{\tau}$ with $\cN=(0,2)$ boundary conditions (such as overlaps $\langle \cB|\Psi\rangle$, interval partition functions, or half-indices). They also appear as elliptic genera of 2d theories \cite{Closset:2019ucb}. 
The state $|\Psi_p\rangle$ can also be mimicked by a half-BPS elliptic boundary condition $\sD_p$, at least for the purposes of BPS computations. It creates a boundary state $|\sD_p\rangle$, which is not normalizable, yet the finite Euclidean evolution turns it into a normalizable half-BPS state $e^{-TH}|\sD_p\rangle$ (see Appendix \ref{app:bdry_states}) that is $\cQ$-cohomologous to the vacuum $|\Psi_p\rangle$. The states $|\Psi_p\rangle$ and $e^{-TH}|\sD_p\rangle$ should have ``the same topology'', i.e., be sections of the isomorphic line bundles over $\cE_{\bfT}\times\cE_{\mathbf{A}'}$. For $e^{-TH}|\sD_p\rangle$, the topology is captured by the boundary 't Hooft anomalies of $\sD_p$ \cite{Dedushenko:2022pem}. This means that the Chern class of the line bundle spanned by $e^{-TH}|\sD_p\rangle$ agrees with the anomaly polynomial of $\sD_p$. Thus the latter must agree with the effective CS term in the vacuum $|\Psi_p\rangle$ (this can also be understood via the inflow).

When the vacuum $|\Psi_p\rangle$ is fully gapped via the real masses, such boundary conditions can be easily constructed. They are sometimes called thimble boundary conditions \cite{Hori:2000ck,Bullimore:2016nji} and are realized via the exceptional Dirichlet boundary conditions in gauge theory \cite{Bullimore:2016nji} (they have also played role in the recent papers \cite{Bullimore:2020jdq,Okazaki:2020lfy}). Since the state $|\Psi_p\rangle$ is an ``in'' state, the corresponding boundary conditions have to be imposed at the past boundary at some $y=y_-<0$. In this case, the boundary conditions are constructed as follows. The matter representation $\cR$ is split into
\begin{equation}
\cR = \cR_+(p)\oplus \cR_-(p) \oplus \cR_0(p),
\end{equation}
where $\cR_\pm(p)$ receive positive/negative real masses in the vacuum $p$ respectively. Then the corresponding chiral multiplets are given boundary conditions according to:\footnote{It is important that we are working with the ``past'' boundary here, i.e., at $y=y_-$ of, say, $(y_-,\infty)$. For the thimble boundary conditions defined at the ``future'' boundary, the Dirichlet and Neumann boundary conditions on $\cR_\pm(p)$ must be swapped.}
\begin{align}
&\cR_+(p),\quad \bar\cR_-(p):\qquad \text{(0,2) Dirichlet},\cr
&\cR_-(p),\quad \bar\cR_+(p):\qquad \text{(0,2) Neumann}.\cr
\end{align}
As for $\cR_0(p)\oplus \bar\cR_0(p)$, they are the matter fields that can receive vevs in the vacuum $p$. More precisely, we split
\begin{equation}
\cR_0(p)=\cR_0^D(p) \oplus \cR_0^N(p),
\end{equation}
such that $\cR_0^D(p)\oplus\bar\cR_0^N(p)$ contains all chirals that receive vevs.\footnote{It might happen that some of the hypermultiplets in $\cR_0(p)$ do not get vevs at all, in which case they are distributed among $\cR_0^D(p)$ and $\cR_0^N(p)$ arbitrarily, as a matter of choice. We will ignore this possibility for now.\label{nonVevR0}} Then we assign boundary conditions:
\begin{align}
&\cR_0^D(p),\quad \bar\cR_0^N(p):\qquad \text{(0,2) Dirichlet with boundary values given by the vevs},\cr
&\cR_0^N(p),\quad \bar\cR_0^D(p):\qquad \text{(0,2) Neumann}.\cr
\end{align}
Finally, the vector multiplets are assigned Dirichlet boundary conditions, with the boundary flat connection corresponding to the vacuum $p$ (i.e., the one consistent with the vevs of matter fields):
\begin{equation}
s = s^{(p)}(x,\hbar).
\end{equation}
In Section \ref{sec:effective} we will discuss such boundary conditions in greater detail. For now, let us compute the boundary 't Hooft anomalies and read off the topology of $|\Psi_p\rangle$.

Since 't Hooft anomalies are RG invariants, it is most convenient to determine them in the UV description. As reviewed in Section \ref{sec:gauge_review}, there is only one CS term present in the UV: The BF coupling \eqref{SBF} between the center of the gauge group and the topological symmetry $\mathbf{A}'$. It contributes to the boundary anomaly via the inflow. The rest of boundary anomalies are computed as in \cite{Dimofte:2017tpi,Bullimore:2020jdq}. Denote the (boundary) gauge field strength by $\mathbf{f}$, the 3d $\cN=2$ R-symmetry gauge field strength by $\mathbf{r}$, the $U(1)_\hbar$, flavor, and topological field strengths by $\mathbf{f}_\hbar$, $\mathbf{f}_x$, and $\mathbf{f}_z$, respectively. Three-dimensional chirals in $\cL_p=\cR_+(p)\oplus\bar\cR_-(p)\oplus \cR_0^D(p) \oplus \bar\cR_0^N(p)$ obey $(0,2)$ Dirichlet boundary conditions and contribute $\frac12{\Tr}_{\cL_p}(\mathbf{f}+\mathbf{f}_x + \frac12\mathbf{f}_\hbar -\frac12\mathbf{r})^2$ to the anomaly polynomial, while the remaining ones in the conjugate representation $\bar\cL_p$ obey Neumann, and contribute $-\frac12{\Tr}_{\bar\cL_p}(\mathbf{f}+\mathbf{f}_x + \frac12\mathbf{f}_\hbar -\frac12\mathbf{r})^2$, where the weights in $\bar\cL_p$ are opposite of those in $\cL_p$. The notation $\Tr_{\cL_p}$ means summation both over the gauge and flavor weights in $\cL_p$. Together, these almost cancel each other, leaving behind only
\begin{equation}
(\mathbf{f}_\hbar -\mathbf{r})\Tr_{\cL_p}(\mathbf{f}+\mathbf{f}_x).
\end{equation}
The adjoint chiral $\Phi$ obeys Dirichlet boundary conditions, and contributes $\frac12 \Tr_{\mathbf{adj}}(\mathbf{f}+\mathbf{f}_\hbar)^2$. Finally, the 3d $\cN=2$ vector multiplet with Dirichlet boundary conditions contributes $-\frac12 \Tr_{\mathbf{adj}}\mathbf{f}^2 -\frac{|G|}2 \mathbf{r}^2$.

Also note that the $G$ symmetry is completely broken at the boundary. First, Dirichlet boundary conditions on the gauge fields only leave $G$ as a global symmetry at the boundary. Second, the boundary vevs (corresponding to the isolated vacuum) further break this global $G$ symmetry. Thus $\mathbf{f}$ should not be present in the anomaly polynomial as the corresponding global symmetry is absent. It should instead be expressed through $\mathbf{f}_x$ and $\mathbf{f}_\hbar$, which we write as $\mathbf{f}=f^{(p)}(\mathbf{f}_x,\mathbf{f}_\hbar)$ (see, e.g., the Appendix of \cite{Bullimore:2020jdq} for such a computation)\footnote{The functional forms of $f^{(p)}(\mathbf{f}_x,\mathbf{f}_\hbar)$ and $s^{(p)}(x,\hbar)$ coincide, i.e., this is the same function.}. This expression is found by solving the system of linear equations of the form:
\begin{equation}
w(\mathbf{f} + \mathbf{f}_x + \mathbf{f}_\hbar)=0,\quad w\in \cR_0^D(p)\oplus \bar\cR_0^N(p),
\end{equation}
where $w$ are the matter weights that receive vevs in the vacuum $p$. In other words, the total field strength must vanish for such weights, which determines $\mathbf{f}$ in terms of $\mathbf{f}_x$ and $\mathbf{f}_\hbar$.

Altogether, the boundary anomaly polynomial is
\begin{equation}
P[\sD_p] = (\mathbf{f}_\hbar -\mathbf{r})\Tr_{\cL_p}(\mathbf{f}+\mathbf{f}_x) + \frac{|G|}2 \mathbf{f}_\hbar^2 - \frac{|G|}2 \mathbf{r}^2 - 2\Tr(\mathbf{f}\mathbf{f}_z),\quad \text{with } \mathbf{f} = f^{(p)}(\mathbf{f}_x,\mathbf{f}_\hbar), 
\end{equation}
where the last term in $P[\sD_p]$ is the inflow effect due to the BF action \eqref{SBF}. The full anomaly polynomial also contains mixed global-gravitational anomalies, which can be deduced from {\cite{Nekrasov:2014xaa}, but we do not need them for our purposes.
	
The four-form $P[\sD_p]$ agrees (via $P=\dd({\rm CS})$) with the effective background CS term in the vacuum $p$ (such CS terms can be found in \cite{Dedushenko:2022pem}). Thus, using the effective CS levels encoded in $P[\sD_p]$, we can use our recipe \eqref{holo_theta} and find how $|\Psi_p\rangle$ transforms under \eqref{large_gauge}. To write the result, let us first introduce some notations. Let us temporarily treat the gauge fugacity (flat connection) $s\in {\rm Hom}(\pi_1(\mathbb{E}_\tau),\mathbf{H})/\cW$ as an independent variable (even though in the vacuum $p$ we have $s=s^{(p)}(x,\hbar)$). It is an elliptic variable as well, $s=e^{2\pi i h}$, $h\in\mathfrak{h}_\C\mod Q^\vee\oplus\tau Q^\vee$ . Let $\mathbf{v}=(\mathbf{v}_1,\dots,\mathbf{v}_{{\rm rk}(G)})$ be some gauge coweight and $\mathbf{w}=(\mathbf{w}_1,\dots,\mathbf{w}_{{\rm rk}(G_H)})$ be a flavor coweight. Also let $c_a$, $a=1..{\rm rk}(\mathbf{A}')$ be gauge group characters picking out the $U(1)$ factors in the gauge group, such that $s^{c_a}$ is a fugacity for the $a$-th abelian factor in $G$. Then the transformation properties are written as:
\begin{align}
	\label{shifts_Psi_p}
	z_a\mapsto q z_a:\quad  &\Psi_p \mapsto s^{c_a} \Psi_p\cr
	s \mapsto q^{\mathbf{v}} s:\quad &\Psi_p \mapsto \hbar^{-\frac12\sum_{\cL_p}\langle\mathbf{v},w\rangle}\prod_{a=1}^{{\rm rk}(\mathbf{A}')}z_a^{\langle \mathbf{v},c_a\rangle} \Psi_p\cr
	x\mapsto q^{\mathbf{w}}x:\quad &\Psi_p \mapsto \hbar^{-\frac12\sum_{\cL_p} \langle\mathbf{w},f\rangle} \Psi_p\cr
	\hbar\mapsto q^2\hbar:\quad &\Psi_p \mapsto \hbar^{-|G|} q^{-|G|} \prod_{(w,f)\in \cL_p}\left(s^{-w} x^{-f}\right) \Psi_p,
\end{align}
from which the true transformations of $|\Psi_p\rangle$ follow simply by the specialization $s=s^{(p)}(x,\hbar)$.
	
It is also not hard to write a model meromorphic section that transforms as \eqref{shifts_Psi_p}. For that, define (following the conventions of \cite{Aganagic:2016jmx})
\begin{equation}
	\label{varth}
	\vartheta(x) = (x^{1/2}-x^{-1/2})\prod_{n>0}(1-q^n x)(1-q^n/x),
\end{equation}
which obeys
\begin{equation}
	\vartheta(q^k x)=(-1)^k q^{-k^2/2}x^{-k}\vartheta(x),\quad k\in\Z.
\end{equation}
Then the following
\begin{equation}
\label{model_section_Dp}
\prod_{(w,f)\in \cL_p} \frac{\vartheta(s^w x^f \hbar^{1/2})}{\vartheta(s^w x^f)\vartheta(\hbar^{1/2})} \times \vartheta(\hbar^{1/2})^{2|G|}\times\prod_{a=1}^{{\rm rk}(\mathbf{A}')} \frac{\vartheta(s^{c_a})\vartheta(z_a)}{\vartheta(s^{c_a}z_a)}
\end{equation}
transforms as \eqref{shifts_Psi_p}. This is an example of $\Theta$ as in \eqref{holo_theta}. It is a section of a certain line bundle $\cL$  on ${\rm E}_\bfT(X) = {\rm Ell}_\bfT(X)\times \cE_{\mathbf{A}'}$. Specialization $s=s^{(p)}(x,\hbar)$ means taking its pullback to the $p$-th irreducible component of ${\rm E}_\bfT(X)$, where we obtain a line bundle that the state $|\Psi_p\rangle$ is valued in.
	
An alternative way to derive the same result \eqref{shifts_Psi_p} is to construct a 2d $(0,2)$ theory that has the same anomaly polynomial $P[\sD_p]$, write its elliptic genus, and see how it transforms. Below we will outline this alternative derivation for the contribution of $\langle \cB|$.
	
\paragraph{Polarization.} Now let us discuss the contribution of $\langle \cB|$. We choose an $\cN=(2,2)$-preserving polarization on the phase space, in the sense of \cite{Dedushenko:2018aox}.\footnote{A polarization $\Pi$ on the phase space is invariant under a symmetry $V$ if the symplectic vector field $V$ obeys $[V,\Pi]\subset\Pi$. In the case of SUSY, $V$ is an odd vector field.} With the description of $\cN=(2,2)$ boundary conditions reviewed in Section \ref{sec:gauge_review}, it is also clear how to obtain $\cN=(2,2)$ invariant polarizations in gauge theory. For the hypermultiplets, to define such polarization on the (infinite-dimensional) phase space, we first choose a polarization on the (finite-dimensional) holomorphic symplectic manifold $T^*\cR$ they are valued in. For simplicity, let us start with a linear polarization, i.e., a Lagrangian splitting $T^*\cR=  \bL\oplus \bL^\perp$ (though we will be more general later):
\begin{equation}
(Q_\bL, \tilde{Q}_\bL),
\end{equation}
where $Q_\bL$ and $\tilde{Q}_\bL$ can be conveniently viewed as 2d $\cN=(2,2)$ chiral multiplets valued in ${\rm Maps}(\R,\bL)$ and ${\rm Maps}(\R, \bL^\perp)$. We then use this splitting to define the polarization on the phase space: pick the ``position'' polarization for $Q_\bL$, forcing the ``momentum'' polarization for $\tilde{Q}_\bL$ (which is slightly deformed by the contribution of $\sigma_{3}+{\ii}A_y$ to covariant derivatives). Modeling the polarization by a boundary condition at $y=0$ means the lowest components of these superfields obey:
\begin{equation}
q_\bL\big|_{y=0} = q_\partial,\quad (D_y+\sigma)\tilde{q}_\bL\big|_{y=0} = \tilde{p}_\partial,
\end{equation}
where $\tilde{p}_\partial$ and $q_\partial$ are some scalar functions on $\bE_\tau$. The rest (i.e., boundary conditions for the hypermultiplet fermions) are completed by SUSY. In particular, fermions residing in $Q_\bL$ and $\tilde{Q}_\bL$ are canonically conjugate to each other, so the natural SUSY completion of the above is to fix boundary values of all fermions in $Q_\bL$. That such boundary conditions on fermions are elliptic and make sense quantum mechanically is not obvious, and we discuss it in the Appendix \ref{app:bdry_states}.
	
One thus obtains a ``boundary multiplet'' $Q_\partial$, which contains boundary restrictions of those fields whose boundary values are kept fixed in the path integral. This $Q_\partial$ is in fact the restriction $Q_\bL\big|_{y=0}$, so it is an $\cN=(2,2)$ chiral multiplet. The auxiliary field of this chiral multiplet is equal on-shell to
\begin{equation}
F_\partial = (D_y+\sigma) \tilde{q}_\bL,
\end{equation}
thus encoding the boundary conditions for $\tilde{q}_\bL$. So we conclude that the polarization for hypermultiplets, corresponding to the Lagrangian splitting $\bL\oplus \bL^\perp$, is implemented by the boundary conditions
\begin{equation}
Q_\bL\big| = Q_\partial,
\end{equation}
where $Q_\partial$ is a background chiral multiplet at the boundary.
	
We can consider more general polarizations on $T^*\cR$, for which the above treatment of $\bL\oplus \bL^\perp$ is a local model. To make contact with \cite{Maulik:2012wi,Aganagic:2016jmx}, the proper choice should be a polarization on the Higgs branch $X$ (viewed as a holomorphic symplectic manifold), which is then lifted to some polarization on $T^*\cR$. In practice, we will be concerned with yet another set of boundary conditions $\langle \sB_p|$ labeled by the fixed points $p$ (massive vacua) on $X$ and a choice of polarization. It is always possible to lift the polarization of $X$ in the vicinity of $p$ to some linear polarization on the hypermultiplets:
\begin{equation}
\bL(p)\oplus \bL^\perp(p),
\end{equation}
and this is what we do. In general, $\bL(p)$ may be different for different $p$, in which case we denote the boundary conditions as $\sB_{\bL(p),p}$.
	
For 3d $\cN=4$ vector multipelts, we will choose the ``Dirichlet'' polarization. We think of a 3d $\cN=4$ vector as a pair of a 2d $\cN=(2,2)$ vector $V^{(2,2)}$ (with the gauge group ${\rm Maps}(\R, G)$), and a 2d $\cN=(2,2)$ chiral $S^{(2,2)}$ in the adjoint of this group. The fermions in $S^{(2,2)}$ are canonically conjugate variables to the fermions in $V^{(2,2)}$. We pick the boundary conditions that prescribe the boundary value of $V^{(2,2)}$,
\begin{equation}
V^{(2,2)}\big| = V_\partial,
\end{equation}
to be the background multipelt $V_\partial$, the ``boundary vector multiplet''. Since the Fermi fields in $S^{(2,2)}$ are canonically conjugate to those in $V^{(2,2)}$, they should stay unconstrained once the entire multiplet $V^{(2,2)}$ is fixed at the boundary. That this leads to sensible elliptic boundary conditions in the quantum theory is again discussed in the Appendix \ref{app:bdry_states}. The lowest component of $S^{(2,2)}$ is $\sigma + iA_y$, and it appears in the auxiliary field of $V_\partial$ as\cite{Dedushenko:2018tgx}
\begin{equation}
iD_\partial = D_y\sigma + iD.
\end{equation}
Therefore, the deformed Neumann boundary condition on $\sigma$ follows from $V^{(2,2)}\big|=V_\partial$ (note that $D$, being an auxiliary field, does not receive an independent boundary condition).
	
Altogether, the states are described as functionals $\Psi[\cB]\equiv\Psi[V_\partial, Q_\partial]$ of the boundary multiplets. If there are $n$ isolated vacua, the corresponding functionals $\Psi_1,\dots,\Psi_n$ span the kernel of the Hamiltonian $\hat{H}$, which is a certain functional differential operator. Such a description is quite formal and hard to make precise in general, but we could certainly imagine computing $\Psi_i[\cB]$ in perturbation theory, either as a solution to $\hat{H}\Psi=0$, or as a partition function on $\bE_\tau\times \R_<$, with a chosen vacuum fixed at $y=-\infty$ and a boundary conditions $\cB$ at $y=0$.
	
If $\cB$ is a supersymmetric boundary condition (picking special values of $V_\partial$ and $Q_\partial$), then $\langle \cB|\Psi_j\rangle$ can often be computed exactly, especially if $|\Psi_j\rangle$ is replaced by a thimble boundary condition $\sD_j$ discussed earlier.
	
Now we define another set of SUSY boundary conditions labeled by the vacua,
\begin{equation}
\sB_{\bL(p),p},
\end{equation}
which additionally depends on the choice of polarization. While the thimble boundary conditions $\sD_p$ were natural in a theory with generic real masses, we use $\sB_{\bL(p),p}$ in the case when real masses are zero, while the vacua are still massive thanks to the equivariant parameters. The main difference between $\sD_p$ and $\sB_{\bL(p),p}$ is what we do to the hypermultiplets that do not develop vev in the vacuum $p$. In the presence of real masses, such hypers receive natural boundary conditions dictated by the sign of their real mass, which was part of the $\sD_p$ definition. In the absence of real masses, there are no such preferred boundary conditions, and it is a choice labeled by the polarization.
	
Hence, we define $\sB_{\bL(p),p}$ as follows. As before, hypermultiplets in the subspace $\cR_0(p)\subset \cR$ are the ones that get vevs.\footnote{Here, it makes sense to \emph{define} $\cR_0(p)$ as the subspace of hypermultiplets that develop vevs. If the situation of the footnote \ref{nonVevR0} occurs, then this $\cR_0(p)$ will be slightly different from the one discussed there in the context of thimble boundary conditions.} More specifically, chirals in $\cR_0^D(p)\oplus \bar\cR_0^N(p)$ have nontrivial vevs. Thus, just like before we assign
\begin{align}
&\cR_0^D(p),\quad \bar\cR_0^N(p):\qquad \text{(0,2) Dirichlet with boundary values given by the vevs},\cr
&\cR_0^N(p),\quad \bar\cR_0^D(p):\qquad \text{(0,2) Neumann}.\cr
\end{align}
In the absence of real masses, it does not make sense to split the remaining hypers into $\cR_+(p)$ and $\cR_-(p)$. Instead, we use a polarization $\bL(p)\oplus \bL^\perp(p)$ to define a split. Only polarization along the complement of $\cR_0(p)\oplus\bar\cR_0(p)$ matters at this point, so we define:
\begin{align}
\label{Ltilde}
\tilde{\bL}(p) = \bL(p) \setminus T^*\cR_0(p),\qquad \tilde{\bL}^\perp(p) = \bL^\perp(p) \setminus T^*\cR_0(p),
\end{align}
such that
\begin{equation}
\cR\oplus\bar\cR = \cR_0(p)\oplus \bar\cR_0(p)\oplus \tilde{\bL}(p)\oplus \tilde{\bL}^\perp(p),
\end{equation}
or more concisely $T^*\cR\cong T^*\cR_0(p)\oplus T^*\tilde{\bL}(p)$, and assign boundary conditions according to
\begin{align}
\tilde{\bL}(p)&:\qquad \text{Dirichlet},\cr
\tilde{\bL}^\perp(p)&:\qquad \text{Neumann}.
\end{align}
On vector multiplets, like in the $\sD_p$ case, we impose the $(2,2)$ Dirichlet boundary conditions with the boundary flat connection $s=s^{(p)}(x,\hbar)$. Such boundary conditions are also of the ``exceptional Dirichlet'' type \cite{Bullimore:2016nji} since they fully break the boundary $G$ symmetry. In particular, they lift the zero modes that otherwise make $(2,2)$ Dirichlet boundary conditions on vector multiplets a bit subtle. Thus we expect that $\sB_{\bL(p),p}$ are elliptic boundary conditions giving rise to reasonably well-behaved boundary states upon any finite Euclidean evolution.

Let us compute the boundary 't Hooft anomalies for $\sB_{\bL(p),p}$, which will determine the first Chern class of the associated line bundle whose section is $\langle \sB_{\bL(p),p}|$. The computation is very similar to the one we did before, except that now $\langle \sB_{\bL(p),p}|$ is the ``out'' state, so it is imposed at the ``future'' boundary $y=y_+>0$, hence the inflow term will have an opposite sign.

Three-dimensional chirals in
\begin{equation}
\label{Lhat}
\hat{\bL}(p)=\tilde{\bL}(p)\oplus \cR_0^D(p)\oplus\bar\cR_0^N(p)
\end{equation}
obey Dirichlet boundary conditions and contribute $\frac12{\Tr}_{\hat{\bL}(p)}(\mathbf{f}+\mathbf{f}_x + \frac12\mathbf{f}_\hbar -\frac12\mathbf{r})^2$ to the anomaly polynomial, while those in the dual representation obey Neumann. Again they almost cancel each other, leaving
\begin{equation}
(\mathbf{f}_\hbar -\mathbf{r})\Tr_{\hat{\bL}(p)}(\mathbf{f}+\mathbf{f}_x).
\end{equation}
Vector multiplets obey $(2,2)$ Dirichlet, so their contribution is the same as before. The inflow term has an opposite sign, so the total boundary anomaly polynomial is
\begin{equation}
\label{anoPol}
P[\sB_{\bL(p),p}] = (\mathbf{f}_\hbar -\mathbf{r})\Tr_{\hat{\bL}(p)}(\mathbf{f}+\mathbf{f}_x) + \frac{|G|}2 \mathbf{f}_\hbar^2 - \frac{|G|}2 \mathbf{r}^2 + 2\Tr(\mathbf{f}\mathbf{f}_z),
\end{equation}
where, as before, one should substitute $\mathbf{f}=f(\mathbf{f}_x,\mathbf{f}_\hbar)$ as $G$ is fully broken.
	
Using \eqref{anoPol}, we can recover the behavior of $\langle \sB_{\bL(p),p}|$ under the flavor and topological large gauge transformations. For illustrative and pedagogical reasons, we will employ a different method here than before. Namely, we use the fact that $\langle \sB_{\bL(p),p}|$ transforms in the same way as the $\mathbb{T}^2$ partition function of a purely 2d system with the same anomaly polynomial $P[\sB_{\bL(p),p}]$. It is easier to construct an auxiliary 2d system with the anomaly polynomial $2P$ first, for example we can build it from: 
\begin{itemize}
	\item a Fermi multiplet with weights in $-\hat{\bL}(p)$, $\hbar$-charge $-\frac12$ and R-charge $+\frac12$;
	\item a chiral multiplet with weights in $-\hat{\bL}(p)$, $\hbar$-charge $+\frac12$ and R-charge $+\frac12$;
	\item a Fermi multiplet of R-charge $0$ and $\hbar$-charge $1$ valued in the adjoint of gauge group;
	\item a chiral multiplet of zero R and $\hbar$-charges in the adjoint of gauge group;
	\item for each $U(1)$ factor in the gauge group, add: (1) a Fermi multiplet (of zero R and $\hbar$ charges) of charge $+1$ under this $U(1)$ and charge $+1$ under the dual topological $U(1)_{\rm top}$; (2) a chiral multiplet (of zero R and $\hbar$ charges) of charge $+1$ under this $U(1)$ and charge $-1$ under the dual topological $U(1)_{\rm top}$.
\end{itemize}
The first two items form together a $-\hat{\bL}(p)$-valued $(2,2)$ chiral multiplet; the first four items on the list form the matter content of the interval reduction of a 3d theory with the same boundary conditions $\sB_{\bL(p),p}$ on the two ends; the last item was cooked up to have the anomaly polynomial $4 \Tr(\mathbf{f}\mathbf{f}_z)$. We write down the flavored elliptic genus of this system following the standard techniques \cite{Benini:2013nda,Gadde:2013dda,Benini:2013xpa}.
	
According to \cite{Benini:2013nda}, the Ramond sector one-loop determinants of a chiral and a Fermi multiplet of weight $x$ are, respectively:
\begin{align}
\label{oneloopdet}
Z_{\rm Ch}(x,q) = \frac{1}{q^{\frac1{12}}\vartheta(x)},\qquad
Z_{\rm F}(x,q) = q^{\frac1{12}}\vartheta(x),
\end{align}
where we used the theta function defined in \eqref{varth}.
	
One can also find slightly different expressions for $Z_{\rm Ch}$ and $Z_{\rm F}$ in the literature \cite{Bobev:2015kza}, the difference originating from the choice of regularization in the infinite products computing the one-loop determinants (\cite{Bobev:2015kza} use the zeta-function regularization):
\begin{equation}
\tilde{Z}_{\rm Ch}(x,q) = \frac{1}{q^{\frac1{12}}e^{\frac{(\log x)^2}{2\log q}}\vartheta(x)},\qquad
\tilde{Z}_{\rm F}(x,q) = q^{\frac1{12}}e^{\frac{(\log x)^2}{2\log q}}\vartheta(x).
\end{equation}
Notice that the $\tilde{Z}_{\rm Ch}$, $\tilde{Z}_{\rm F}$ and $Z_{\rm Ch}$, $Z_{\rm F}$ are related by the modular transformation. Indeed, if we first rewrite in the Jacobi's notations:
\begin{equation}
q^{\frac1{12}}\vartheta(x) = \frac{i\theta_1(a;\tilde\tau)}{\eta(\tilde\tau)},
\end{equation}
where $x=e^{2\pi i a}$ and $q=e^{2\pi i \tilde\tau}$, then we observe:
\begin{equation}
\frac{\theta_1(\frac{a}{\tilde\tau};-\frac1{\tilde\tau})}{\eta(-\frac1{\tilde\tau})} = -i e^{\frac{i\pi a^2}{\tilde\tau}} \frac{\theta_1(a;\tilde\tau)}{\eta(\tilde\tau)},
\end{equation}
where $\frac{i\pi a^2}{\tilde\tau} = \frac{(\log x)^2}{2\log q}$ is exactly the factor that distinguishes $Z$ and $\tilde{Z}$. While in the purely 2d theories one could attempt to argue for one expression over the other, (since the two 1-cycles of $\mathbb{T}^2$ are not entirely equivalent: one is thought of as the spatial circle and another is the temporal circle), in our case it is not so. Our $\mathbb{T}^2$ is the spatial torus, and there is no reason to choose say $Z_{\rm Ch}$ over $\tilde{Z}_{\rm Ch}$. This is of course a manifestation of mixed global-gravitational anomaly. We can just make a choice to work with $Z_{\rm Ch}$ and $Z_{\rm F}$, and proceed to compute the partition function of an auxiliary system described above as:
\begin{align}
\label{aux_ell_gen}
\prod_{(w,f)\in \hat{\bL}(p)} \frac{\vartheta(s^w x^f \hbar^{1/2})}{\vartheta(s^w x^f \hbar^{-1/2})} \times \prod_{\alpha\in\mathbf{adj}(G)}\frac{\vartheta(s^\alpha \hbar)}{\vartheta(s^\alpha)}\times\prod_{a=1}^{{\rm rk}(\mathbf{A}')} \frac{\vartheta(s^{c_a}z_a)}{\vartheta(s^{c_a}z_a^{-1})}.
\end{align}
Here $(w,f)\in \hat{\bL}(p)$ denotes gauge and flavor weights, so $s$ and $x$ are gauge and flavor fugacities (flat connections on $\mathbb{T}^2$). The last product is over the $U(1)$ factors in the gauge group, and $s^{c_a}$ are the corresponding fugacities as before.
	
As written, the expression \eqref{aux_ell_gen} is divergent, since the second product includes a factor of $\left(\frac{\vartheta(\hbar)}{\vartheta(1)} \right)^{{\rm rk}(G)}$, and $\vartheta(1)=0$. This problem is resolved once we substitute $s=s^{(p)}(x,\hbar)$, after which the numerators in the first product in \eqref{aux_ell_gen} include precisely ${\rm rk}(G)$ factors of $\vartheta(1)$ that cancel the divergence. This precise cancellation may be traced back to the vacua being isolated and massive. Thus we can safely use the expression \eqref{aux_ell_gen} with $s=s^{(p)}(x,\hbar)$, and the state $\langle\sB_{\bL(p),p}|$ transforms as its square root. We write down the transformation rules like before:
\begin{align}
\label{qshifts}
z_a\mapsto q z_a:\quad  &\Psi \mapsto s^{-c_a} \Psi\cr
s \mapsto q^{\mathbf{v}} s:\quad &\Psi \mapsto \hbar^{-\frac12\sum_{\hat{\bL}(p)}\langle\mathbf{v},w\rangle}\prod_{a=1}^{{\rm rk}(\mathbf{A}')}z_a^{-\langle \mathbf{v},c_a\rangle} \Psi\cr
x\mapsto q^{\mathbf{w}}x:\quad &\Psi \mapsto \hbar^{-\frac12\sum_{\hat{\bL}(p)} \langle\mathbf{w},f\rangle} \Psi\cr
\hbar\mapsto q^2\hbar:\quad &\Psi \mapsto \hbar^{-|G|} q^{-|G|} \prod_{(w,f)\in \hat{\bL}(p)}\left(s^{-w} x^{-f}\right) \Psi,
\end{align}
from which the transformation rules of $\langle\sB_{\bL(p),p}|$ follow simply by the specialization $s=s^{(p)}(x,\hbar)$. In equation \eqref{qshifts}, $\mathbf{v}=(\mathbf{v}_1,\dots,\mathbf{v}_{{\rm rk}(G)})$ is some gauge coweight and $\mathbf{w}=(\mathbf{w}_1,\dots,\mathbf{w}_{{\rm rk}(G_H)})$ is a flavor coweight.
	
The $\Psi$ in \eqref{qshifts} is a section of some line bundle $\cF$  on ${\rm E}_\bfT(X) = {\rm Ell}_\bfT(X)\times \cE_{\mathbf{A}'}$. Specialization $s=s^{(p)}(x,\hbar)$ means taking its pullback to the $p$-th irreducible component of ${\rm E}_\bfT(X)$, where we find a line bundle that the state $\langle\sB_{\bL(p),p}|$ is valued in. We can easily construct an explicit model expression for $\Psi$ that is meromorphic and transforms as in \eqref{qshifts}:
\begin{align}
\label{genPsi}
{\Psi} \ =\  \Xi \times \prod_{(w,f)\in \hat{\bL}(p)} \frac{\vartheta \left (s^w x^f \hbar^{\frac 12} \right)}{\vartheta(s^w x^f)\vartheta(\hbar^{\frac 12})} \times \vartheta \left( \hbar^{\frac 12} \right)^{2|G|}\times\prod_{a=1}^{{\rm rk}(\mathbf{A}')} \frac{\vartheta \left( s^{c_a}z_a \right) }{\vartheta(s^{c_a})\vartheta(z_a)},
\end{align}
where $\Xi$ is an undetermined meromorphic function (section of a trivial line bundle) on ${\rm E}_\bfT(X)$, which is thus invariant under \eqref{qshifts}. Any meromorphic section of $\cF$ has the form \eqref{genPsi} with some meromorphic function\footnote{The same statement about the elliptic meromorphic ambiguity $\Xi$ also applies to the earlier expression \eqref{model_section_Dp}.} $\Xi$. Note that for $\Psi$ to be meromorphic with respect to its arguments, we treat $\hbar^{\frac 12}$ as a symbol denoting a coordinate on the double cover, which is natural, given that the fields with half-integral $U(1)_\hbar$ charges are present in the theory.
	
Now we know that states $|\Psi_p\rangle$ are sections of the line bundle characterized by a model section \eqref{model_section_Dp}, while the bundle that $\langle \sB_{\bL(p),p}|$ is valued in is characterized by a model section \eqref{genPsi}. The overlap $\langle \sB_{\bL(p_1),p_1}|\Psi_{p_2}\rangle$ is thus characterized by the product of \eqref{model_section_Dp} and \eqref{genPsi}, with $s=s^{(p_2)}(x,\hbar)$ in the former and $s=s^{(p_1)}(x,\hbar)$ in the latter. We see that if $p_1=p_2$, the terms of the form $\frac{\vartheta(sz)}{\vartheta(s)\vartheta(z)}$ containing topological fugacities $z_a$ cancel. If $p_1\neq p_2$, they do not cancel, leading to a nontrivial dependence on the $z_a$ (and thus nontrivial monodromies along the cycles of $\cE_{\mathbf{A}'}$).

\section{Janus interfaces}\label{sec:Januses}
A family of quantum field theories depending on a continuous parameter $g$ can often be studied in a $(d+1)$-dimensional background, in which $g$ varies along a chosen spatial direction $y\in\R$. Such backgrounds break part of the Poincare group, preserving the $d$-dimensional translations and Lorentz transformations. Of special interest are the configurations where $g$ stays constant in most of the spacetime, and only changes in a small or finite region $y\in(a,b)$. Following \cite{Bak:2003jk}, such a region is often referred to as a ``Janus interface,'' and we will adopt this nomenclature. Janus interfaces for the gauge coupling (and their AdS gravity dual solutions) were first studied in the 4d $\cN=4$ SYM. The non-supersymmetric interface appeared in \cite{Bak:2003jk,Freedman:2003ax,Clark:2004sb}, and the SUSY-preserving generalization was studied in \cite{Clark:2005te,DHoker:2006vfr,DHoker:2006qeo,DHoker:2007zhm}. A generalization that includes a spatially varying theta angle was introduced in \cite{Gaiotto:2008sd}, which is an important ingredient in the construction of the S-duality kernel $T[G]$ \cite{Gaiotto:2008ak}.

One can of course consider varying parameters other than the gauge coupling. For example, \cite{Gaiotto:2009fs} constructs Janus interfaces for twisted chiral couplings in 2d $\cN=(2,2)$ theories, and \cite{Goto:2018bci} studies such interfaces in 2d $\cN=(2,2)$ gauged linear sigma models (GLSM), where the coupling of interest is the complexified FI parameter. The latter type is closely related to one of the interfaces we discuss below. Much more central to our story is an interface implementing the change of mass parameters, more precisely, real masses in 3d $\cN=2$ gauge theories. Such interfaces, or more generally, backgrounds with spatially modulated masses, have recently attracted attention in the literature as well, see \cite{Anderson:2019nlc,Gran:2008vx,Arav:2018njv,Kim:2018qle,Kim:2019kns,Arav:2020asu,Arav:2020obl,Arav:2021tpk}, and \cite{DHoker:2009lky,Bobev:2013yra} for Janus solutions in M-theory on AdS$_4$. Mass Janus in the context closely related to the current paper has also played role in \cite{Bullimore:2017lwu}.

By adjusting one or several parameters of a given quantum field theory, we can land in different phases. This naturally leads to a special class of Janus interfaces that can be called phase interfaces, or phase walls. They interpolate between distinct phases in different spacetime regions, and understanding their properties allows to establish mapping of various objects across the phases (e.g., study the brane transport, see \cite{Herbst:2008jq,Clingempeel:2018iub,Brunner:2021cga,Brunner:2021ulc}, especially the recent work \cite{Galakhov:2021omc}). The interfaces we study in this paper will be of such type: they will interpolate between Higgs, Coulomb, mixed, as well as the CFT phases of the given quiver gauge theory.

Before proceeding with concrete constructions, note that in a Euclidean theory, the interface can be oriented in an arbitrary way due to the rotational symmetry of spacetime. In Lorentzian signature, however, as is always the case for extended objects, there exist physically distinct configurations depending on whether some of the interface directions are time-like, space-like, or null. The standard physical treatment of interfaces assumes that they are time-like, i.e., bound spatial regions. We do not consider null interfaces here (which can be thought of as infinite-boost limits of either time-like or space-like cases). The space-like interfaces, on the other hand, are more commonly understood as making couplings of the theory time-dependent. We can consider the limiting case, where the parameters of the theory change abruptly at a given moment of time, and stay constant afterwards. Such a discontinuous change of parameters, --- known as a quench, --- is actually under a better analytical control than a continuous change. The reason is that the wave function of the system cannot change immediately, and stays constant across the quench (or jumps in a controllable way if the Hamiltonian contains delta-functions), so the wave function right after the quench is known and serves as an initial condition for the Schr{\"o}dinger equation with the new values of parameters. This defines a codimension-one operator (interface) acting between the spaces of states of theories before and after the quench.

\subsection{FI Janus in 3d $\cN=2$}\label{sec:FIJanus}
We start with the simplest case of Janus interface corresponding to a $y$-dependent FI parameter $\zeta(y)$ in a 3d $\cN=2$ gauge theory. We can define it by a term in the Lagrangian:
\begin{equation}
\label{FIJanus}
\cL_{\rm FI} = \Tr (i\zeta(y) D - \zeta'(y) \sigma),
\end{equation}
which only differs from the usual real FI coupling by a derivative term $\zeta'(y)\sigma$. This coupling is consistent with 2d $\cN=(0,2)$ SUSY, and in the case of $\cN=4$ theories, -- with the entire 2d $\cN=(2,2)$ subalgebra. After we act with SUSY, to see that \eqref{FIJanus} is indeed invariant, we must integrate by parts and assume that fermions vanish at infinity. 

We could imagine integrating by parts before acting with SUSY, to replace $-\zeta'(y)\sigma$ by $\zeta(y) D_y\sigma$. However, this produces a boundary term at $y=\pm\infty$, unless $\zeta(y)\sigma$ takes the same vale at both infinities. The latter does not have to be the case in general, as the vacua at $y\to\pm\infty$ might be different. We could of course contemplate defining the $y$-dependent FI coupling as $\Tr(i\zeta(y) D + \zeta(y)D_y\sigma)$, in which case it is supersymmetric on the nose, without any extra integration by parts. We do not do so, because \eqref{FIJanus} has a more natural property that it coincides with the standard FI term whenever $\zeta(y)$ is constant.

Let us vary $\zeta(y) \mapsto \zeta(y) + \delta\zeta(y)$, while keeping its asymptotic values intact, $\delta\zeta(\pm\infty)=0$. The variation of the action is
\begin{equation}
\delta S_{\rm FI} = \int \dd^3 x\, \Tr (i\delta\zeta(y) D - \delta\zeta'(y) \sigma) = \int \dd^3 x\, \Tr i\delta\zeta(y)(D - iD_y\sigma),
\end{equation}
where now we integrate by parts without earning a boundary term. Furthermore, using our definition of the elliptic supercharge $\cQ$ (and conventions in the Appendix \ref{app:conv}), we find that
\begin{equation}
D - iD_y\sigma = \cQ \left[\frac12 (\lambda_- - \bar\lambda_-) \right].
\end{equation}
In other words, $\delta S_{\rm FI}$ is $\cQ$-exact, showing that we can vary $\zeta(y)$ arbitrarily, as long as the asymptotic (or boundary) values are fixed.

One could worry whether this naive argument might fail due to some sort of singularities and monodromies in the space ${\rm Lie}\,{Z\left( {}^L G \right)}$ of FI parameters, such that the partition function would depend on the homotopy class of $\zeta(y)$ (viewed as a map from $\R$ to the space of FI parameters). After all, such  things do happen in 2d, for example in the space of twisted chiral couplings, such as the FI-theta terms (see, e.g., \cite{Goto:2018bci} and references therein). Such phenomena would be relevant for computations of the $\cQ_A$-invariant quantities, since they are meromorphic in the $\C^\times$-valued ``K\"ahler'' parameter in 3d (which reduces to the usual complex FI-theta parameter in 2d). This is a complex combination of the 3d FI parameter with the topological symmetry holonomy along the B-cycle, and in the space of such complex parameters, one naturally finds singularities and monodromies (they are important to the story of \cite{Okounkov:2016sya,Aganagic:2016jmx}).

In the case of $\cQ$-invariant observables, however, the real FI parameters do not admit a natural complexification. 
As explained in Section \ref{sec:prel}, the holonomies on $\bE_\tau$ for topological symmetries combine to form the elliptic K{\"a}hler parameter $z$ in 3d, while the FI term remains real. More precisely, the $\cQ$-invariant quantities are meromorphic in the elliptic equivariant and K{\"a}hler parameters, but they are locally constant in the real FI parameter (the local constancy remains true after reduction to 2d and 1d as well). We observe quite a different phenomenon: 
the space of real FI parameters is subdivided by the walls into chambers. Generically, FI parameters land inside some chamber, but we can adjust them to be on the wall. What matters is in what chamber (or on what wall) $\zeta(y)$ starts at $y=-\infty$, and where it ends at $y=+\infty$.

One can understand walls physically as those values of parameters, at which the mixed branches open up. For generic values of the FI parameters, the Coulomb branch is often lifted,\footnote{At least this is true in the class of theories we study. One can easily imagine theories that do not have enough FI parameters, for example if we replace some $U(N)$ gauge groups by $SU(N)$. In that case one can still consider turning on the available FI parameter as generic as possible, thus lifting some
part of the Coulomb branch.} except for a discrete set of points where it intersects the Higgs branch (they all sit at the origin when the masses vanish). From the Coulomb branch point of view, the FI parameters $\zeta$ are seen as 
generators of the topological symmetries, i.e. the maximal torus $\mathbf{A}'$ of $G_C$. These unlifted points are simply the fixed points, denoted by $\cM_C^{\mathbf{A}'}$, of the torus $\mathbf{A}'$ acting on the Coulomb branch $\cM_C$. By tuning the FI parameters $\zeta$ to non-generic values, we may un-lift certain directions of the Coulomb branch. In other words, we pass to a sub-torus of $\mathbf{A}'$, which leaves a larger subset of $\cM_C$ fixed, which we may call $\cM_C^\zeta$, to indicate its dependence on $\zeta$. Assuming that we do not turn on masses, we end up with a mixed branch: we can explore both the subset of Coulomb branch that is not lifted, and the corresponding Higgs branch (partly resolved by the non-generic FI parameters).

We thus see how to determine the walls. Thinking of $\zeta$ as the generators of topological symmetries, the walls correspond to those non-generic values, at which the fixed locus $\cM_C^\zeta$ becomes larger. In the absence of masses, $\cM_C = {\rm Spec}(R_C)$, and we can find an ideal in $R_C$, which we denote as $(\zeta R_C)$, generated by the images of elements in $R_C$ under the action of $\zeta$. This is the ideal of functions vanishing along $\cM_C^\zeta$, and $R_C / (\zeta R_C) \cong \C[\cM_C^\zeta]$. Therefore, simply looking at how the generators of $R_C$ are acted on by $\zeta$, is enough to determine $\cM_C^\zeta = {\rm Spec}\left(R_C /(\zeta R_C) \right)$. Again, the walls are located at those values of $\zeta$, at which $R_C /(\zeta R_C)$ jumps up in size. More formally, this can be described in a way similar to the Higgs branch case in \cite{Maulik:2012wi}. Namely, the Coulomb analog of Definition 3.2.1 from \cite{Maulik:2012wi} is: the $\mathbf{A}'$-roots are the $\mathbf{A}'$-weights $\{\alpha_i'\}$ occurring in the normal bundle to $\cM_C^{\mathbf{A}'}$ inside $\cM_C$. 
Then the walls are just the root hyperplanes $(\alpha_i')^\perp$, defined by the vanishing of ${\alpha}_{i}'$. The space of FI parameters ${\rm Lie}\,{Z\left( {}^L G \right)}\cong \mathfrak{a}'_\R$ (where $\mathfrak{a}'_\R$ is the Cartan subalgebra of $G_C$, or equivalently a Lie algebra of $\mathbf{A}'$) is partitioned into the chambers:
\begin{equation}
\mathfrak{a}_\R' \setminus \bigcup (\alpha_i')^\perp = \bigsqcup \mathfrak{C}'_i.
\end{equation}

\subsection{Mass Janus in 3d $\cN=2$}
The FI Janus described above can be understood as a supersymmetric background for the vector multiplets gauging the topological symmetry. We can turn on precisely the same SUSY configuration for the background vector multiplets gauging flavor symmetries. This results in a solution that we call a mass Janus. More precisely, it is characterized by the following vevs for the flavor vector multiplet:
\begin{equation}
\sigma^{\rm f} = m (y),\quad D_\R^{\rm f} = {\ii}\frac{\dd m(y)}{\dd y}.
\end{equation}
In other words, we make real masses $y$-dependent, and also turn on the auxiliary field $D_\R^{\rm f}$ in the background vector multiplet. Position dependent masses alone completely break SUSY, while the presence of $D_\R^{\rm f}$ vev allows to preserve half of supersymmetry. 

If we only consider $\cN=4$ real masses, this background preserves 2d $\cN=(2,2)$ SUSY, which is then broken down to $\cN=(0,2)$ by the $\hbar$ equivariant parameter. We never turn on the $U(1)_\hbar$ real mass, but if we did, it would also preserve $\cN=(0,2)$.

Let us explicitly write down the extra terms that appear in the Lagrangian due to this deformation. They form a supersymmetric mass term for the chiral multiplets:
\begin{equation}
\cL_m = \bar\phi m(y)^2 \phi + 2 \bar\phi\sigma m(y)\phi - \bar\phi \frac{\dd m(y)}{\dd y}\phi + i\bar\psi m(y)\psi,
\end{equation}
where the dynamical vector multiplet scalar $\sigma$ appears due to the term $\bar\phi (m+\sigma)^2\phi$ in the full action. Note that $\sigma$ acts in the gauge group representation, while $m(y)$ acts in the flavor group representation. Now let us vary the mass profile, while keeping the asymptotic values intact:
\begin{equation}
m(y) \mapsto m(y) + \delta m(y),\quad \delta m(\pm\infty)=0.
\end{equation}
The latter condition allows integration by parts, and we find:
\begin{equation}
\delta S_m = \int\dd^3 x\, \delta \cL_m = \int \dd^3 x\,\delta m^a(y) \left( 2\bar\phi (m(y)+\sigma)f^a\phi + \frac{\dd}{\dd y}(\bar\phi f^a \phi) + {\ii}\bar\psi f^a \psi \right),
\end{equation}
where we write $f^a$ for the generators of the flavor symmetry maximal torus, so $\delta m = \delta m^a f^a$. Once we attempt to check whether this expression is $\cQ$-exact, we find:
\begin{equation}
\label{mass_vari}
\delta S_m = \int\dd^3 x\, \delta m^a(y) \cQ (\bar\phi f^a \bar\rho \psi + \rho\bar\psi f^a\phi) + {\ii} \int\dd^3 x\, \delta m^a(y) (\bar{F}f^a \phi + \bar\phi f^a F),
\end{equation}
where $(\bar\rho, \rho)$ parameterize the two supersymmetries of the 3d $\cN=2$ algebra that are broken by the background, and $(F,\bar{F})$ are auxiliary fields in the chiral multiplets. So we see that $\delta S_m$ is not $\cQ$-exact off shell.

To improve the situation, we can put the auxiliary field on shell in the path integral. Let us see why it is legal to do so. Suppose we study some partition or correlation function:
\begin{equation}
Z = \int \cD[\text{fields}]\, e^{-S} (\dots),
\end{equation}
and we want to understand how it depends on the mass profile $m(y)$. Assuming that the insertions $(\dots)$ do not depend on masses, the variation $\delta m(y)$ only affects the action. If we are off shell, we can drop the $\cQ$-exact term in \eqref{mass_vari} without affecting the answer, and take the other term down from the exponent:
\begin{equation}
\delta_m Z = -{\ii}\int \cD[\text{fields}]\, e^{-S} (\dots)\int \dd^3x\, \delta m^a(y) (\bar\phi f^a F + \bar{F} f^a\phi).
\end{equation}
Let us assume that the insertions $(\dots)$ do not include fields $(F, \bar{F})$. We integrate out $(F,\bar{F})$ by replacing them with their on shell values, simply because the action is quadratic in $(F,\bar{F})$, so they cannot pick up any extra contractions with $(\dots)$. In other words, we apply the Dyson-Schwinger equations following from $\int \cD F \cD \bar{F}\, \frac{\delta}{\delta F}(\dots)=\int \cD F \cD \bar{F}\, \frac{\delta}{\delta\bar{F}}(\dots)=0$. As we know, the on shell values are expressed in terms of the superpotential: $F = \frac{\partial\bar{W}}{\partial\bar{\phi}}$ and $\bar{F}=-\frac{\partial W}{\partial\phi}$. Thus we find:
\begin{equation}
\label{dZeqdW}
\delta_m Z = -{\ii}\int \cD[\text{fields}]\, e^{-S} (\dots)\int \dd^3x\, \delta m^a(y) \left(\bar\phi f^a \frac{\partial\bar{W}}{\partial\bar{\phi}} - \frac{\partial W}{\partial\phi} f^a\phi\right).
\end{equation}
This expression vanishes for arbitrary insertions $(\dots)$,  since the superpotential $W$ is invariant under the flavor symmetry, in particular:
\begin{equation}
\delta m^a\frac{\partial W}{\partial \phi} f^a \phi = \delta m\circ W=0,
\end{equation}
where $\delta m\circ$ denotes action by the flavor algebra element $\delta m=\delta m^a f^a$.

We thus obtain that $\delta_m Z=0$ under the variation $m(y)\mapsto m(y)+\delta m(y)$ of the real mass profile. Therefore, the Janus interface for real masses has the same universality property as the Janus for real FI parameters: The precise shape of $m(y)$ does not matter in the $\cQ$-cohomology, only the asymptotic values are relevant for the $\cQ$-invariant computations. We could anticipate this from the mirror symmetry, given that the same property holds for the FI Janus, but that, strictly speaking, would apply to the $\cN=4$ case only, while here we clearly see that the 3d $\cN=2$ SUSY is enough to prove this. There is a caveat, though, related to the fact that the mass Janus background violates unitarity if $y$ is treated as (Euclidean) time, and violates positivity of the Hamiltonian if $y$ is treated as space (the violation occurs in the region where $m'(y)\neq 0$). We will get back to this in the next section.

Just like in the FI case, we do not expect to encounter any monodromies as we vary the real mass profile $m(y)$, at least when we study $\cQ$-invariant observables. Instead, there are real codimension-1 walls, again as in the FI case. At generic values of masses, the Higgs branch $X$ of the theory is lifted, except for a certain locus $X^{\mathbf A}$ fixed by the flavor torus $\mathbf{A}$. Its $\mathbf{A}$-roots $\{\alpha_i\}$ are defined as the $\mathbf{A}$-weights that occur in the normal bundle to $X^{\mathbf A}$ in $X$, which is the content of Definition 3.2.1 from \cite{Maulik:2012wi}. Then the walls are the root hyperplanes $\alpha_i^\perp$, and the space $\mathfrak{a}_\R$ of real masses (which is the Cartan of the $\cN=4$ flavor symmetry group $G_H$) is partitioned into chambers:
\begin{equation}
\mathfrak{a}_\R \setminus \bigcup \alpha_i^\perp = \bigsqcup \mathfrak{C}_i.
\end{equation}
Inside the chambers, the Higgs branch is lifted, except for the fixed locus $X^{\mathbf A}$, which often is a discrete set of points corresponding to massive vacua. Specializing real masses to the walls makes some of the matter multiplets massless, partially unlifting the Higgs branch and restoring some of its flat directions. The full Higgs branch is restored at the origin of the mass space, where all the walls intersect. If we specialize to the walls both in the real mass and real FI parameter spaces, the mixed branches (involving the Higgs and Coulomb directions) open up.

Therefore, the mass Janus is also determined (in the $\cQ$-cohomology, or, equivalently, up to a quasi-isomorphism) by the choice of masses at $y=\pm\infty$. At each of the two infinities, or boundaries, we either pick a chamber or a particular component of the wall in the space of real masses. In the relation to stable envelopes later in this paper, we will often take masses to infinity, which really means taking them to infinity within a chosen chamber.

To finish this subsection, let us summarize what we have so far. There are two classes of half-BPS interfaces in 3d $\cN=2$ theories preserving 2d $\cN=(0,2)$ SUSY. They are given by Janus interfaces for real masses and real FI parameters. We will wrap these interfaces on the elliptic curve $\bE_\tau$, and view them as operators acting in the Hilbert space $\cH[\bE_\tau]$ on $\bE_\tau$. When working in the $\cQ$-cohomology of $\cH[\bE_\tau]$, both interfaces become ``canonical'', in the sense that they do not depend on the profiles of masses/FI parameters, and at most depend on their asymptotic values. We will denote these operators as
\begin{equation}
\cJ_m(m_1,m_2),\quad \cJ_\zeta(\zeta_1,\zeta_2),
\end{equation}
for the mass and FI cases respectively, where the arguments represent values of the corresponding parameters at $y=-\infty$ and $+\infty$. In fact, the operators will only capture the chamber or wall in which the asymptotic values lie. Thus they provide certain chamber-dependent linear maps acting between the spaces of vacua on $\bE_\tau$. Reducing from 3d to 2d and 1d preserves all these statements, and only changes the periodicity of equivariant and K\"ahler parameters, $(x,\hbar)$ and $z$, respectively. Janus interfaces, considered as operators in the $\cQ$-cohomology, obey the following natural identities:
\begin{align}
\label{Janus_comp}
\cJ_m(m_1,m_2)\cJ_m(m_2,m_3)&=\cJ_m(m_1,m_3),\cr \cJ_\zeta(\zeta_1,\zeta_2)\cJ_\zeta(\zeta_2,\zeta_3)&=\cJ_\zeta(\zeta_1,\zeta_3),
\end{align}
which are proved by stacking the interfaces one after another, and using their independence of the intermediate values of masses or FI parameters.

\section{Analysis via supersymmetric quantum mechanics}\label{sec:SQM_section}
It is illuminating to have an alternative approach to the mass Janus based on the operator formulation. In this section, we develop it in the case of quantum mechanics only, as generalizations to higher dimensions are conceptually straightforward. Passing to 2d corresponds to replacing the target space by its loop space, and working equivariantly with respect to the loop rotations (i.e., treating them as part of flavor symmetries). Passing to 3d, likewise, corresponds to working with the double loops.
\subsection{Time-dependent Morse function}
Let us start with a system of free chiral multiplets $\phi^i$ with the real mass matrix $m(y)={\rm diag}(m_1(y),\dots,m_n(y))$ that depends on the Euclidean time $y$. We switch off the equivariant parameters first for simplicity, so the action is:
\begin{equation}
L = |\dot\phi^i|^2 + \bar\phi^i (m^2-\dot{m})\phi^i - {\ii}\bar\psi^i \gamma^3 \dot\psi^i + {\ii}\bar\psi^i m\psi^i + \bar{F}^i F^i,
\end{equation}
where the dot denotes the $y$ derivative. This action preserves two supercharges, and their sum is our $\cQ$. 
Let us rewrite this in terms of the real coordinates $X^\mu$, $\mu=1, \ldots, 2n$ on the target, such that $\phi^j=\frac1{\sqrt 2}(X^j + {\ii} X^{n+j})$. For the fermions $(\psi^i_\alpha,\bar\psi^i_\alpha)$, the components $\psi^i_2$ and $\bar\psi^i_2$ are relabeled as the tangent fermion $\psi^\mu$, while the components $\psi^i_1, \bar\psi^i_1$ are relabeled as the cotangent fermion $\chi_\mu$. The action becomes:
\begin{align}
L &= \frac12 g_{\mu\nu} \dot{X}^\mu \dot{X}^\nu + \frac12 g^{\mu\nu}\partial_\mu f \partial_\nu f + \frac12 g^{\mu\nu}F_\mu F_\nu +{\ii}\chi_\mu (\dot\psi^\mu + g^{\mu\nu}D_\nu \partial_\lambda f \psi^\lambda)- \frac{\partial f}{\partial y},
\end{align}
where $g_{\mu\nu}=\delta_{\mu\nu}$, and
\begin{equation}
f(X^\mu,y) = \frac12 X m(y) X=\frac12 \sum_j m_j(y)(X^j X^j + X^{n+j} X^{n+j}),
\end{equation}
and $\frac{\partial f}{\partial y}=\frac12 X \dot{m} X$ only includes the derivative of $m(y)$ due to its explicit $y$-dependence. The supercharge $\cQ$ acts as follows:
\begin{align}
\cQ X^\mu &= \psi^\mu,\qquad \cQ \chi_\mu = {\ii} \dot{X}_\mu + F_\mu + {\ii}\partial_\mu f,\cr
\cQ\psi^\mu&=0,\qquad\ \ \cQ F_\mu = -{\ii} \dot{\psi}_\mu - {\ii} D_\mu\partial_\nu f \psi^\nu.
\end{align}
The action can also be written as:
\begin{equation}
\label{qm_Qex-lagr}
L = \cQ\left[-\frac{\ii}{2}\chi_\mu(\dot{X}^\mu + \partial^\mu f + {\ii} F^\mu) \right] - \frac{\dd f}{\dd y}.
\end{equation}
If the mass is constant, i.e., $\partial f/\partial y=0$, then this is just the well-known quantum mechanics of \cite{Witten:1982im}, with $f$ playing the role of Morse function. The essence of our problem is, therefore, in making the Morse function time-dependent. We could also be slightly more general, and instead of starting with chiral multiplets in a theory with four supercharges, simply consider a general $\cN=(1,1)$ quantum mechanics with time-dependent Morse function $f$.

It is straightforward to find expressions for the Hamiltonian $H$ and the supercharge $\cQ$:
\begin{align}
H &= \frac12 p^\mu p_\mu + \frac12 \partial^\mu f \partial_\mu f + i\chi_\mu D^\mu \partial_\nu f \psi^\nu - {\ii} \frac{\partial f}{\partial t},\cr
\cQ &= -{\ii}\psi^\mu p_\mu + \psi^\mu \partial_\mu f.
\end{align}
Here $p_\mu$ is a conjugate momentum to $X^\mu$, and $\chi_\mu$ is canonically conjugate to $\psi^\mu$. We also temporarily Wick-rotated to the real time $t=-{\ii}y$. After quantization, as usual identifying $\psi^\mu=\dd x^\mu$, $\chi_\mu={\ii}\iota_{\partial_\mu}$, and choosing an ordering, we get:
\begin{align}
H &= \frac12\left( \{\dd,\dd^*\}   + \partial^\mu f \partial_\mu f - D^\mu\partial_\nu f \left[\iota_{\partial_\mu},\dd x^\nu\right] \right) - {\ii}\frac{\partial f}{\partial t}, \\
\label{Q-def}
\cQ &= \dd + \dd f\wedge. 
\end{align}
If $f$ were time-independent, there would be one more conserved supercharge $\cQ^\dagger$ given by
\begin{equation}
\cQ^\dagger = \dd^* + \iota_{\nabla f}.
\end{equation}
Using this expression as the definition of $\cQ^\dagger$ even in the time-dependent case, we have:
\begin{equation}
\label{HeqQQ}
H = \frac12 \{\cQ, \cQ^\dagger\} - {\ii}\frac{\partial f}{\partial t}.
\end{equation}
The operator $\cQ$ as written in \eqref{Q-def} explicitly depends on time through $f$. Therefore, when we say that $\cQ$ is conserved, we mean that the following equation holds:
\begin{equation}
\frac{\dd \cQ}{\dd t}\equiv {\ii} [H,\cQ] + \frac{\partial \cQ}{\partial t}=0.
\end{equation}
In general, we let the Morse function interpolate between two asymptotics:
%   , so it becomes manifest that variations of $f$ give $\rQ$-exact deformations of $\rH$. More precisely, we keep the asymptotics fixed:
\begin{align}
f(X,t) \longrightarrow \begin{cases}
f_-(X), \text{ as } t\to-\infty,\\
f_+(X), \text{ as } t\to+\infty,
\end{cases}
\end{align}
which determines a $\cQ$-closed ``interface'' between the two quantum mechanics with Morse functions $f_-$ and $f_+$. This operator is clearly not unitary (we do not expect a generic operator to be unitary anyways), which is manifested in our construction by $H$ in \eqref{HeqQQ} being non-Hermitian. Hermiticity of the Hamiltonian is violated precisely in the region where $\frac{\partial f}{\partial t}\neq 0$.

We can perform a non-unitary ``canonical transformation'' on wave functions:
\begin{equation}
\Psi \mapsto \Omega=e^{f}\Psi,
\end{equation}
after which we obtain:\footnote{Here one has to take into account that $H$ transforms as connection, i.e., it gets shifted by $i\frac{\partial f}{\partial t}$, in addition to the conjugation by $e^f$.}
\begin{align}
\rQ  &=e^{f}\cQ e^{-f} = \dd,\qquad \rG=e^f \cQ^\dagger e^{-f} =\dd^* + 2\iota_{\nabla f},\cr
\rH &= \frac12 \{\rQ,\rG\}= \frac12 \{\dd,\dd^*\} + \cL_{\nabla f},
\end{align}
and the inner product becomes:
\begin{equation}
\label{OmegaProd}
\langle \Omega_1, \Omega_2\rangle = \int e^{-2f} \star\bar\Omega_1\wedge \Omega_2,
\end{equation}
This transformation corresponds to dropping the ``topological term'' in the Lagrangian \eqref{qm_Qex-lagr}:
\begin{align}
\label{LocLagr}
L &= \cQ\left[-\frac{i}{2}\chi_\mu(\dot{X}^\mu + \partial^\mu f + {\ii}F^\mu) \right]\cr 
&=\frac12 g_{\mu\nu} (\dot{X}^\mu+\partial^\mu f)(\dot{X}^\nu+\partial^\nu f) + \frac12 g^{\mu\nu}F_\mu F_\nu +{\ii}\chi_\mu (\dot\psi^\mu + g^{\mu\nu}D_\nu \partial_\lambda f \psi^\lambda).
\end{align}

In such a formulation, $\rQ=\dd$ is a time-independent operator, and the Hamiltonian $\rH$ is $\rQ$-exact. Because of that, the de Rham cohomology class of the state $\Omega$ does not change under the time evolution (and in particular is not sensitive to how exactly $f$ interpolates between $f_-$ and $f_+$). This seems like a bad news: The interface between $f_-$ and $f_+$ appears transparent (i.e., equal to the identity) in the cohomology.

This argument is totally true when the target space of quantum mechanics is compact. When it is non-compact, however, there is a caveat. Our interface is defined via a non-unitary deformation of the theory, and under such a non-unitary evolution, an $L^2$-normalizable wave function on the non-compact space may turn unnormalizable. Then we conclude that the corresponding state ``quits'' the Hilbert space.

\subsubsection{A toy example}
This can be illustrated in a simple example of the target space $\C$, with the Morse function $f=\frac12 m |z|^2$, where $m$ is a real coefficient that may depend on time. For time-independent $m$, we can write two states annihilated by the Hamiltonian:
\begin{equation}
\psi_0=e^{-f},\qquad \psi_2=e^f \dd z\wedge\dd\bar{z},
\end{equation}
where $\psi_0$ is $L^2$ for $m>0$, and $\psi_2$ is $L^2$ for $m<0$.

Now if the mass $m$ depends on time, the function $\psi_0$ still solves the Schr{\"o}dinger equation (with our non-unitary deformation), while $\psi_2$ does not. We can thus start with $m>0$ and the normalizable vacuum $\psi_0$ in the far past, and change the mass to a negative value $-m$ in the far future. The state $\psi_0$, still being a solution, will remain equal to $e^{-f}$, which is not $L^2$ for the negative mass in the far future. Thus the quantum state $\psi_0$ ``quits'' the Hilbert space due to the non-unitary evolution, while $\psi_2$ ``enters'' it. What does make sense, however, is to compute the overlap of $\psi_0$ with an $L^2$ state $\psi_2$ in the future. In the case at hands, we simply get zero,
\begin{equation}
\langle \psi_2|\psi_0\rangle=0,
\end{equation}
as $\psi_0$ and $\psi_2$ have fermion numbers zero and two, respectively (and the fermion number is conserved).

Despite the appearance of unnormalizable states, the path integral computing the (possibly non-unitary) evolution is perfectly well-defined, since the Hilbert space norm plays no role in such computations. Let us elaborate this in slightly more detail. The initial state $\psi_0=e^{-f}$, after the similarity transformation, corresponds to a form
\begin{equation}
\Omega=1,
\end{equation}
which in the path integral (with the ``topological term'' dropped),
\begin{equation}
\int \cD X \cD F \cD \psi \cD\chi \, e^{-\int \dd y \frac12 (\dot{X} + \nabla f)^2 +\dots}
\end{equation}
can be taken as the initial condition, realized via the supersymmetric Neumann boundary:
\begin{equation}
\label{bcN1}
\chi_\mu\big|=0,\quad (\dot{X} + \nabla f)\big|=0.
\end{equation}
To compute the wave function that is produced at the output, we need to impose Dirichlet boundary conditions at the other end (in the far future):
\begin{equation}
\label{bcD1}
\psi^\mu\big|=0,\quad X^\mu\big|=x^\mu.
\end{equation}
This will compute the super wave function $\Omega(x,\theta)$ at the output. In principle, it only gives $\Omega(x,\theta)\big|_{\theta=0}$, but the conservation of the fermion number (and because $\Omega=1$ in the far past) guarantees that there are no terms proportional to $\theta^\mu\equiv \dd x^\mu$. One can easily perform localization with \eqref{bcN1} and \eqref{bcD1} at the two ends: All the Fermi zero modes are eliminated; the BPS equation $\dot{X}+\nabla f=0$, subject to $X^\mu\big|=x^\mu$ in the far future, has a unique solution; and the one-loop determinants cancel. One finds the expected answer $\Omega(x,\theta)=1$. Note that this computation does not depend on the normalizability of state at any point in time, and in fact the norm simply plays no role.

Using the inner product \eqref{OmegaProd} written in terms of $\Omega$, we indeed see that $1$ has zero overlap with $\Omega_2=e^{2f}\dd z\wedge\dd\bar{z}$ corresponding to $\psi_2$:
\begin{equation}
\langle 1, \Omega_2\rangle = \int e^{-2f} \star 1\wedge \Omega_2=0.
\end{equation}
If we want to have a non-zero overlap, i.e., a non-trivial vacuum-vacuum transition, turning on equivariant parameters helps.

\subsection{Equivariant extension}
It is straightforward to reintroduced equivariance into the problem, which we neglected at first. Namely, suppose that $V(\epsilon)$ generates an isometry of the target (with $\epsilon$ an equivariant parameter), and that $f$ is invariant, $\cL_{V(\epsilon)} f=0$. Then the conserved supercharge is extended in the usual way:
\begin{equation}
\cQ = \dd + \dd f\wedge + \iota_{V(\epsilon)},
\end{equation}
so that $\cQ^2=\cL_{V(\epsilon)}$. The conjugate supercharge (broken by the time-dependence of $f$) is
\begin{equation}
\cQ^\dagger = \dd^* + \iota_{\nabla f} + V^\flat(\bar\epsilon)\wedge,
\end{equation}
where $V^\flat$ is the one-form dual to $V$ via the metric. The Hamiltonian is still given by the same formula:
\begin{equation}
H = \frac12 \{\cQ, \cQ^\dagger\} - {\ii}\frac{\partial f}{\partial t},
\end{equation}
so that $\cQ$ is indeed conserved,
\begin{equation}
{\ii}[H,{\cQ}] + \frac{\partial\cQ}{\partial t}=0.
\end{equation}
We can still perform the similarity transformation and obtain:
\begin{align}
\rQ=e^f \cQ e^{-f}=\dd + \iota_{V(\epsilon)},\quad \rG = e^f \cQ^\dagger e^{-f}=\dd^* + 2\iota_{\nabla f} + V^\flat(\bar\epsilon)\wedge,\quad
\rH=\frac12 \{\rQ, \rG\}.
\end{align}
The variation of $f(X,t)$ by $\delta f(X,t)$ is still a $\rQ$-exact deformation, therefore, everything we said about the interpolation between $f_-$ and $f_+$ defining the interface is still true.

\subsubsection{Toy example continued}
We can now resume discussing the toy example with $\C$ as a target and $f=\frac12 m |z|^2$, but now also including the equivariant parameter $\epsilon$ for the $U(1)$ rotation of the complex plane. With some work, we can identify the unique normalizable ground state of this system:
\begin{align}
\label{OmegaM}
\Psi=e^{-f} \Omega^{(m)}, \quad \Omega^{(m)}=\frac{\epsilon}{\sqrt{2\pi(\omega-m)}} e^{-\frac12(\omega-m)|z|^2 - {\ii}\frac{\omega-m}{2\epsilon}\dd z\wedge\dd\bar{z}}, \text{ where } \omega=\sqrt{m^2 + |\epsilon|^2}.
\end{align}
This solution is $L^2$-normalizable for any $m\in \R$, circumventing the problem we had before.

We can use this solution to compute the transition amplitude across the mass-changing interface. However, the explicit solution $\Omega^{(m)}$ as presented above does not solve the Schrodinger equation with time-dependent mass. Instead of finding such a more general solution, we may simply use the independence on the precise mass profile, and assume that it behaves as a step function, remaining constant most of the time and experiencing a jump at $t=0$:
\begin{equation}
m(t)=\begin{cases}
m_1, &t<0\\m_2, &t>0
\end{cases}
\end{equation}
This jump, due to the term $-{\ii}\frac{\partial f}{\partial t}$, adds a delta-function potential in $H$, while the Hamiltonian $\rH$ remains finite at all times. Thus the wave function $\Omega$, which evolves according to $\rH$, stays continuous across the jump, while $\Psi$ jumps. Suppose the system was in the vacuum state $\Omega^{(m_1)}$ for $t<0$. Right after the jump, it still is described by the wave function $\Omega^{(m_1)}$, but it is no longer a vacuum. We have to compute its overlap with the new vacuum $\Omega^{(m_2)}$, using the inner product as in \eqref{OmegaProd}, with $f$ corresponding to the region $t>0$:
\begin{equation}
R[m_1 \to m_2]=\int e^{-m_2 |z|^2} \star\bar{\Omega^{(m_2)}}\wedge \Omega^{(m_1)}.
\end{equation}
Where we denoted by $R[m_1 \to m_2]$ the overlap that computes the vacuum-vacuum transition amplitude across the mass-changing Janus interface, with the mass $m_1$ at the input and $m_2$ at the output. It is straightforward to compute the integral:
\begin{equation}
R[m_1\to m_2]=\sqrt{\frac{\omega_2-m_2}{\omega_1-m_1}}.
\end{equation}
Notice that the transitivity property $R[m_2\to m_3]\circ R[m_1\to m_2]=R[m_1\to m_3]$ is obeyed, which confirms the expectation that only the asymptotic values of masses are important.

The object like $R[m\to -m]$ is the prototype of R-matrices in our future applications, while $R[m\to 0]$ is the prototype of stable envelopes.

Another observation is that $\Omega^{(m)}$ from \eqref{OmegaM} reduces to $1$ as $m\to+\infty$, up to a divergent factor of $\frac{\epsilon}{\sqrt{2\pi(\omega-m)}}$. The support of $1$ is the entire $\C$, which happens to be the attractor of the origin for $m>0$. Likewise, if we send $m\to-\infty$ (and ignore the same pre-factor $\frac{\epsilon}{\sqrt{2\pi(\omega-m)}}$ that now vanishes in this limit), $\Omega^{(m)}$ tends to $\delta^2(z)\dd z\wedge \dd\bar{z}$. Its support is just the origin of $\C$, which is the only point in the attractor for $m<0$. The singular pre-factors appear as an irrelevant nuisance, but they might become more problematic in higher-dimensional QFT. However, if we have twice as many supercharges, we are talking about the hypermultiplet that consists of a chiral $I$ of mass $m$ (and equivariant parameter $\epsilon$) and a chiral $J$ of mass $-m$ (and equivariant parameter $\tilde\epsilon$). In this case, the problematic prefactors combine into:
\begin{equation}
\frac{\epsilon\tilde\epsilon}{2\pi\sqrt{(\omega-m)(\tilde{\omega}+m)}} \to \begin{cases}
\frac{\epsilon\tilde\epsilon}{2\pi|\epsilon|} & \text{if } m\to +\infty\\
\frac{\epsilon\tilde\epsilon}{2\pi|\tilde\epsilon|} & \text{if } m\to -\infty
\end{cases}
\end{equation}
and have finite limits. Thus in the $m\to \pm\infty$ limits, the product of two wave functions $\Omega^{(m)}\tilde{\Omega}^{(-m)}$ has a reasonable behavior:\footnote{Here $\tilde\Omega$ is the same as $\Omega$ with the equivariant parameter $\epsilon$ replaced by $\tilde\epsilon$.} it simply becomes proportional to the delta-form supported on the attractor of the origin.

It is also instructive to look at the R-matrix of one hypermultiplet, which is given by $R[m\to -m]\cdot\tilde{R}[-m\to m]$ (with tilde signifying that the second factor contains the equivariant parameter $\tilde\epsilon$). It equals:
\begin{equation}
\sqrt{\frac{\omega-m}{\omega+m}}\sqrt{\frac{\tilde\omega+m}{\tilde\omega-m}} \approx \frac{|\epsilon|}{|\tilde\epsilon|}\quad \text{as } m\to\infty.
\end{equation}
Of course, this computation only produces the answer up to a phase (as the wave function \eqref{OmegaM} was only defined up to a phase), hence we might as well write
\begin{equation}
\frac{\epsilon}{\tilde\epsilon}
\end{equation}
as an answer. To get such a simple result, it was important to take the limit $m\to\infty$. 

In this limit, the vacuum wave functions $\Omega^{(m)}\tilde{\Omega}^{(-m)}$ become delta-forms, as we now know, and can be mimicked by the appropriate boundary conditions given by the complex Lagrangian submanifolds (on which the delta-form was supported). If we factorize the R-matrix as
\begin{equation}
R[0\to -m]\cdot\tilde{R}[0\to m]\times R[m\to 0]\cdot\tilde{R}[-m\to 0],
\end{equation}
the factor $R[m\to 0]\cdot\tilde{R}[-m\to 0]$ corresponds to the jump of masses from $(m,-m)$ to $(0,0)$. This jump is where the state $\Omega^{(m)}\tilde{\Omega}^{(-m)}$ enters the massless region, and in the limit $m\to +\infty$, it can be replaced by the $\frac12$-BPS boundary conditions for the pair $(I,J)$ of chirals:\footnote{Such boundary conditions will be justified in Section \ref{sec:effective}.}
\begin{equation}
\label{example_interval}
I\big|=0,\quad \partial_y J\big|=0.
\end{equation}
Likewise, the jump $R[0\to -m]\cdot\tilde{R}[0\to m]$ can be replaced by exactly the same boundary conditions at the location were masses jump from $(0,0)$ to $(-m,m)$ (in the limit $m\to+\infty$). Hence, we end up with an interval partition function with the same boundary conditions \eqref{example_interval} on both ends. Such a partition function is computed by identifying zero modes: the chiral multiplet $I$ subject to $I\big|=0$ on both ends leads to one Fermi zero mode, while $J$ leads to one bosonic zero mode, and together they give the expected answer ${\epsilon}/{\tilde\epsilon}$. The interval non-zero modes completely cancel out in the one-loop determinants (see, e.g., \cite{Sugiyama:2020uqh}), which is an expected features of theories that do not break SUSY dynamically: Changing the size of the interval is realized via the Hamiltonian evolution, which is Q-exact and does not affect BPS observables. Thus SUSY computations should not depend on the length of the interval, and, consequently, can only capture the contribution of the interval zero modes.

One of the many lessons we learned from this example is the importance of $m\to\infty$ limit. Without it, the answer is a complicated function of $m$, which is not what we want. We are interested in quantities that have natural interpretation in the equivariant cohomology, and thus only contain equivariant parameters (and K\"ahler parameters in the higher-dimensional generalizations).

\subsection{Including holomorphic superpotential}
In the case of K\"ahler target space, it is straightforward to also include the holomorphic superpotential. Namely, suppose we pick a holomorphic function of chiral multiplets $W$, which is invariant with respect to the gradient flow,
\begin{equation}
g^{\mu\nu}\partial_\mu f \partial_\nu W=0,
\end{equation}
and preserves the symmetry, $\cL_{V(\epsilon)}W=0$.
More generally, one may allow $W$ to be multi-valued, while $\partial W$ must be a globally defined $(1,0)$-form. Then the theory admits a deformation $f\mapsto f + {\rm Re}(\kappa W)$ that preserves all the symmetries, where $\kappa$ is a complex number. In our conventions $\kappa=-2i$, and the preserved supercharge becomes:
\begin{equation}
\cQ = \dd + \dd f\wedge + \iota_{V(\epsilon)} + 2\dd (\Im W)\wedge,
\end{equation} 
while the supercharge that is broken by the time-dependence of $f$ is:
\begin{equation}
{\cQ}^\dagger = {\dd}^* + \iota_{\nabla f} + V^\flat(\bar\epsilon)\wedge + 2\iota_{\nabla \Im(W)}.
\end{equation}
As expected, we still have
\begin{equation}
H=\frac12 \{\cQ,\cQ^\dagger\} - {\ii}\frac{\partial f}{\partial t},
\end{equation}
ensuring that the conservation equation ${\ii} [H,\cQ]+\frac{\partial\cQ}{\partial t}=0$ holds.

\subsection{Gauged quantum mechanics}
In the general setting of $\mathcal{N}=(1,1)$ quantum mechanics, gauging is performed by adding vector multiplets in the adjoint of $\mathfrak{g}$. The appropriate vector multiplet contains a gauge field, a complex scalar, and a pair of fermions, $(A_t, \sigma_\C, \bar\sigma_\C, \lambda, \bar\lambda)$, with the SUSY:
\begin{align}
\cQ\sigma_{\C} \ &\ =\ 0, \quad \cQ\bar\sigma_\C=\bar\lambda,\quad \cQ A_t={\ii}{\lambda},\cr
\cQ\lambda\ &\ = \ {\ii} D_t\sigma_{\C},\quad \cQ\bar\lambda={\ii}[\sigma,\bar\sigma],
\end{align}
and the action:
\begin{align}
\label{gaugeActionQM}
\cL&=\frac1{e^2}\cQ\, {\Tr} \left( -{\ii}\lambda D_t\bar\sigma_\C -{\ii}\bar\lambda[\bar\sigma_\C,\sigma_\C] \right)\cr
&=\frac1{e^2}{\Tr} \left( D_t {\bar\sigma}_{\C} D_{t} {\sigma}_{\C} + {\ii}\lambda D_t\bar\lambda -[{\sigma}_{\C},{\bar\sigma}_{\C}]^2 +{\ii}[\lambda,\bar\sigma_\C]\lambda - {\ii}[\bar\lambda,\sigma_\C]\bar\lambda\right).
\end{align}
The SUSY Noether current for the vector multiplet is $\cQ_{\mathfrak{g}}=\bar\lambda D_t\sigma + [\sigma_\C,\bar\sigma_\C]\lambda$, which quantum mechanically becomes
\begin{equation}
\cQ_{\mathfrak{g}} = \bar\partial + \Tr [\sigma_\C,\bar\sigma_\C]\iota_{\partial/\partial\bar\sigma_\C}\qquad \text{acting on } \Omega^{(0,\cdot)}(\mathfrak{g}\otimes\C).
\end{equation}
This immediately tells us how to gauge some symmetry $G$ of a general quantum-mechanical sigma model with target $X$. For that, we simply promote the equivariant parameter $\epsilon$ to a complex coordinate $\sigma_\C$ on $\mathfrak{g}_\C\equiv \mathfrak{g}\otimes\C$. Now the Hilbert space is identified with (the square-integrable part of)
\begin{equation}
\left[\Omega^\cdot(X)\otimes \Omega^{(0,\cdot)}(\mathfrak{g}_\C)\right]^{\mathfrak{g}},
\end{equation}
where we take the $\mathfrak{g}$-invariant subspace, due to the Gauss law constraint enforced by the gauge field $A_t$. The total $\cQ$ operator is:
\begin{align}
\cQ = \dd_X + \dd f\wedge + \iota_{V(\sigma_\C)} + \bar\partial_{\mathfrak{g}} + \Tr [\sigma_\C,\bar\sigma_\C]\iota_{\partial/\partial\bar\sigma_\C},
\end{align}
where $V(\sigma_\C)$ generates the $G$-action on $X$, and we ignored the obvious possibility of additional (ungauged) flavor symmetries. We used $\dd_X$ to denote the de Rham differential on $X$, and $\bar\partial_{\mathfrak{g}}$ -- the Dolbeault differential on $\mathfrak{g}_\C$.

One can then identify:
\begin{align}
\cQ^\dagger = \dd_X^* + \iota_{\nabla f} + V^\flat(\bar\sigma_\C)\wedge + \bar\partial_{\mathfrak{g}}^* + \Tr [\sigma_\C,\bar\sigma_\C]\dd\bar\sigma_\C
\end{align}
The Hamiltonian is still given by $H=\frac12 \{\cQ,\cQ^\dagger\}-{\ii}\frac{\partial f}{\partial t}$.

\subsection{General setting}
Our general setup in 1d is a gauged quantum mechanics with $X=\cR\times \bar\cR\times \mathfrak{g}_\C\times\mathfrak{g}$, where the $\cR\times\bar\cR$ factor describes hypermultiplets, $\mathfrak{g}_\C$ corresponds to the adjoint-valued chiral $\Phi$, and $\mathfrak{g}$ -- to the real scalar $\sigma$. As usual, $\cR$ is a complex representation of the gauge group $G$, so that $X$ admits an action of $G$. Note that, in addition to $\mathfrak{g}_\C\times\mathfrak{g}$, we add yet another factor of $\mathfrak{g}_\C$ to the target space in the process of gauging, as explained earlier.

Then we include the Morse function (real superpotential)
\begin{equation}
f=\bar\phi (\sigma + m(y))\phi - \zeta_\R\cdot\sigma \equiv m(y)\cdot\mu^{\rm f}_\R + {\sigma}\cdot{\tilde\mu}^{\rm g}_\R,
\end{equation}
where $\phi$ runs over all chiral multiplets, $\zeta_\R$ is the real FI parameter, and $\mu^{\rm f}_\R$, $\tilde\mu^{\rm g}_\R$ are the real flavor and gauge moment maps, with the latter including the real FI term. In theories with eight supercharges, it is customary to separate the contributions of $\zeta_\R$ and the adjoint chiral $\Phi$ in $\tilde\mu^{\rm g}_\R$:
\begin{equation}
\tilde\mu^{\rm g}_\R = \mu^{\rm g}_{\R} + \frac1{e^2}[\Phi,\bar\Phi] - \zeta_\R.
\end{equation}
Finally, the holomorphic superpotential is as in the 3d theory:
\begin{equation}
W=\tilde{Q}\Phi Q,\quad Q\in\cR,\quad \tilde{Q}\in\bar\cR,
\end{equation}
and we include the equivariant deformation by $V(\epsilon)$ for all flavor symmetries that were not gauged.

At low energies, assuming $\zeta_\R$ is generic enough, the theory is effectively described as a nonlinear sigma model (NLSM) into the Higgs branch, with the Morse potential and the equivariant deformation present. The Morse potential simply follows from the Morse potential of the gauge theory by setting $\sigma=0$:
\begin{equation}
\label{HiggsMorse}
f=m(y)\cdot \mu^{\rm f}_\R = \bar\phi m(y)\phi,
\end{equation}
where the latter expression is restricted to the zero level of the real and complex gauge moment maps, which then naturally descends to the quotient $\cM_H$. The equivariant deformation is still encoded in a vector field $V(\epsilon)$ on $\cM_H$.

\subsection{Localization and gradient trajectories}
The quantum mechanics described above localizes to the gradient flows for $f$, 
either in the gauge description or in the NLSM description, 
which is already manifested by the first term in \eqref{LocLagr}.

In the gauge theory description, the metric of the target space is $\frac1{e^2}\Tr \dd\sigma^2 + \sum_{\phi\in\text{chirals}}|\dd\phi|^2$, and the corresponding gradient flow is
\begin{align}
\label{gradflow}
D_y\phi&=-(\sigma + m(y))\phi,\quad \text{for each chiral } \phi,\cr
D_y\sigma&=-e^2\tilde\mu^{\rm g}_\R.
\end{align}
%Additionally, equivariant deformation enforces \emph{almost everywhere} the localization on the fixed locus of the corresponding symmetry,
%\begin{equation}
%\label{equiLoc}
%V(\epsilon)=0,
%\end{equation}
%where $\epsilon$ collectively denotes equivariant parameters for flavor symmetries (which were called $x$ before), and the parameter $\hbar$ for the $U(1)_\hbar$ symmetry. Condition \eqref{equiLoc} should be understood as equivariant localization under the symmetry acting on the moduli space of solutions to the gradient flow equations \eqref{gradflow}.

In the NLSM description (cf. \cite{Frenkel:2006fy, Frenkel:2007ux}), one instead finds gradient flow equations for the Morse function on the Higgs branch \eqref{HiggsMorse},
\begin{equation}
\label{gradflowNLSM}
\partial_y{X}^\mu = -g^{\mu\nu}\partial_\nu f,
\end{equation}
where $g$ is the classical metric on the Higgs branch (which is known to receive no quantum corrections).

Both types of equations admit solutions that connect $\mathbf{A}$-fixed points on the Higgs branch. Suppose $p_1$ and $p_2$ are two such points. In the NLSM description, the solution to \eqref{gradflowNLSM} is a flow from $p_1$ to $p_2$ that manifestly belongs to a single $\mathbf{A}_\C$-orbit on the Higgs branch. Gradient flows that solve the gauge-theoretic BPS equations \eqref{gradflow}, however, do not remain on the Higgs branch: While at the end points $D_y\sigma=0$, so the real moment map equation $\tilde\mu^{\rm g}_\R=0$ is obeyed, the intermediate region has $D_y\sigma\neq 0$, so $\tilde\mu^{\rm g}_\R\neq 0$ and the trajectory gets off the Higgs branch, viewed as a subquotient inside some bigger space. More precisely, it leaves the Higgs branch corresponding to the fixed value of the FI parameter $\zeta_\R$, but still remains within the union of complexified gauge orbits of the Higgs branch points, at least if the value of $D_y\sigma$ is generic (so the trajectory stays away from the unstable locus). At some non-generic moments of time, the trajectory could, in principle, pass through the unstable region. In truth, however, this does not happen, because the first equation in \eqref{gradflow} simply describes the flow as the complexified gauge and flavor transformation that depends on $y$. We can illustrate the relationship between the gauge and NLSM descriptions in the following pictorial way:
\begin{figure}[h]
	\centering
	\includegraphics[scale=0.1]{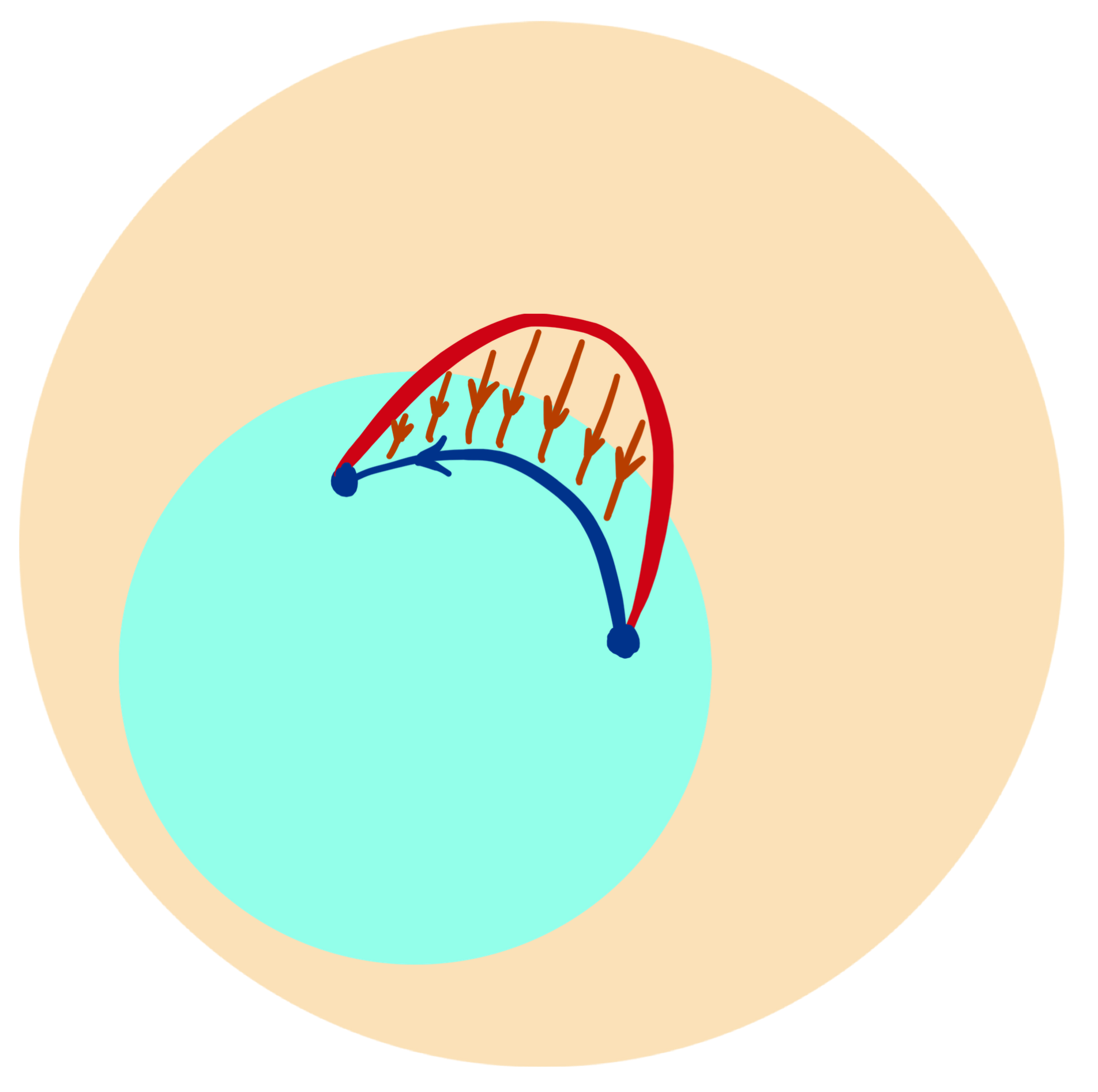}
	\caption{The blue curve represents the gradient trajectory on the Higgs branch solving \eqref{gradflowNLSM}, the red curve represents the gauge-theoretic gradient trajectory that solves \eqref{gradflow}, and the orange arrows represent the complexified gauge transformation that relates the two. Since all points along the blue curve are stable, so are the points of the red curve.}
\end{figure}\\

It is therefore clear that the gauge theoretic and NLSM gradient trajectories are in one to one correspondence with each other. It could be more intuitive to work with the latter (especially since in both approaches trajectories start and end on the Higgs branch), but the gauge theoretic description is more useful for explicit computations of the matrix elements.

\subsubsection{Example of a solution}
It is instructive to write down an explicit solution of the GLSM gradient flow equations in at least one example. To this end, consider a $U(1)$ gauge theory with a pair of charge $+1$ hypers, $(I_1, J_1)$ and $(I_2, J_2)$, where the chirals $I_i$ both have gauge charge $+1$, while their flavor charges are opposite and taken to be $\pm1$. For a non-zero FI parameter $\zeta$ (chosen to be positive), the theory has $T^*\C P^1$ as its Higgs branch. The $U(1)$ flavor symmetry has two fixed points given by the North and South poles of the zero section $\C P^1$. We turn on a real mass $m>0$ for the flavor symmetry. Let us assume that it takes the special value obeying
\begin{equation}
e^2\zeta = 2m^2.
\end{equation}
In this case we can write an elementary analytic solution to the flow equations (the solution also exists for general $m$, but takes a more complicated form). It has $J_1=J_2=0$, as well as $\Phi=0$ for the adjoint chiral. The non-zero fields are given by:
\begin{align}
I_1 &= \frac{\sqrt{\zeta}}{1+ e^{2my}},\cr
I_2 &= \frac{\sqrt{\zeta}}{1+ e^{-2my}},\cr
\sigma&= m\tanh(my).
\end{align}
We see that at $y=-\infty$, $(I_1,I_2)=(\sqrt{\zeta},0)$, while at $y=+\infty$, $(I_1,I_2)=(0,\sqrt{\zeta})$. At the same time, $\sigma$ interpolates between $-m$ and $m$. Thus, the trajectory connects the two fixed points in the base of the Higgs branch, which obeys $|I_1|^2 + |I_2|^2=\zeta$. At the intermediate times we have
\begin{equation}
0<|I_1|^2 + |I_2|^2<\zeta,
\end{equation}
clearly showing that the trajectory gets off the Higgs branch without leaving the stable locus. This solution demonstrates some of the key features of the gradient trajectories: (1) it takes infinite time to complete the transition between two critical points; (2) the trajectory stays close to the critical points most of the time, only significantly diverging from them in a time interval of order $1/m$; (3) the transition is effectively instantaneous in the $m\to \infty$ limit.

\subsubsection{Broken, or concatenated, trajectories}
Back to the general discussion, an important phenomenon is that of ``broken trajectories'': in addition to simple one-component flows connecting two fixed points, there exist composite trajectories obtained by concatenating several simple segments. They must be taken into account due to the following reason. The leading term in the localizing action is given by:
\begin{equation}
s\int\dd y (D_y X + \nabla f)^2.
\end{equation}
As we take $s\to\infty$, the path integral localizes to integration over those trajectories, on  which $\int\dd y (D_y X + \nabla f)^2$ is vanishingly small. The space of such trajectories has many components: (1) they can be either of the form $X_0(y) + \delta X(y)$, where $\delta X(y)$ is a small fluctuation around a simple gradient trajectory $X_0(y)$ (obeying $D_y X_0 + \nabla f(X_0)=0$) that connects just two fixed points $p_1$ and $p_2$; (2) or they can be trajectories that approximate the concatenated gradient flows going through multiple fixed points, $p_1 \to p_2\to \dots \to p_n$. Now, the concatenated gradient trajectories are not, strictly speaking, solutions: Passing from $p_1$ to $p_2$ requires an infinite amount of time, so a solution that goes from $p_1$ to $p_2$, and then to $p_3$, already requires ``twice-infinite'' amount of time, and cannot be written as a single function $X: \R\to {\rm Target}$. However, there exist approximate solutions $X(y)$, which get arbitrarily close to the concatenated trajectory $p_1\to\dots\to p_n$, and on which the action $\int\dd y (D_y X + \nabla f)^2$, therefore, is arbitrarily small:\footnote{To construct an example of an approximate solution, consider $n$ gradient trajectories $X_i: \R \to {\rm Target}$, such that $X_i(+\infty) = X_{i+1}(-\infty)$. Let us cut off the domain of $X_1(y)$ to $y\in (-\infty,T)$, the domain of $X_n$ -- to $y\in (-T,\infty)$, and the domains of $X_2(y),\dots,X_{n-1}(y)$ -- to $y\in(-T,T)$. Next we glue trajectories together in a smooth way, such that $X_{i}(T)$ and $X_{i+1}(-T)$ are connected by a small curve $s_i$ (minimal in some convenient sense). In the $T\to\infty$ limit, this glued trajectory becomes an exact solution, while at finite $T$, the interpolating curves $s_i$ can be interpreted as Brownian motion in the vicinity of the fixed points that allows us to jump between different components of the concatenated trajectory.}\\
\begin{figure}[h]
	\centering
	\includegraphics[trim={0 0 12cm 0},scale=0.2]{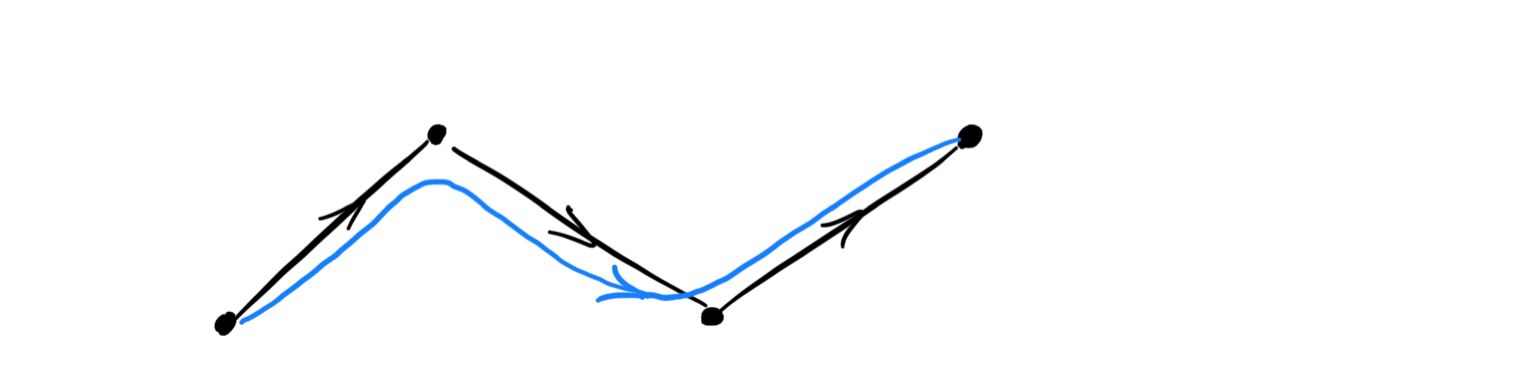}
	\caption{A concatenated gradient trajectory (in black) and an approximate solution (in blue), which is described by a continuous function of time $y\in\R$.}
\end{figure}\\

Integration over such approximate solutions is expected to produce the one-loop determinant over the fluctuations around the concatenated trajectory, even though such a trajectory itself does not belong to the field space of the theory. This suggests that we may also consider a partial compactification of the field space that includes the concatenated trajectories as ``points at infinity''. Then the path integral over such a space admits the usual localization, both on simple and concatenated gradient trajectories. The multi-component instantons, also known as ``multi-instantons'', are long-known in the literature, see for example \cite{Coleman:1978ae} for a similar claim that they lie at infinity. A point of view advocated in \cite{Nekrasov:2018pqq} is that they should be thought of as complex saddle points of the analytically continued theory. In the current work, we will not directly address these issues. Instead we will develop an effective description of the mass Janus in the next section, where this question does not even arise, yet the effect of ``broken trajectories'' is clearly present.

The necessity to include contributions from the concatenated trajectories is, in a sense, forced upon us by the non-compactness of spacetime: There exist smooth field configurations on $\R$ that are close to such trajectories, on which the action is vanishingly small, implying that they contribute in the supersymmetric localization. Other subtleties of similar nature might arise, which also require some sort of compactification of the field space. Rather than trying to plow through all these technical difficulties, one may take an alternatively route, where we cut off the non-compact ends of spacetime, replacing them by the appropriate boundary conditions that mimic the removed non-compact pieces. After this, we are left with compact spacetime, albeit with boundaries. Such a system is more straightforward to analyze.

Another way to argue for the need to take the broken trajectories into account, is to deform the theory by turning on an infinitesimally small complex FI parameter $\zeta_{\C}$. The ordinary gradient trajectories would limit, as ${\zeta}_{\C} \to 0$, to a
union of trajectories, broken at the intermediate critical points.

\subsubsection{Flavor equivariant parameters}
In the above discussion, we have been temporarily ignoring equivariant parameters, so now we turn them on. In the NLSM description, the equivariant deformation enforces \emph{almost everywhere} the fixed point condition:
\begin{equation}
\label{equiLoc}
V(\epsilon)=0,
\end{equation}
where $\epsilon$ collectively denotes equivariant parameters for flavor and $U(1)_\hbar$ symmetries, and $V(\epsilon)$ is the corresponding vector field on the NLSM target. 

Saying that \eqref{equiLoc} holds almost everywhere means that it can in fact be violated for very short periods of time. This is crucial for transitions between different fixed points to be possible, for otherwise, if \eqref{equiLoc} holds identically, it simply freezes us into a chosen fixed point. Fortunately, this equation is enforced via a term $s\int\dd y\, |V(\epsilon)|^2$ in the localizing action, which only has to vanish in the $s\to\infty$ limit. Thus it is enough to demand $\int \dd y\, |V(\epsilon)|^2 = o(s^{-1})$, which is a weaker condition than being identically zero. Suppose the Morse function $f$ on the Higgs branch corresponds to a real mass $m$. The transition between two consecutive fixed points takes an infinite amount of time, but most of that time the trajectory lingers close to one of the two fixed points. It only spends time of order $\frac1{m}$ significantly far from the fixed point. Therefore, on such a trajectory:
\begin{equation}
\int\dd y\, |V(\epsilon)|^2 \sim \frac1{m}\times {\rm const} \to 0,\quad \text{as } m\to\infty.
\end{equation}
Since we indeed take the $m\to\infty$ limit, the localization condition is obeyed, but the equation \eqref{equiLoc} is violated at time instances at which jumps between the critical points happen. We see that this is the order of limits issue: it is important to take $m\to\infty$ first (or at least much faster than $s\to\infty$). This reflects the fact that $m\to\infty$ is not just a computational trick: We really are interested in sending the real mass to infinity, such that the physical theory produces a delta-form state supported on the attractor of the fixed point.

If we try to include equivariant deformation in the GLSM description, however, we start running into all sorts of issues that ultimately have to do with the attempts to perform SUSY localization on the non-compact spacetime. Indeed, the equation \eqref{equiLoc} gets replaced by its linear version:
\begin{equation}
(\epsilon + \sigma_\C)\phi=0,
\end{equation}
where the quantum-mechanical gauge equivariant parameter $\sigma_\C$ gets replaced by flat connections in higher dimensions, in particular, the flat connection $A_{\bar z}$ on $\bE_\tau$ in the 3d case. It follows from this equation that at the different fixed points, $\sigma_\C$ takes different values, partly screening $\epsilon$ and allowing for $\phi$ to develop a vev. So $\sigma_\C$ has to change its values along the gradient flow to allow transitions between the fixed points, but is it possible under the BPS equations? If we use the gauge kinetic term \eqref{gaugeActionQM} for the localization (in the $e^2\to0$ limit), it appears that $\sigma_\C$ must be constant. The trick that allowed to relax the condition \eqref{equiLoc} does not work here: if $\sigma_\C$ changes by $\Delta \sigma_\C$ over time $m^{-1}$, it leads to the action of order $\int |D_y\sigma_\C|^2 \dd y \sim m |\Delta \sigma_\C|^2$, which is large in the $m\to\infty$ limit. On the other hand, one can find field configurations ``at infinity'' of the field space, such that $\int_{-\infty}^{\infty} |D_y \sigma_\C|^2\dd y=0$, yet $\int_{-\infty}^\infty D_y\sigma_\C\neq 0$. An easy example is
\begin{equation}
\sigma_\C =A \tanh\frac{y}{T},
\end{equation}
which is very slowly-changing for large values of $T$, and $\int |\partial_y \sigma_\C|^2\dd y$ vanishes in the $T\to\infty$ limit, yet $\int_{-\infty}^\infty \dd y\, \partial_y\sigma_\C=2A$. This suggests that it \emph{is} possible to change $\sigma_\C$ with time, as long as $\partial_y\sigma_\C$ is small enough. Again, this phenomenon clearly has its roots in the non-compactness of the time direction $\R$.

%The resolution of issues in matching the NLSM and GLSM descriptions (or rather, of the extra difficulties we encounter in the latter) most likely has to do with the fact that the localization limit $e^2\to 0$ and the NLSM limit $e^2\to\infty$ are different, corresponding to the UV and IR limits. One therefore has to properly identify parameters between these two limits. Since all mass-like parameters (including the equivariant parameters for flavor symmetries) grow in the IR, they have to be taken small in the UV, which likely helps to resolve the problem. Again, we are not going to pursue this direction, instead we will use localization on the interval in the next section.

It is expected that the detailed analysis of localization in GLSM on a non-compact spacetime would eventually resolve all the aforementioned problems. For example, it is conceivable that proper compactification of the field space contains BPS solutions that connect different fixed points, which we saw to be the case in the NLSM description. While it would be interesting to clarify this point, we instead describe a simpler approach in Section \ref{sec:effective}.

\section{Relation to stable envelopes}\label{sec:StabEnv}
As alluded to earlier, the Janus interface interpolating between zero and non-zero masses,
\begin{equation}
\cJ_m(0,m),
\end{equation}
plays the role of a building block in our story, especially due to the relations \eqref{Janus_comp}, $\cJ_m(m_1,m_2)=\cJ_m(m_1,0)\cJ_m(0,m_2)+\{\cQ,\dots\}$. We would like to have a more quantitative understanding of its properties, in particular, compute its matrix elements between the supersymmetric vacua in the limit of infinite masses. This will allow to make contact with the ideas of \cite{Maulik:2012wi,Aganagic:2016jmx}. A preview of what we find is: the stable envelopes introduced in those references (or, more precisely, the ``pole subtraction matrix'', which is proportional to stable envelopes) are realized via such mass Janus interfaces. In the following we focus primarily on the 3d case, with the 2d and 1d specializations following by the dimensional reduction as explained in Section \ref{sec:gauge_review}.

\subsection{Janus background and gradient flows}
Consider the mass Janus $\cJ_m(m, 0)$ wrapped on $\bE_\tau$ in the $\bE_\tau\times \R$ geometry, where the real mass changes from $0$ to $m$ in a small neighborhood of $y=0$. At $y=-\infty$ and $y=+\infty$ we fix the Higgs branch vacua (the FI terms $\zeta$ are taken to be $y$-independent), which may be different, and we would like to compute the transition amplitude as $m\to\infty$.

Even though we do not perform a full localization analysis of such transition amplitudes, in this subsection we provide a qualitative discussion, relating it to some ideas in the math literature. The path integral localizes onto the BPS configurations, identified in the previous section with (concatenated) gradient trajectories. In the terminology often used in the literature on SUSY gauge theories, we use the Higgs branch localization scheme. Let us list all the BPS equations here for convenience: %The equations can either be derived in flat space, or one can take the $\theta\to\frac{\pi}{2}$ limit of the squashed sphere equations (in which case we should remember that $D_y = \frac{1}{f(\theta)}D_\theta$), giving:
\begin{align}
\label{JBPS1}\partial W = \bar{\partial W}&=0,\\
\label{JBPS2}D_\alpha\phi = D_\varphi\phi&=0,\\
\label{JBPS3}D_\alpha\sigma=D_\varphi\sigma&=0,\\
\label{JBPS3.1}F_{\mu\nu}&=0,\\
\label{JBPS4}(D_y+\sigma + m(y))\phi&=0,\\
\label{JBPS5}D_y \sigma - e^2\mu_\R&=0.
\end{align}
The most important equations here are written in the last two lines. They are our familiar gradient flow equations describing an $\mathbf{A}_\C$-flow on the Higgs branch, which were extensively discussed earlier. Equations \eqref{JBPS3} play just a technical role, ensuring that $\sigma$ is constant along $\bE_\tau$, and equations \eqref{JBPS1} contain the complex moment map constraint. Equations \eqref{JBPS2} seem to imply that the chirals $\phi$ are convariantly constant along $\bE_\tau$, while \eqref{JBPS3.1} says that the gauge connection is flat. Both of the latter two are problematic on the noncompact space, as we already explained in the previous section, and lead to the wrong conclusion that transitions between fixed points are impossible. For that reason, we will develop an approach that avoids the noncompactness issues in the next subsection. For now, let us focus on the gradient flow part once more.

%Focusing on the equations
%\begin{align}
%(D_y+\sigma + m(y))\phi&=0,\\
%D_y \sigma - e^2\mu_\R&=0,
%\end{align}
%their Higgs branch image is a flow for the complexified flavor torus $\mathbf{A}_\C$. To see that, we drop the second equation in favor of a stability condition, and interpret the first equation as a flow for $\mathbf{A}_\C$, with $y$ playing the role of time. The variable $\sigma$ generates (together with the gauge field $A_y$) a complexified gauge transformation along the flow, which of course is immaterial due to the quotient by $G_\C$. 

To each fixed point $p\in X^{\mathbf{A}_\C}$ of the $\mathbf{A}_\C$ action on the Higgs branch $X$, one associates an attracting submanifold
\begin{equation}
{\rm Attr}(p)\subset X,
\end{equation}
swept by the simple gradient trajectories that end at $p$. One also introduces the full attractor
\begin{equation}
\bar{\rm Attr}(p)\subset X,
\end{equation}
swept by all the gradient flows, including the concatenated, or ``broken'', trajectories. The latter means that for a point $x\in \bar{\rm Attr}(p)$, there is a sequence of fixed points $(p_1,\dots,p_n)$, with each $p_i\in X^{\mathbf{A}_\C}$, such that there exists a concatenated flow
\begin{equation}
x\to p_1\to p_2 \dots\to p_n\to p,
\end{equation}
where ``$a\to b$'' means that the points $a$ and $b$ are connected by the gradient flow.  By analogy, we can consider trajectories that start at $p$ and define the repelling and the full repelling submanifolds:
\begin{equation}
{\rm Rep}(p)\subset X,\qquad \bar{\rm Rep}(p)\subset X.
\end{equation}
The attracting and repelling submanifolds depend on the choice of chamber $\mathfrak{C}$ in the space of real masses, which determines the pattern of gradient flows. It is clear that ${\rm Attr}(p)$ (or $\bar{\rm Attr}(p)$) for $\mathfrak{C}$ is the same as ${\rm Rep}(p)$ (or $\bar{\rm Rep}(p)$, respectively) for the chamber $-\mathfrak{C}$.

${\rm Attr}(p)$ and $\bar{\rm Attr}(p)$ play important role in the construction of stable envelopes \cite{Maulik:2012wi}, and they also show up in the description of Janus interfaces due to the presence of gradient flow equations. If we fix a massive vacuum $\langle p|$ (the ``out'' state) corresponding to $p\in X^{\mathbf{A}_\C}$ in the far future $y=+\infty$, where the real masses $m$ are present, then the corresponding state in the far past at $y=-\infty$, where masses are zero, is by definition captured in the $\cQ$-cohomology by:
\begin{equation}
\langle p|\cJ_m(m, 0).
\end{equation}
This state is supported, in the appropriate sense, on the full attractor $\bar{\rm Attr}(p)$. Likewise, if the real masses are turned on in the past and are zero in the future, and we fix a state $|p\rangle$ in the past, then at $y=+\infty$ we find the state
\begin{equation}
\cJ_m(0,m)|p\rangle,
\end{equation}
which is supported, in the appropriate sense, on the full repellent $\bar{\rm Rep}(p)$.

The sense in which these states have a certain support has already been worked out in the previous section. There we saw (in the case of quantum mechanics) that a vacuum wave function was $\Psi = e^{-f}\Omega$, with the equivariant differential form $\Omega$ given by the path integral that admits localization to the gradient trajectories. In the limit of infinite masses (that enter the Morse function $f$), this $\Omega$ becomes, up to a normalization factor, a delta-form supported on the union of gradient trajectories. Hence we claim that in the $m\to\infty$ limit, $\langle p|\cJ_m(m,0)$ with $e^{-f}$ stripped off becomes a distribution supported on $\bar{\rm Attr}(p)$ (representing class in the appropriate cohomology theory). Likewise, $\cJ_m(0,m)|p\rangle$, with the pre-factor $e^{-f}$ removed, becomes a distribution supported on $\bar{\rm Rep}(p)$.

We see from the above discussion that the matrix of $\cJ_m(0,m)$ is upper-triangular, with the matrix elements
\begin{equation}
\label{matrp1p2}
\langle p_2|\cJ_m(0,m)|p_1\rangle
\end{equation}
being non-zero when $p_2 \in \bar{\rm Rep}(p_1)$. The off-diagonal elements in this matrix are attributed to the flows that connect different fixed points, and the matrix elements that are at a distance more than $1$ from the diagonal are due to the ``broken trajectories''. We will see that the same exact structure emerges from the effective description of the next subsection.

Figures \ref{fig:flow_left} and \ref{fig:flow} below illustrate our setups. In general, we pick an arbitrary Higgs phase vacuum state $|\psi\rangle$ in the massless region and a fixed point $p$ at infinity of the massive region. The path integral then computes overlaps, either $\langle p|\cJ_m(m,0)|\psi\rangle$ or $\langle \psi|\cJ_m(0,m)|p\rangle$.
\begin{figure}[h]
	\centering
	\includegraphics[scale=0.7]{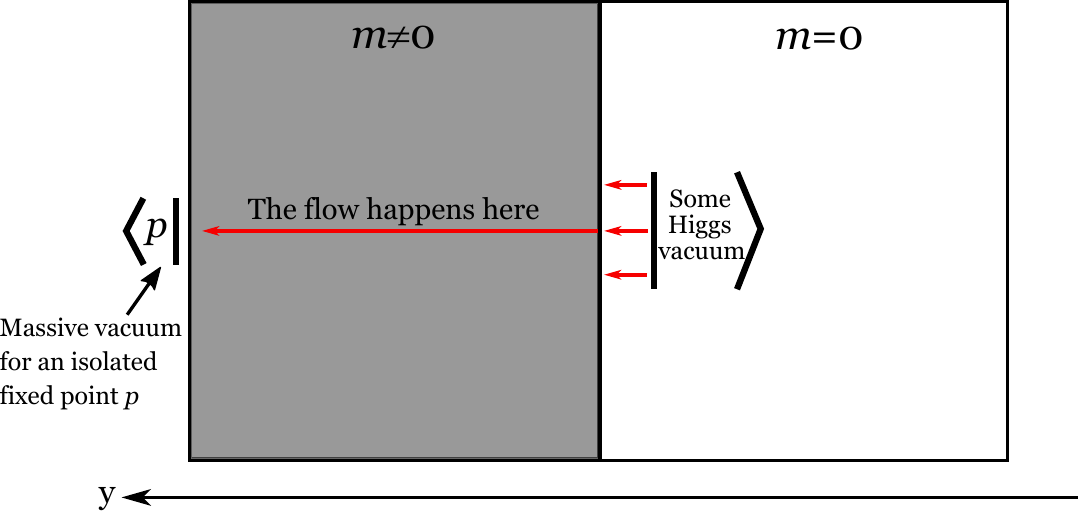}
	\caption{The setup defining $\langle p|\cJ_m(m,0)$. Here $\langle p|$ is a massive vacuum taken as  an ``out''-state in the far future $y\to+\infty$. In the region with zero masses, we start with an arbitrary Higgs phase vacuum wave function $\psi$. The overlap is computed by counting flows that start somewhere in the support of $\psi$ and end at the fixed point $p$.}\label{fig:flow_left}
\end{figure}\\
\begin{figure}[h]
	\centering
	\includegraphics[scale=0.7]{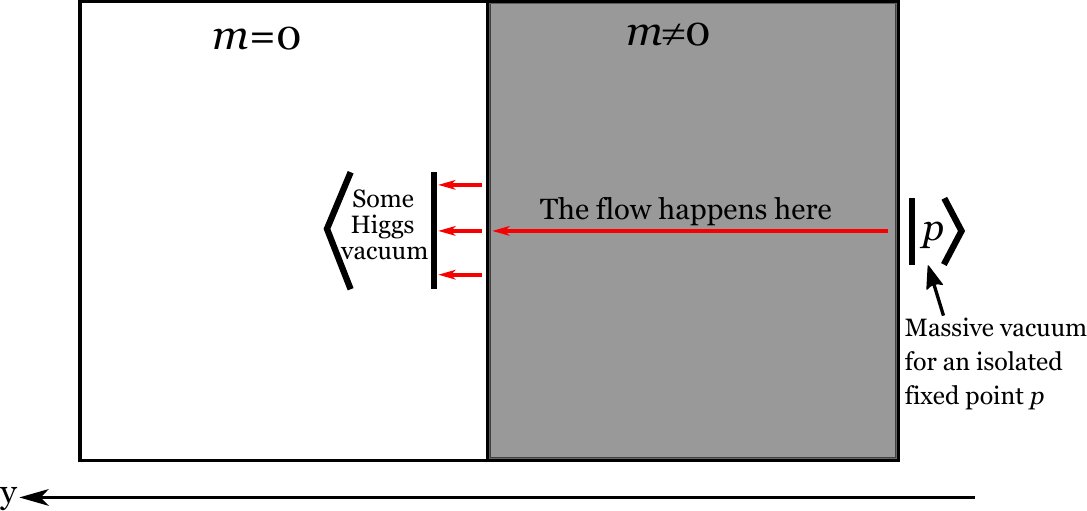}
	\caption{The setup defining $\cJ_m(0,m)|p\rangle$. Here $|p\rangle$ is a massive vacuum taken as  an ``in''-state in the far past $y\to-\infty$. In the massless region, it is capped off with an arbitrary Higgs phase vacuum wave function $\psi$. The overlap is computed by counting flows that start at the fixed point $p$ and end somewhere in the support of $\psi$.}\label{fig:flow}
\end{figure}\\
In the equation \eqref{matrp1p2}, we also used the fixed point basis $\langle p|$ on the left (that is, in the massless region). This is understood in the sense of equivariant localization: even though general vacuum wave functions in the massless region are not localized at the fixed points $p$, they can still be decomposed in a ``fixed point basis''. The latter consists of special smooth wave functions that are appropriately centered at the fixed points. They can be made more sharply localized there if we start increasing the equivariant parameters (flavor holonomies on $\bE_\tau$). They also admit certain delta-function-like representatives in the $\cQ$-cohomology (the analogs of de Rham currents in differential geometry), as we will discuss later.

The above clearly suggests an interpretation of $\cJ_m(0,m)$ as the map:
\begin{equation}
\cJ_m(0,m): \{\text{Vacua at }m\neq 0\} \to \{\text{Vacua at }m=0\}.
\end{equation}
According to Section \ref{sec:VacAndCoh}, the space of vacua is identified with the appropriate equivariant cohomology theory of the vacuum manifold. The latter refers to the whole Higgs branch if $m=0$, or the fixed locus $X^{\mathbf{A}}$ if masses $m\neq 0$ are generic. In 1d, we study our theory on $\R$, and $\cJ_m(0,m)$ gives a map in the $\bfT$-equivariant cohomology:
\begin{align}
H^\bullet_{\bfT}(X^{\mathbf{A}}) \to H^\bullet_{\bfT}(X)\quad \text{(1D case.)}
\end{align}
In two dimensions, we study the theory on $\R\times S^1$, and $\cJ_m(0,m)$ realizes a map in the equivariant complex K-theory:
\begin{equation}
K_\bfT(X^{\mathbf{A}}) \to K_\bfT(X)\quad \text{(2D case.)}
\end{equation}
In our main three-dimensional setting on $\R\times \bE_\tau$, the operator $\cJ_m(0,m)$ realizes
\begin{equation}
\Gamma({\rm E}_\bfT(X^{\mathbf{A}}),\cL) \to \Gamma({\rm E}_\bfT(X),\cL),\quad \text{(3D case,)}
\end{equation}
which is a map between the spaces of sections of sheaves on ${\rm E}_\bfT(\cdot)={\rm Ell}_\bfT(\cdot)\times \cE_{\mathbf{A}'}$. Here ${\rm Ell}_\bfT(\cdot)$ is the scheme representing equivariant elliptic cohomology, and $\cE_{\mathbf{A}'}$ is the Abelian variety of K\"ahler parameters. The line bundle $\cL$ describing the topology of vacua in the elliptic case was introduced in Section \ref{sec:ell_vacua}.

\subsection{Effective description}\label{sec:effective}
We will now explain how to compute matrix elements like \eqref{matrp1p2} using an effective gauge-theoretic description. The main idea is that as we send masses $m$ in the settings of Figures  \ref{fig:flow_left} and \ref{fig:flow} to infinity (in the chamber $\mathfrak{C}$), certain $\mathfrak{C}$-dependent boundary conditions emerge at the interface where masses change. On the massless side, we have our full gauge theory, let us call it $\cT$. On the massive side, we obtain the theory $\cT^{\mathfrak{C}}$, which we now describe.

As real masses $m$ increase, the theory $\cT$ decomposes into light and heavy degrees of freedom. The light ones are either massless or have masses of order $e\sqrt{\zeta}$ generated on the Higgs branch (here $\zeta$ represents FI parameters), and they describe low-energy excitations around each massive vacuum.  The heavy degrees of freedom have masses of order $m$, they describe high-energy processes such as tunneling between vacua, and they effectively go away (are integrated out) as $m\to\infty$. In this limit, we are left exactly with $\cT^{\mathfrak{C}}$, which by definition describes light degrees of freedom. It clearly decomposes into sectors labeled by the massive vacua, which do not talk to each other:
\begin{equation}
\cT^{\mathfrak{C}}=\bigoplus_{p\in\left\{\substack{\rm massive\\ \rm vacua}\right\}} \cT_p^{\mathfrak{C}}.
\end{equation}
One can think of $\cT_p^{\mathfrak{C}}$ as an effective QFT describing the tower of light excitations above the vacuum $p$. This notation is better suited to the situation when $p$ are \emph{isolated} vacua, and it is the case in all our examples, but it does not have to be so: in general $p$ labels components of the fixed locus on the Higgs branch that remain unlifted in the presence of masses $m$.

To be specific, suppose we work in the setting of Figure \ref{fig:flow_left}. In the massless region, we start with a state $|\psi\rangle$ as the input data. After the masses $m$ are turned on at $y=0$, the state $\psi$ gets quickly projected to the light sectors, and then slowly gets further projected to the vacuum subspace, both projections happening due to the Euclidean evolution:
\begin{equation}\label{relaxation}
|\psi\rangle \longmapsto \sum_{p\in\left\{\substack{\rm massive\\ \rm vacua}\right\}} |\psi_p\rangle \longmapsto \sum_{p\in\left\{\substack{\rm massive\\ \rm vacua}\right\}} c_p|p\rangle,\quad \text{where } |\psi_p\rangle \in \cH[\cT_p^{\mathfrak{C}}].
\end{equation}
The relaxation time of the first process is of order $1/m$, which becomes effectively instantaneous in the infinite mass limit. The relaxation time of the second process is determined by masses of the light modes, and can be quite long.

We want to pick out the term corresponding to the vacuum $|p\rangle$ in the above sum. This is done by fixing the vacuum at $y\to+\infty$ or, equivalently, by imposing the appropriate boundary conditions at a finite distance $y=y_+$. Assuming this has been done, in most of the massive region (except a vanishingly thin layer at $y=0$) the system is described by the theory $\cT_p^{\mathfrak{C}}$. Hence we effectively obtain the interface between $\cT$ and $\cT_p^{\mathfrak{C}}$:
\begin{equation}
\cT_p^{\mathfrak{C}}\ \Big|\  \cT.
\end{equation}
We claim that this interface admits a simple description in gauge theory. By further computing interval partition functions with such interfaces inserted, we are able to extract the coefficients of $|p\rangle$ in \eqref{relaxation} and henceforth produce matrix elements of the Janus interface.

Before proceeding, let us also note that we will be often switching between the viewpoints of Figure \ref{fig:flow_left} and Figure \ref{fig:flow}. They are conceptually and computationally similar, and we construct the $\cT^{\mathfrak{C}}_p\Big|\cT$ interface in both cases. They are both related to stable envelopes, the only difference being that in the Figure \ref{fig:flow} one works with the repelling subspaces of fixed points, rather than attractors in the setting of Figure \ref{fig:flow_left}.

\subsubsection{Theory $\cT_p^{\mathfrak{C}}$}
First, let us determine the effective theory labeled by the vacuum $p$. In this vacuum, the real scalars $\sigma$ from the vector multipelt ``adjust'' themselves to the special values $\sigma^{(p)}$, which partly screen real masses, such that the hypermultiplets developing vevs have zero effective real masses. Note that we study theories on $\R_y\times \mathbb{E}_\tau$, which are macroscopically one-dimensional. Therefore the statements about fields developing vevs should be understood in terms of the wave function being centered around the corresponding value. For example, when we say that the fields $\sigma$ take special values $\sigma^{(p)}$, we really mean that the vacuum wave function on the field space of the theory, considered as a functional of $\sigma$, is centered around $\sigma^{(p)}$. The corresponding probability distribution around $\sigma^{(p)}$ has some finite width, even in the $m\to\infty$ limit: The field $\sigma$ can still fluctuate around $\sigma^{(p)}$. However, in the $m\to\infty$ limit,  different values corresponding to different massive vacua, say $\sigma^{(p_1)}$ and $\sigma^{(p_2)}$, become infinitely separated. As a result, the overlap between vacua, and in fact between the whole light sectors $\cT^{\mathfrak{C}}_{p_1}$ and $\cT^{\mathfrak{C}}_{p_2}$, clearly vanishes. This is the sense in which we single out the theory $\cT^{\mathfrak{C}}_p$.

If the theory $\cT$ has a non-abelian gauge group $G$, the vev $\sigma=\sigma^{(p)}$ Higgses it to a certain subgroup
\begin{equation}
    G_p\subset G
\end{equation}
by giving large masses to the roots that do not commute with $\sigma^{(p)}$ (it is reasonable to refer to them as W-bosons). Assuming that $\sigma^{(p)}$ is a generic element of a torus $\mathbb{S}\subset \mathbf{H}$, which does not necessarily coincide with the maximal torus $\mathbf{H}$, we may say that
\begin{equation}
G_p=C_G(\mathbb{S})
\end{equation}
is a centralizer of $\mathbb{S}$. While $G_p$ contains the maximal torus, $\mathbf{H}\subset G_p\subset G$, it does not necessarily coincide with it if $\mathbb{S}$ is strictly smaller than $\mathbf{H}$. This $G_p$ is identified as the gauge group of $\cT^{\mathfrak{C}}_p$. Note that in the abelian case, we simply have $G_p=G=\mathbf{H}$.

As is usual in the Higgs mechanism, the group $G$ is not truly broken: The wave functional is still $G$-invariant, i.e., constant along the $G$-orbit of $\sigma^{(p)}$. Expanding around the chosen value of $\sigma^{(p)}$ and ``breaking''  $G$ is just a repackaging of degrees of freedom, with $G_p$ being the new gauge group. As a remnant of the full $G$ gauge symmetry, the wave functional is not an arbitrary function of $\sigma^{(p)}$.  It is invariant under the action of the Weyl group relative to $\mathbb{S}$:
\begin{equation}
\cW(G,\mathbb{S}):=N_G(\mathbb{S})/C_G(\mathbb{S}),
\end{equation}
where $N_G(\mathbb{S})$ is the normalizer of $\mathbb{S}$ in $G$, and $C_G(\mathbb{S})$ is the centralizer of $\mathbb{S}$. In many interesting examples, $\mathbb{S}$ coincides with the maximal torus $\mathbf{H}$, in which case $G_p=\mathbf{H}$, and $\cW(G,\mathbf{H})$ is the ordinary Weyl group. Working with the gauge group $\mathbf{H}$ instead of $G$ and the \emph{same} matter content is referred to as the \emph{abelianization} \cite{ShenThesis,Aganagic:2016jmx}. Note that our theory $\cT^{\mathfrak{C}}_p$ is \emph{not} the abelianization, since we truncate both the gauge group (not always to the maximal torus) and the matter content, as will be explained shortly. Also, it will be enough to work with a given $\sigma^{(p)}$, leaving Weyl-averaging to the very end of most computations.

Now let us determine the matter content of $\cT^{\mathfrak{C}}_p$. Recall that the hypers of $\cT$ are valued in $\cR\oplus \bar\cR$, where $\cR$ is a complex representation of gauge and flavor groups. As discussed in Section \ref{sec:ell_vacua},  $\cR$ splits into vector subspaces (and, in fact, $G_p$-subrepresentations) according to the value of the effective real mass:
\begin{equation}
\label{R_decomp}
\cR \cong \cR_0(p) \oplus \cR_+(p) \oplus \cR_-(p),
\end{equation}
where $\cR_0(p)$ contains components (viewed as $G_p$-weights) with vanishing effective real masses, which can develop vevs in the vacuum $p$. They may still obtain small masses on the Higgs branch, but such masses are not controlled by the large parameters $m$. In other words, the hypermultipelts in $\cR_0(p)$ are light degrees of freedom, and hence part of the theory $\cT^{\mathfrak{C}}_p$. The rest of the hypers, valued in $\cR_+(p)$ and $\cR_-(p)$, have very large positive and negative effective real masses, respectively, and ought to be integrated out.

We have to take care of the Chern-Simons (CS) terms generated in the process.\footnote{Other possibly generated terms vanish in the $m\to\infty$ limit.} Integrating out a 3d $\cN=2$ chiral of real mass $m$ produces a CS term at the level $\frac12{\rm Sign}(m)$ for the corresponding symmetry (this is really the contribution of the Dirac fermion that is part of the multiplet). Since a hyper consists of two chirals of \emph{opposite} flavor/gauge charges (and hence opposite masses), their respective CS terms cancel, at least when the full $\cN=4$ SUSY is preserved. In our case, we break $\cN=4$ down to $\cN=2$ by the $U(1)_\hbar$ holonomy, which leaves a possibility to generate a mixed CS term between the $U(1)_\hbar$ and flavor/gauge symmetries. Such a CS term plays role in the anomaly inflow, so let us write it as a four-form term in the anomaly polynomial. Consider a hyper $(I,J)$, with the gauge field strength coupled to $I$ written as $\mathbf{f}$, the flavor field strength -- as $\mathbf{f}_x$, and the $U(1)_\hbar$ field strength -- as $\frac12\mathbf{f}_\hbar$. Assuming the real mass of $I$ is positive, the contribution of $(I,J)$ is
\begin{equation}
\frac12(\mathbf{f}+\mathbf{f}_x + \frac12\mathbf{f}_\hbar-\frac12 \mathbf{r})^2 - \frac12(-\mathbf{f}-\mathbf{f}_x + \frac12\mathbf{f}_\hbar-\frac12\mathbf{r})^2 = (\mathbf{f}+\mathbf{f}_x)(\mathbf{f}_\hbar-\mathbf{r}),
\end{equation}
where we also included the contribution of the R-symmetry field strength $\mathbf{r}$, for better comparison with the boundary anomalies, even though we do not turn on any $U(1)_R$ background. More generally, the extremely massive hypers in $\cR_+(p)\oplus \cR_-(p)$ contribute
\begin{equation}
\label{CSinflowH}
(\mathbf{f}_\hbar-\mathbf{r})\left[ \Tr_{\cR_+(p)}(\mathbf{f}+\mathbf{f}_x)- \Tr_{\cR_-(p)}(\mathbf{f}+\mathbf{f}_x)\right].
\end{equation}
The corresponding CS term can be written as
\begin{equation}
\frac1{4\pi}\int (A_\hbar - A_R) \wedge\left[ \Tr_{\cR_+(p)}(F+F_x)-\Tr_{\cR_-(p)}(F+F_x)\right]+\dots,
\end{equation}
where the ellipsis represents the $\cN=2$ completion.

Now that we have determined the fate of hypermultiplets, --- those in $\cR_0(p)$ give the matter content of $\cT^{\mathfrak{C}}_p$, and those in $\cR_+(p)\oplus \cR_-(p)$ generate the effective CS terms, --- it remains to understand what happens to the massive components of vector multiplets, that is W-bosons. In abelian theory there are no W-bosons, so we are done.

In a non-abelian theory, integrating out the massive W-bosons and their superpartners valued in $\mathfrak{g}\setminus\mathfrak{g}_p$ (here $\mathfrak{g}_p={\rm Lie}(G_p)$) also contributes to the effective CS term. More precisely, 3d $\cN=2$ vector multipelts do \emph{not} contribute: Their massive components come in pairs of roots $\alpha$ and $-\alpha$, which receive opposite masses $\langle\alpha,\sigma^{(p)}\rangle$ and $-\langle\alpha,\sigma^{(p)}\rangle$. Integrating out the corresponding gaugini results in opposite effective CS terms that cancel each other. The adjoint chiral $\Phi$, however, avoids this cancellation, which is again due to the nonzero $U(1)_\hbar$ charge of $\Phi$. Let $\Delta_+(p)$ denote the ``positive roots'' $\alpha$, such that the corresponding components $\langle \alpha, \Phi\rangle$ have positive real masses in the vacuum $p$. Note that $\Delta_+(p)\cup (-\Delta_+(p))$ is the set of all massive roots, which are in bijection with a basis of $\mathfrak{g}\setminus \mathfrak{g}_p$. Then we can write down the contribution of $\Phi$ as
\begin{align}
\label{CSinflowV}
\sum_{\alpha\in\Delta_+(p)} \frac12 \left[(\langle\alpha,\mathbf{f}\rangle+\mathbf{f}_\hbar)^2-(-\langle\alpha,\mathbf{f}\rangle+\mathbf{f}_\hbar)^2\right]=2\mathbf{f}_\hbar \sum_{\alpha\in\Delta_+(p)}\langle \alpha, \mathbf{f}\rangle.
\end{align}
The corresponding effective $\cN=2$ CS term is then
\begin{equation}
\frac1{2\pi}\int A_\hbar\wedge \sum_{\alpha\in\Delta_+(p)} \langle \alpha, F\rangle + \dots
\end{equation}

\subsubsection{Interface between $\cT$ and $\cT^{\mathfrak{C}}_p$}
As real masses are sent to infinity, a certain natural interface emerges between $\cT$ and $\cT^{\mathfrak{C}}_p$. Since the field content of $\cT^{\mathfrak{C}}_p$ is identified with the light fields in $\cT$ (up to a Weyl permutation), the interface must be transparent for such fields. Their masses do not change across the interface, except in a thin layer of size $1/m$ where the relaxation takes place (and $\sigma$ does not properly screen $m$ just yet). This layer has vanishingly small size and simply cannot be probed by the low-energy modes. Hence the statement that the light fields do not see the interface.

To be more precise, the \emph{low-energy excitations} of the light fields do not sense the interface. Indeed, some of the light fields develop a vev across the interface, like the $\mathfrak{g}_p$-valued part of $\sigma$, which is $0$ for $y<0$ and $\sigma^{(p)}$ for $y>0$ (more precisely, for $y\gtrsim m^{-1}$). In this case, a more accurate statement is that the low-energy excitations of $\sigma$ above the $\sigma^{(p)}\Theta(y)$ background\footnote{Here $\Theta(y)$ is the Heaviside theta function, such that $\Theta(y>0)=1$ and $\Theta(y<0)=0$.} cannot resolve distances as short as $m^{-1}$, hence they see the interface as transparent. Every time we say ``low-energy'' here, we mean compared to $m$, so this restriction goes away in the $m\to\infty$ limit.

The comment about Weyl permutations briefly made above means the following. A state of $\cT^{\mathfrak{C}}_p$, as a function of $\sigma\in\mathfrak{t}$, has to be averaged over $\cW(G,\mathbb{S})$ to give a state from the low-energy sector of $\cT$ with large masses. Hence the definition of the interface secretly involves the Weyl-averaging (which would take place anyways on the $\cT$ side due to the non-abelian gauge symmetry). For example if $G=U(N)$ and $G_p=U(1)^N$, then a sate of $\cT^{\mathfrak{C}}_p$, as a function of $\sigma\in \mathfrak{t} = \R^N$, has no symmetry properties, while on the $\cT$ side it has to be permutation-symmetric. Thus the fields of $\cT^{\mathfrak{C}}_p$ are identified with the light fields of $\cT$ only up to Weyl permutations. This is one of the ways to see why in the computations of matrix elements of interfaces we will have to perform Weyl-symmetrization.

Heavy fields, on the other hand, are not part of the $\cT^{\mathfrak{C}}_p$ theory. Therefore, they ought to terminate at the interface, subject to certain boundary conditions. Such boundary conditions are naturally induced by the mass jumping from $0$ to a very large value in the supersymmetric way. Also note that boundary conditions are part of the UV data of the theory, and all our theories are UV-free. Therefore, it is enough to analyze the problem in the free case.

%Boundary conditions are part of the UV data, and all our theories are free in the UV. We have to send masses to infinity so fast that in the UV limit, the free chiral multiplets still have infinite masses in one half of the spacetime. This is related to the order of limits discussed in the previous section: masses should be sent to infinity faster than any other parameter in the problem (that grows in the UV, such as $1/e^{2}$).

\paragraph{Scalar boson with jumping mass.} We start by analyzing a free complex scalar (in a chiral multiplet) with spatially modulated mass that is part of the half-BPS Janus profile. Suppose the mass is zero for $y<0$ and $m$ for $y>0$, and do not forget about the $-\bar\phi m'(y) \phi$ term required by the SUSY.

There are several ways to proceed with the analysis. Let us start in the Euclidean signature, thinking of the system as defined by the path integral
\begin{equation}
\int \cD \phi\ e^{-\int \dd^3 x\, \bar\phi \left[-\partial^\mu\partial_\mu + m^2 \Theta(y) - m\delta(y) \right]\phi},
\end{equation}
where $\Theta(y)$ is the Heaviside theta (step) function, and the delta function comes from the $\bar\phi m'(y)\phi$ term. Definition of such a path integral involves expanding in the eigenfunctions of the kinetic operator, such that the fields behave reasonably at infinity (do not blow up). Thus we look at
\begin{equation}
\left[-\partial^\mu\partial_\mu + m^2 \Theta(y) - m\delta(y) \right]\phi = E\phi,
\end{equation}
where $E$ is real since the operator is Hermitian. At this point we may use the unbroken translation invariance in the plane parallel to the interface, and Fourier transform in those directions. This replaces $\partial^\mu\partial_\mu$ by $\partial_y^2 - p^2$, where $p$ is the momentum parallel to the interface:
\begin{equation}
\left[-\partial_y^2 + p^2 + m^2\Theta(y) -m\delta(y) \right]\phi = E \phi.
\end{equation}
Equivalently, we get equations in the two regions:
\begin{equation}
\label{eigEqns}
\begin{cases}
\qquad\ (p^2-\partial_y^2)\phi^< = E\phi^<, & y<0\\
(p^2+m^2-\partial_y^2)\phi^> = E\phi^>, & y>0
\end{cases}
\end{equation}
subject to the sewing conditions:
\begin{equation}
\label{sewing}
\phi^<\big|_{y=0}=\phi^>\big|_{y=0},\qquad \partial_y\phi^<\big|_{y=0} - \partial_y\phi^>\big|_{y=0}=m\phi\big|_{y=0}.
\end{equation}
It is easy to see that \eqref{eigEqns}, \eqref{sewing} have no solutions for $E<p^2$, while for $E\geq p^2$ solutions exist and fall in two classes:
\begin{enumerate}
	\item For $p^2 \leq E < p^2 + m^2$, the solution is
	\begin{align}
	\label{class1Sol}
	\phi^< = A e^{i y \sqrt{E-p^2}} + B e^{-iy\sqrt{E-p^2}},\qquad \phi^> = C e^{-y \sqrt{p^2 + m^2 -E}},
	\end{align}
	where $C=A+B$, and
	\begin{equation}
	\label{ABrelation}
	B = A \frac{i\sqrt{E-p^2} + \sqrt{p^2+m^2-E}-m}{i\sqrt{E-p^2}-\sqrt{p^2+m^2-E}+m}.
	\end{equation}
	In the $y>0$ region we picked the solution that does not blow up as $y\to+\infty$. Note that the coefficients $A$, $B$, $C$ are of course functions of $p$ and $E$, which was suppressed in the notations to reduce clutter.
	\item For $E\geq p^2 + m^2$, the solution is
	\begin{align}
	\phi^< = A e^{i y \sqrt{E-p^2}} + B e^{-iy\sqrt{E-p^2}},\qquad \phi^> = C e^{iy \sqrt{E-p^2 -m^2}}+D e^{-iy \sqrt{E-p^2 -m^2}},
	\end{align}
	where $A+B=C+D$ and $i\sqrt{E-p^2}(A-B)-i\sqrt{E-p^2-m^2}(C-D)=m(A+B)$.
\end{enumerate}

As we tend $m^2\to\infty$, all solutions fall into the class 1, while those in the class 2 go away to infinity. Focusing on the class 1, it immediately follows from \eqref{ABrelation} that we get $B=\pm A$ in this limit, depending on the sign of $m$:
\begin{align}
m\to +\infty &\Rightarrow B=A,\cr
m\to -\infty &\Rightarrow B=-A.
\end{align}
Solution $\phi^<$ with $B=A$ obeys Neumann boundary condition at $y=0$, while $B=-A$ implies that $\phi^<$ obeys Dirichlet boundary condition at $y=0$. Solution for $\phi^>$ from \eqref{class1Sol} decays fast: It has a relaxation time of order $1/|m|$, which tends to zero in the $m^2\to\infty$ limit. Thus we confirm that the field $\phi$ lives in the $y<0$ half-space in the $m\to\infty$ limit, with the boundary conditions at $y=0$ determined by the mass sign:
\begin{align}
\label{answerBC}
\boxed{m\to +\infty \Rightarrow \partial_y \phi^<\big|_{y=0}=0,}\qquad \boxed{m\to-\infty \Rightarrow \phi^<\big|_{y=0}=0.}
\end{align}

We could reach the same conclusion in several other ways. For example, we could work in the Mikowski signature instead, thinking about the field $\phi$ in the on-shell operator formalism, i.e., subject to the equation of motion $\left[ -\partial^\mu\partial_\mu + m^2\Theta(y) -m\delta(y)\right]\phi=0$. Then we solve the scattering problem, with an incoming and reflected plane waves in the region $y<0$, and some solution in the region $y>0$. These solutions are subjected to the same sewing conditions \eqref{sewing}, and the answer we get from such an analysis is the same as in \eqref{answerBC}. Another way to derive \eqref{answerBC} is by using supersymmetry. The BPS equation in the $y>0$ region reads $\partial_y \phi = -m\phi$. For $m>0$ this gives a decaying solution, which thus admits an arbitrary initial value at $y=0$, consistent with the Neumann boundary conditions. For $m<0$, the BPS equation gives a solution that blows up, which can only be avoided by the Dirichlet boundary condition at $y=0$, again consistent with \eqref{answerBC}.

Finally, let us comment that if we consider the mirror-reflected configuration, with $m\neq 0$ for $y<0$ and $m=0$ for $y>0$, then the relation between the sign of mass and the boundary conditions gets flipped:
\begin{align}
\label{answerBCalt}
\boxed{m\to -\infty \Rightarrow \partial_y \phi^>\big|_{y=0}=0,}\qquad \boxed{m\to+\infty \Rightarrow \phi^>\big|_{y=0}=0.}
\end{align}

\paragraph{Fermion with jumping mass.} Supersymmetry preserved by the Janus interface allows to determine the fermionic boundary conditions for free. We could, however, run the same analysis and see how the boundary conditions for the fermions emerge in the $m\to\infty$ limit. Such an analysis is even simpler than in the bosonic case due to the absence of the delta-function term in the equations of motion, which are simply the Dirac equation with $y$-dependent mass. In the massive region, one gets solutions that either decay or blow up:
\begin{equation}
\psi_+ \sim e^{-my},\quad \psi_- \sim e^{+my},
\end{equation}
and we have to kill the solution that blows up at infinity. Depending on the sign of $m$, that would be either $\psi_+$ or $\psi_-$, providing exactly the SUSY completion of the boundary conditions obeyed by the bosons.

In principle, the most economic way to derive boundary conditions would be to start with the fermionic equations of motion as above, and then argue that bosonic boundary conditions follow by the SUSY. We chose a different path and started with the bosons here, as it is somewhat interesting and pedagogical to see explicitly how the Dirichlet/Neumann boundary conditions for the bosons emerge.

\paragraph{W-bosons.} In the case of non-abelian gauge group, as mentioned earlies, non-zero vevs $\sigma^{(p)}$ give masses to the W-bosons. Thus components of the vector multipelt corresponding to the root $\alpha$ have their masses jump from $0$ to $\langle\alpha,\sigma^{(p)} \rangle$ at $y=0$. This, likewise, generates certain boundary conditions for such components. We could study them by analyzing the Proca equation, similar to our analysis of the Klein-Gordon equation above. To save some time, though, we take a shortcut based on supersymmetry now. A 3d $\cN=4$ vector multiplet consists of the 3d $\cN=2$ vector $V$ and the adjoint chiral $\Phi$. As reviewed in Section \ref{sec:gauge_review}, the 2d $\cN=(2,2)$ SUSY preserved at the boundary relates the boundary conditions on $\Phi$ to those on $V$. Namely, 2d $\cN=(0,2)$ Dirichlet boundary conditions on $\Phi$ imply $(0,2)$ Dirichlet on $V$, and Neumann on $\Phi$ imply Neumann on $V$.

Since we already understand the case of chiral multiplets, we know that the components of $\Phi$ receive Dirichlet/Neumann boundary conditions depending on the sign of mass. This then implies boundary conditions on the whole $\cN=4$ vector multiplet as summarized below. Denoting the mass of root $\alpha$ by $m_\alpha=\langle \alpha,\sigma\rangle$, we have:
\begin{align}
\label{vm_bndry}
m_\alpha(y<0)=0,\quad m_\alpha(y>0)\to +\infty \quad &\Longrightarrow \text{Neumann},\cr
m_\alpha(y<0)=0,\quad m_\alpha(y>0)\to -\infty \quad &\Longrightarrow \text{Dirichlet},\cr
m_\alpha(y<0)\to +\infty,\quad m_\alpha(y>0)=0 \quad &\Longrightarrow \text{Dirichlet},\cr
m_\alpha(y<0)\to -\infty,\quad m_\alpha(y>0)=0 \quad &\Longrightarrow \text{Neumann},
\end{align}
where Neumann/Dirichlet refers to the $\cN=(2,2)$ boundary conditions on the $\cN=4$ vector multiplet.

\paragraph{Anomalies.} By construction, our interface between $\cT$ and $\cT^{\mathfrak{C}}_p$ is not expected to carry any 2d anomalies, since it is defined simply as a limit of a certain mass profile in 3d. The effective boundary conditions described earlier, however, look like they could support boundary anomalies. Here we note that such anomalies indeed cancel, via the inflow, against the effective CS terms discussed earlier. Assuming we have the theory $\cT$ for $y<0$ and $\cT^{\mathfrak{C}}_p$ for $y>0$, the hypers in $(\cR_+(p),\bar\cR_+(p))$ and $(\cR_-(p),\bar\cR_-(p))$ receive (Neumann, Dirichlet) and (Dirichlet, Neumann) boundary conditions, respectively.\footnote{In the following, we will often abbreviate such boundary conditions as (N,D) or (D,N).} Boundary anomaly of each hyper with the (D,N) boundary conditions reads $\frac12(\mathbf{f}+\mathbf{f}_x + \frac12 \mathbf{f}_\hbar -\frac12 \mathbf{r})^2-\frac12(-\mathbf{f}-\mathbf{f}_x + \frac12 \mathbf{f}_\hbar -\frac12 \mathbf{r})^2=(\mathbf{f}_\hbar -\mathbf{r})(\mathbf{f}+\mathbf{f}_x)$, and the opposite for the (N,D) case. In total, we have the boundary anomaly of hypers,
\begin{equation}
(\mathbf{f}_\hbar-\mathbf{r})\Tr_{\cR_-(p)}(\mathbf{f} + \mathbf{f}_x) - (\mathbf{f}_\hbar-\mathbf{r})\Tr_{\cR_+(p)}(\mathbf{f} + \mathbf{f}_x),
\end{equation}
which is clearly canceled by the inflow term \eqref{CSinflowH}.

The 3d $\cN=4$ vector multiplet contribution to the boundary anomaly, for our boundary conditions, is completely due to the adjoint chiral $\Phi$, and given by:
\begin{equation}
\sum_{\alpha\in\Delta_+(p)} \left[\frac12 (-\langle\alpha,\mathbf{f}\rangle + \mathbf{f}_\hbar)^2-\frac12 (\langle\alpha,\mathbf{f}\rangle + \mathbf{f}_\hbar)^2 \right]=-2\mathbf{f}_\hbar\sum_{\alpha\in\Delta_+(p)}\langle\alpha,\mathbf{f}\rangle,
\end{equation}
which also clearly cancels against the inflow term \eqref{CSinflowV}. Thus we see that the effective CS terms present on the massive side cancel the boundary anomalies, in a sense transporting them to the opposite end of the massive region, which might be important when we have another boundary at $y=y_+$.

\paragraph{Mini summary.} Let us briefly summarize our description of the interface between $\cT$ and $\cT^{\mathfrak{C}}_p$. Recall that $\cT$ is a gauge theory with the gauge group $G$ and hypermultiplets $(I,J)$, where the chiral multiplet $I$ is valued in a unitary $G\times G_H$-representation $\cR$, and $J$ -- in its conjugate $\bar\cR$. The gauge group of $\cT^{\mathfrak{C}}_p$ is $G_p=C_G(\mathbb{S})$, the subgroup of $G$ that commutes with $\sigma=\sigma^{(p)}$ in the massive vacuum $p$. The matter content of $\cT^{\mathfrak{C}}_p$ is given by the hypermultiplets valued in the $G_p$-submodule $\cR_0(p)\subset\cR$ determined as follows. Under the embedding $G_p\subset G$, the representation $\cR$ breaks into $G_p$-submodules, $\cR=\cR_0(p)\oplus \cR_+(p)\oplus \cR_-(p)$, where $\cR_0(p)$ is singled out as the subspace of hypers that remain light in the vacuum $p$ as $m\to\infty$ in the chanber $\mathfrak{C}$.

Hypers valued in $\cR_\pm(p)$ receive large positive/negative real masses in the vacuum $p$, respectively (here we refer to the signs of masses of the $\cR_\pm(p)$-valued components of the chiral $I$). These hypers are not present in the $\cT^{\mathfrak{C}}_p$ theory, thus they terminate at the interface via the boundary conditions, which in the case when $\cT^{\mathfrak{C}}_p$ occupies the $y>0$ half-space are constructed as follows: $\cR_+(p)$-valued and $\bar\cR_-(p)$-valued chirals are given $(0,2)$ Neumann boundary conditions; $\bar\cR_+(p)$-valued and $\cR_-(p)$-valued chirals are given $(0,2)$ Dirichlet boundary conditions. Together these define $(2,2)$ boundary conditions for the $\cR_\pm(p)$-valued hypermultiplets. The $\cR_0(p)$-valued part of the hypermultiplets in $\cT$ are simply identified (up to Weyl permutations) with the hypermultipelts of the theory $\cT^{\mathfrak{C}}_p$ along the interface. That is, for the $\cR_0(p)$-valued hypers the interface is transparent.

As for the gauge multiplets, the $\mathfrak{g}_p$-valued vector multiplets of $\cT^{\mathfrak{C}}_p$ are identified at the interface with the $\mathfrak{g}_p$-valued part of the vectors in $\cT$ (up to the same Weyl permutations). The remaining $\mathfrak{g}\setminus\mathfrak{g}_p$-valued components of the vector multiplets in $\cT$ are given $(2,2)$ boundary conditions, Dirichlet or Neumann, depending on whether the massive W-boson multiplets receive negative or positive masses respectively. We denote the set of roots that have positive masses in the vacuum $p$ by $\Delta_+(p)$.

If the theory $\cT^{\mathfrak{C}}_p$ lives in the $y<0$ half-space instead, then all the Dirichlet and Neumann boundary conditions must be swapped. Finally, the theory $\cT^{\mathfrak{C}}_p$ also contains some mixed Chern-Simons terms generated by the very massive multiplets that have been integrated out, as explained around \eqref{CSinflowH} and \eqref{CSinflowV}.

\subsection{Boundary conditions for vacua}\label{sec:bndry_vacua}
Now that we have a detailed description of the interface between $\cT$ and $\cT^{\mathfrak{C}}_p$ that emerges in the $m\to\infty$ limit of the Janus $\cJ$, we would like to compute its vacuum matrix elements. The interface is wrapped on $\mathbb{E}_\tau \times \{0\}$ in the $\mathbb{E}_\tau\times \R_y$ Euclidean geometry, with the  vacua $\alpha$ and $\beta$ fixed at $y=-\infty$ and $y=+\infty$, respectively. Recall that we also turn on flat connections for the flavor, $U(1)_\hbar$ and topological symmetries along $\mathbb{E}_\tau$, denoted as $x$, $\hbar$, and $z$, respectively. With $\R_y$ treated as the Euclidean time, this setup computes the equivariant overlap $\langle \beta|\cJ|\alpha\rangle$ that we are after. Treating one of the cycles in $\mathbb{E}_\tau$ as the Euclidean time instead, the same quantity can be interpreted as a BPS index in the $\alpha\beta$ soliton sector, with the flat connections playing the role of fugacities.

Of course one would like to compute such indices exactly using the supersymmetric localization. Non-compactness of $\R_y$, however, leads to some technical issues that were mentioned earlier. It is convenient, therefore, to replace $\R_y$ by an interval $(y_-, y_+)$, with $y_+>0$ and $y_-<0$, such that the boundary conditions at its endpoints are $\cQ$-cohomologous to the vacua $\alpha$ and $\beta$, thus mimicking them for the purposes of BPS computations. We already discussed some aspects of such boundary conditions in Section \ref{sec:ell_vacua}. The interface is still located at $0\in (y_-, y_+)$.

As in the Section \ref{sec:ell_vacua}, such boundary conditions are more straightforward to construct if in the presence of generic masses, $\cT$ only has isolated massive vacua corresponding to the isolated $\mathbf{A}$-fixed points on the Higgs branch. Let us start with the boundary conditions in the theory $\cT^{\mathfrak{C}}_p$ that governs light excitations above an isolated vacuum $p$. By definition, $\cT^{\mathfrak{C}}_p$ has only one SUSY vacuum, namely $p$ itself. Recall from Section \ref{sec:ell_vacua} that it can be identified with a $\cQ$-cohomology class in $\cH[\mathbb{E}_\tau]$. Since there is only one vacuum, any normalizable state of $\cT^{\mathfrak{C}}_p$ that is $\cQ$-closed and not $\cQ$-exact can serve as a proxy for it, at least in the BPS computations. According to the proposal of the Appendix \ref{app:bdry_states}, an elliptic boundary condition creates a boundary state $|\cB\rangle$, which after finite Euclidean evolution becomes a normalizable state $e^{-TH}|\cB\rangle$. If such a boundary condition preserves $\cQ$, then the corresponding state will be $\cQ$-closed, as we want. Therefore, in principle, any reasonable $\cQ$-invariant boundary condition can be used to generate a $\cQ$-closed state that mimics the vacuum.

To check that it is not $\cQ$-exact, it is enough to compute its inner product with some $\cQ$-closed covector $\langle \cB'|$ and get a non-zero answer (which would also imply that the covector is not $\cQ$-exact):
\begin{equation}
\langle \cB'| e^{-TH}|\cB\rangle \neq 0 \Rightarrow \left[ e^{-TH}|\cB\rangle \right]\neq 0 \text{ in the cohomology}.
\end{equation}

In practice, however, we will construct certain special classes of boundary conditions that better fit our applications. To construct a vacuum boundary condition in $\cT^{\mathfrak{C}}_p$, first observe how the $\cR_0(p)$-valued hypers get vevs. For every weight $w\in\cR_0(p)$, denote the corresponding hypermultiplet as $(I_w,J_w)$. In the vacuum $p$, either the chiral $I_w$ or the chiral $J_w$ develops a vev (or neither), but not the two simultaneously. This can be easily proven\footnote{One might notice that in the Section \ref{sec:ell_vacua} we have already used this statement a few times without a proof.} using the following steps: (1) every isolated $\mathbf{A}$-fixed point is also fixed by $U(1)_\hbar$, for otherwise one would find a continuous family of $\mathbf{A}$-fixed points; (2) $I_w$ and $J_w$ have opposite gauge and flavor charges but the same $U(1)_\hbar$ charge, thus if both of them had a nonzero vev in an $\mathbf{A}$-fixed point, acting by $U(1)_\hbar$ would again generate a continuous family of $\mathbf{A}$-fixed points.

Now define the following two types of boundary conditions on the $\cR_0(p)$-valued hypers:
\begin{enumerate}
	\item The boundary conditions $D_p$. For a hyper $(I_w, J_w)$, the chiral that develops a vev in the vacuum $p$ is given $(0,2)$ Dirichlet boundary conditions (with the boundary value equal to the vacuum value), while the other one is given $(0,2)$ Neumann. If neither develops vev, which one is given Dirichlet and which one is Neumann is an arbitrary choice.
	\item The boundary conditions $N_p$. In each pair $(I_w, J_w)$, the chiral that develops a vev in the vacuum $p$ is given $(0,2)$ Neumann boundary conditions, while the other one is given $(0,2)$ Dirichlet with zero boundary vev.
\end{enumerate}
It is also convenient to further split $\cR_0(p)$ (which of course was already anticipated in Section \ref{sec:ell_vacua}):
\begin{equation}
\cR_0(p) \cong \cR_0^D(p)\oplus \cR_0^N(p),
\end{equation}
according to the type of boundary conditions that chirals in $\cR_0(p)\oplus \bar\cR_0(p)$ receive. Namely, we use the following convention. For the $D_p$ type boundary conditions, chirals in $\cR_0^D(p)\oplus \bar\cR_0^N(p)$ receive Dirichlet boundary conditions (because they contain components that develop vevs in the vacuum $p$), and those in $\cR_0^N(p)\oplus \bar\cR_0^D(p)$ are given Neumann boundary conditions. For the $N_p$ type, the Dirichlet and Neumann are swapped. 

The $\cN=4$ vector multiplets in $\cT^{\mathfrak{C}}_p$ are also given either $(2,2)$ Neumann ($\cN_p$) or $(2,2)$ Dirichlet ($\cD_p$) boundary conditions. In the Dirichlet case, a flat connection $s \in {\rm Hom}(\mathbb{E}_\tau, \mathbf{H})/\cW$ is fixed along the boundary. When all boundary vevs of the hypers vanish, $s$ can be arbitrary. However, if the hypers are given boundary vevs, as in the boundary conditions of type $D_p$ above, then $s$ should take a specific value,
\begin{equation}
s = s^{(p)}(x,\hbar),
\end{equation}
which screens flavor and $U(1)_\hbar$ flat connections and allows for non-trivial vevs of the matter fields.\footnote{This is the same $s^{(p)}(x,\hbar)$ that was introduced in Section \ref{sec:ell_vacua}.} This condition originates from the equations
\begin{equation}
D_{\alpha}\phi = D_\varphi\phi =0,
\end{equation}
obeyed by every chiral multiplet scalar $\phi$ in the vacuum. A non-trivial vev for such a $\phi$ requires that the gauge flat connection cancels the flavor and $U(1)_\hbar$ flat connections (when acting on $\phi$), resulting in the condition $s=s^{(p)}(x,\hbar)$.

One more requirement on the boundary conditions mimicking massive vacua pertains to anomalies. As we explored in detail in Section \ref{sec:ell_vacua}, boundary 't Hooft anomalies must match the effective CS terms generated in the massive vacuum. Additionally, if the gauge fields obey Neumann boundary conditions, gauge and chiral anomalies should be absent at the boundary. The absence of gauge anomalies is the usual consistency requirement. The chiral anomalies should vanish because the vacuum does not break (the maximal torus of) flavor, $U(1)_\hbar$, or topological symmetries, hence the boundary conditions should not either.

Define the following two types of boundary conditions in $\cT^{\mathfrak{C}}_p$:\footnote{Such $\sN_p$ can be problematic in non-abelian gauge theories, as we will explain later.}
\begin{align}
\sD_p &= \left\{ \begin{matrix}
D_p \text{ boundary conditions on the hypers}\\
\cD_p \text{ boundayr conditions on the vectors, with } s=s^{(p)}
\end{matrix} \right\},\\
\sN_p &= \left\{ \begin{matrix}
N_p \text{ boundary conditions on the hypers}\\
\cN_p \text{ boundayr conditions on the vectors}\\
\text{Boundary theory } \Upsilon_p
\end{matrix} \right\}.
\end{align}
As the notation suggests, $\sD_p$ is just the thimble boundary condition discussed in Section \ref{sec:ell_vacua}. We noted there that its 't Hooft anomaly agrees with the effective CS terms, so there is no need to add any degrees of freedom at the boundary. To be more precise, here we define it for $\cT^{\mathfrak{C}}_p$, but if we collide it with the interface between $\cT$ and $\cT^{\mathfrak{C}}_p$, we will get precisely the thimble boundary conditions in $\cT$.

The $\sN_p$ boundary conditions, unlike $\sD_p$, support dynamical gauge fields and might require extra boundary degrees of freedom $\Upsilon_p$. They are necessary to cancel the chiral anomalies (gauge anomalies are absent due to $(2,2)$ SUSY), and help adjust the boundary 't Hooft anomaly to the desired value. Instead of (perhaps naively) asking that $\sN_p$ had the same 't Hooft anomaly as $\sD_p$, we would like the following conditions to hold:
\begin{align}
\label{dual_states}
\langle \sD_p | \sN_p\rangle &= \langle \sD_p | e^{-TH} | \sN_p\rangle=1,\cr
\langle \sN_p | \sD_p\rangle &= \langle \sN_p | e^{-TH} | \sD_p\rangle=1.
\end{align}
This will ensure that $|\sD_p\rangle \langle \sN_p|$ and $|\sN_p\rangle \langle \sD_p|$ act as projectors on the $p$-th vacuum (up to a $\cQ$-exact piece). In particular, if we have a vacuum state $\Psi$ described by a collection of overlaps $\langle \sD_p|\Psi\rangle$, then we can write $\cJ |\Psi\rangle = \sum_p \cJ|\sN_p\rangle \langle \sD_p|\Psi\rangle$ in the $\cQ$-cohomology, showing that it is enough to know how $\cJ$ acts on $|\sN_p\rangle$.

Equations \eqref{dual_states} imply that the left $\sD_p$ and the right $\sN_p$ boundary conditions (or the left $\sN_p$ and the right $\sD_p$) have opposite 't Hooft anomalies. Notice that if we omitted $\Upsilon_p$ from the definition of $|\sN_p\rangle$, the relations \eqref{dual_states} would still hold,\footnote{Overlaps are computed by the interval partition functions. The non-zero modes cancel out in such computations \cite{Sugiyama:2020uqh}. Because $\sD_p$ and $\sN_p$ impose opposite boundary conditions on each field, there are no zero modes on the interval with $\sD_p$ and $\sN_p$ at the opposite ends. Thus the overlap is simply $1$. Cancellation of non-zero modes also guarantees that the partition function is independent of the length of the interval, which is only true in theories that do not break SUSY spontaneously. This is why we were free to insert $e^{-TH}$ in \eqref{dual_states} regularizing the otherwise singular boundary states.} however, such $\sN_p$ carries chiral anomalies breaking the $U(1)_\hbar$ and topological symmetries. Therefore, $\Upsilon_p$ should have the property that it cancels the unwanted anomalies, but does not contribute in the overlaps like $\langle \sD_p|\sN_p\rangle$. Thus, the elliptic genus of $\Upsilon_p$ should be a non-trivial meromorphic non-elliptic function of $s,x,\hbar,z$ (non-ellipticity is what cancels the anomaly), such that at $s=s^{(p)}(x,\hbar)$ it becomes $1$.

It is relatively easy to achieve this. Suppose the future boundary that engineers $\langle \sD_p|$ has the anomaly polynomial $P_+[\sD_p](\mathbf{f},\mathbf{f}_x,\mathbf{f}_\hbar,\mathbf{r})$, in which we substitute $\mathbf{f}=f_p(\mathbf{f}_x,\mathbf{f}_\hbar)$. The anomaly polynomial associated with $|\sD_p\rangle$ is likewise denoted $P_-^{\sD_p}$. What is the anomaly of $|\sN_p\rangle$? If we allow ourselves to abuse the notations and formally include the chiral anomalies (to be canceled) in the polynomial, then we associate the following with $|\sN_p\rangle$:
\begin{equation}
\label{sNp_anom}
-P_+[\sD_p](\mathbf{f},\mathbf{f}_x,\mathbf{f}_\hbar,\mathbf{r}) + P[\Upsilon_p](\mathbf{f},\mathbf{f}_x,\mathbf{f}_\hbar,\mathbf{r}).
\end{equation}
The first term is $-P_+[\sD_p]$ because in $|\sN_p\rangle$, all the Dirichlet/Neumann boundary conditions have been flipped compared to $\langle\sD_p|$, and the inflow terms have opposite signs as well. Cancellation of chiral anomalies implies that all the terms proportional to $\mathbf{f}$ must cancel in \eqref{sNp_anom}, and we would like to obtain $-P_+[\sD_p]\big|_{\mathbf{f}=f_p(\mathbf{f}_x,\mathbf{f}_\hbar)}$ in the end. Thus, the role of $P[\Upsilon_p]$ is to replace $\mathbf{f}$ by $f_p(\mathbf{f}_x,\mathbf{f}_\hbar)$, and we want:
\begin{equation}
\label{expected_PUps}
P[\Upsilon_p] \sim \Tr (\mathbf{f}-f_p(\mathbf{f}_x,\mathbf{f}_\hbar))(\dots),
\end{equation}
where $(\dots)$ is the linear expression in $\mathbf{f}_x$, $\mathbf{f}_\hbar$, and $\mathbf{r}$ that enters the chiral anomaly.

To make this more concrete, let us evaluate the boundary anomalies of $\sD_p$, which are the 't Hooft anomalies. The anomaly polynomials consist of two terms:
\begin{equation}
P^{\sD_p}_\pm = \pm P_{\rm CS} + P_{\partial},
\end{equation}
where $\pm P_{\rm CS}$ is the inflow term due to the CS terms \eqref{CSinflowH}, \eqref{CSinflowV}, and the BF term \eqref{SBF}, all of which are present in $\cT^{\mathfrak{C}}_p$. The sign in front of $\pm P_{\rm CS}$ depends on whether the boundary conditions are ``left'' or ``right'', i.e., which end of the interval $(0, y_+)$ we are looking at. At $y=y_+$, the inflow term is $+P_{\rm CS}$, and it can be read off from \eqref{CSinflowH}, \eqref{CSinflowV}, \eqref{SBF} as
\begin{equation}
P_{\rm CS}=(\mathbf{f}_\hbar - \mathbf{r})\left[ \Tr_{\cR_-(p)}-\Tr_{\cR_+(p)} \right](\mathbf{f} + \mathbf{f}_x) - 2\mathbf{f}_\hbar\sum_{\alpha\in\Delta_+(p)}\langle \alpha, \mathbf{f}\rangle + 2 \Tr(\mathbf{f}\mathbf{f}_z).
\end{equation}
At $y=0$, the inflow term is $-P_{\rm CS}$. As for the term $P_\partial$, it is the contribution of boundary conditions alone, and is given by:
\begin{equation}
P_\partial = \underbrace{(\mathbf{f}_\hbar - \mathbf{r})\left[ \Tr_{\cR_0^D(p)}-\Tr_{\cR_0^N(p)}\right](\mathbf{f}+\mathbf{f}_x)}_{\text{hypers in } \cR_0(p)} + \underbrace{\frac{|G_p|}{2}(\mathbf{f}_\hbar^2 - \mathbf{r}^2)}_{G_p \text{ vector multiplet}}.
\end{equation}
We have to collect the coefficients in front of $\mathbf{f}$ in $-P_+[\sD_p]$, which will give the expression for $(\dots)$ in \eqref{expected_PUps}. With \eqref{expected_PUps} known, it is fairly easy to construct $\Upsilon_p$ meeting our requirements. Its general structure can be cumbersome (and is not unique), so we will rather construct it in examples.

Note that here we do not assume $\langle \sD_p|$ and $\langle \sN_p|$ to be the conjugates of $|\sD_p\rangle$ and $|\sN_p\rangle$, and in fact they are not. For example, $|\sD_p\rangle$ is $\cQ$-closed, but, unlike the exact vacuum $|p\rangle$, is not $\cQ^\dagger$-closed. The conjugate of $|\sD_p\rangle$ is thus some $\cQ^\dagger$-closed covector that we never use. What we call $\langle \sD_p|$ is a different covector that is $\cQ$-closed and works as a proxy for $\langle p|$. For the same reason, the overlaps like $\langle \sD_p|\sD_p\rangle$ and $\langle \sN_p|\sN_p\rangle$ are \emph{not} the norms of $|\sD_p\rangle$ and $|\sN_p\rangle$. It is possible to compute them via localization. In particular, $\langle \sN_p|\sN_p\rangle$ admits an expression in terms of the Jeffrey-Kirwan residues obtained in \cite{Sugiyama:2020uqh}. The expression for $\langle \sD_p|\sD_p\rangle$ then follows by the relation $1=\frac{|\sN_p\rangle \langle \sN_p|}{\langle \sN_p|\sN_p\rangle}$ in the cohomology:
\begin{equation}
\langle \sD_p| \sD_p\rangle = \langle \sD_p|\frac{|\sN_p\rangle \langle \sN_p|}{\langle \sN_p|\sN_p\rangle}|\rangle \sD_p\rangle= \frac{\langle \sD_p|\sN_p\rangle \langle \sN_p|\sD_p\rangle}{\langle \sN_p|\sN_p\rangle} = \frac1{\langle \sN_p|\sN_p\rangle}.
\end{equation}

\textbf{Remark: } The $\sN_p$ boundary conditions may be problematic in some non-abelian gauge theories. They involve the $N_p$ boundary conditions on the hypermultiplets, which may explicitly break gauge symmetry by giving the Dirichlet and Neumann boundary conditions to components of the same irreducible representation of the non-abelian $G$. In such cases, a modification is required. Instead of using the $N_p$ boundary conditions, we simply give the Neumann boundary conditions to all chiral multiplets in $\cR_0(p)\oplus \bar\cR_0(p)$. This introduces the gauge anomaly, which should be further cancelled by the boundary theory $\Upsilon_p$.

Now, after describing the vacuum boundary conditions in $\cT^{\mathfrak{C}}_p$, let us move to $\cT$. The corresponding boundary conditions $\sB_{\bL(p),p}$ were constructed in Section \ref{sec:ell_vacua}, so we briefly remind their definition here. The weight subspace $\cR_0(p)$ still contains the hypermultiplets that develop vevs in the vacuum $p$. The remaining weights are split according to the polarization $T^*\cR=\bL(p)\oplus \bL^\perp(p)$ (which can be chosen separately for each fixed point). Namely, we induce polarization $\tilde{\bL}(p)\oplus \tilde{\bL}^\perp(p)$ on the complement of $T^*\cR_0(p)$ as in \eqref{Ltilde}, and obtain:
\begin{equation}
\cR\oplus \bar\cR = \cR_0(p) \oplus \bar\cR_0(p) \oplus \tilde{\bL}(p) \oplus \tilde{\bL}^\perp(p).
\end{equation}

It is still true that for every weight $w\in \cR_0(p)$, only one of the two chirals in the hyper $(I_w, J_w)$ develops a vev. Thus the definition of $D_p$ boundary conditions for the hypermultiplets in $\cR_0(p)$ still makes sense, and we have:
\begin{align}
\sB_{\bL(p),p} = \left\{ \begin{matrix}
D_p \text{ boundary conditions on the hypers in }\cR_0(p)\\
\text{Dirichlet boudnary conditions on the chirals in }\tilde{\bL}(p)\\
\text{Neumann boundary conditions on the chirals in }\tilde{\bL}^\perp(p)\\
\text{Dirichlet boundayr conditions on the vector multiplets, with } s=s^{(p)}
\end{matrix} \right\}.
\end{align}
We will be interested in the overlaps of the sort $\Psi[\sB_{\bL(p),p}]=\langle \sB_{\bL(p),p}|\Psi\rangle$. Note that the line bundle that $\langle \sB_{\bL(p),p}|$ is valued in was studied, among other things, in Section \ref{sec:ell_vacua}.

Our main object of interest in the following is
\begin{equation}
\label{matr_el_to_compute}
\langle \sB_{\bL(p_2),p_2}| \cJ(0, m\to \infty_{\mathfrak{C}})|\sN_{p_1}\rangle,
\end{equation}
where the notation $\infty_{\mathfrak{C}}$ means that we take the infinite mass limit within the chamber $\mathfrak{C}$. The main statement is that this object, up to normalization, coincides with the matrix element of the elliptic stable envelope.

\subsection{Computation of matrix elements}
Now we can put all the ingredients together. Using the effective description of the interface between $\cT$ and $\cT^{\mathfrak{C}}_{p}$ developed in Section \ref{sec:effective}, and the boundary conditions defined in Sections \ref{sec:ell_vacua} and \ref{sec:bndry_vacua}, we are able to compute the matrix elements \eqref{matr_el_to_compute}.

According to Section \ref{sec:effective}, to compute \eqref{matr_el_to_compute} we consider the theory $\cT$ for $y>0$ and $\cT^{\mathfrak{C}}_{p_1}$ for $y<0$, with the natural interface in between. The boundary conditions $\sN_{p_1}$ are imposed at $y=y_-<0$ (``in the past''), and $\sB_{\bL(p_2),p_2}$ -- at $y=y_+>0$ (``in the future''). Because the gauge fields obey Dirichlet boundary conditions at $y=y_+$, the flat connection is not integrated over, and is simply fixed to $s=s^{(p_2)}$ everywhere on the interval. As usual, it is enough to assume that it belongs to the maximal torus, $s\in \mathbf{H}$. Since $\mathbf{H}\subset G_{p_1}$, this flat connection is unchanged as we cross the interface.

Now, some vector multiplets have no interval zero modes as they obey Dirichlet boundary conditions on one end and Neumann on another. All the $\mathfrak{g}_{p_1}$-valued components are like that: they live on the entire interval, obey $(2,2)$ Dirichlet boundary conditions at $y=y_+$ and $(2,2)$ Neumann at $y=y_-$. The remaining components in $\mathfrak{g}\setminus \mathfrak{g}_{p_1}$ only live on $(0,y_+)$ and obey $(2,2)$ Dirichlet boundary conditions at $y=y_+$, while their boundary conditions at $y=0$ are determined according to \eqref{vm_bndry}. We see that components corresponding to the roots $\alpha\in -\Delta_+(p_1)$ obey Neumann boundary conditions at $y=0$ and thus have no zero modes. Components in $\Delta_+(p_1)$, on the other hand, obey Dirichlet boundary conditions at $y=0$ and do have interval zero modes.

Recall how the $(2,2)$ Dirichlet boundary conditions for a 3d $\cN=4$ vector multiplet $\cV=(V,\Phi)$ work \eqref{Dir22V}. They leave unfixed the adjoint $(0,2)$ chiral $S$ and the adjoint $(0,2)$ Fermi $\Psi_\Phi$ at the boundary. When $(2,2)$ Dirichlet boundary conditions are imposed on both ends, these multiplets become the interval zero modes. Together, $(S,\Psi_\Phi)$ form a $(2,2)$ chiral, which is broken up into the $(0,2)$ multiplets $S$ and $\Psi_\Phi$ by the $U(1)_\hbar$ background, as $S$ has zero $U(1)_\hbar$ charge while $\Psi_\Phi$ has charge $+1$. The interval zero mode multiplets $(S,\Psi_\Phi)$ are labeled by the roots $\Delta_+(p_1)$, and we can easily compute their contribution to the partition function:
\begin{equation}
\label{WbosEll}
\mathbb{V}_{p_1}(s,\hbar):=\prod_{\alpha\in\Delta_+(p_1)} \frac{\vartheta(s^{-\alpha}\hbar)}{\vartheta(s^\alpha)}.
\end{equation}

To find the contribution of hypermultiplets, we identify boundary conditions on their constituent chiral multiplets. At the $\sB_{\bL(p_2),p_2}$ boundary, the ones that obey Dirichlet boundary conditions have weights in \eqref{Lhat},
\begin{equation}
\hat{\bL}(p_2)=\cR_0^D(p_2) \oplus \bar\cR_0^N(p_2) \oplus \tilde{\bL}(p_2),
\end{equation}
and the complementary weights in $\cR_0^N(p_2) \oplus \bar\cR_0^D(p_2) \oplus \tilde{\bL}^\perp(p_2)$ obey Neumann. At the interface between $\cT$ and $\cT^{\mathfrak{C}}_{p_1}$, the ones obeying Dirichlet are (since $\cT$ lives on the $y>0$ side):
\begin{equation}
\cR_+(p_1)\oplus\bar\cR_-(p_1),
\end{equation}
with $\cR_-(p_1)\oplus\bar\cR_+(p_1)$ obeying the Neumann.
Finally, at the $\sN_{p_1}$ boundary, the remaining hypers are broken into
\begin{equation}
\cR_0^N(p_1) \oplus \bar\cR_0^D(p_1)
\end{equation}
obeying Dirichlet, and $\cR_0^D(p_1) \oplus \bar\cR_0^N(p_1)$ obeying Neumann.

For the purpose of this computation, the $\cT|\cT^{\mathfrak{C}}_{p_1}$ interface and the $\sN_{p_1}$ boundary conditions in the theory $\cT^{\mathfrak{C}}_{p_1}$ can be collided together to give a single set of boundary conditions in $\cT$ (that could be called stable boundary conditions). Identifying the interval zero modes (i.e., those originating from the fields obeying identical boundary conditions on the two ends), we then straightforwardly compute the elliptic genus contribution from the matter fields:
\begin{align}
\mathbb{M}_{p_1, p_2}(s,\hbar,x) := \prod_{(w,f)\in Z_{p_1,p_2}}\frac{\vartheta(s^w x^f \hbar^{1/2})}{\vartheta(s^w x^f \hbar^{-1/2})},
\end{align}
where $w$ and $f$ denote gauge and flavor weights, respectively, and
\begin{align}
Z_{p_1,p_2}:=\hat{\bL}(p_2)\cap \left(\cR_0^N(p_1)\oplus \bar\cR_0^D(p_1)\oplus \cR_+(p_1)\oplus\bar\cR_-(p_1)\right).
\end{align}

The final contribution that must be included is the elliptic genus of the boundary degrees of freedom $\Upsilon_{p_1}$, which may be denoted as
\begin{equation}
\mathbb{W}[\Upsilon_{p_1}](s,x,\hbar,z).
\end{equation}
We will write it explicitly in the examples. The final answer for the overlap is obtained by multiplying these three factors and averaging them over the relative Weyl group $\cW(G,\mathbb{S})$. We define what may be called an off-shell overlap:
\begin{equation}
\mathbf{S}_{p_1}(s,x,\hbar,z) := {\rm Sym}_{\cW(G,\mathbb{S})} \Big(\mathbb{V}_{p_1}(s,\hbar) \mathbb{M}_{p_1, p_2}(s,\hbar,x) \mathbb{W}[\Upsilon_{p_1}](s,x,\hbar,z)\Big).
\end{equation}
This expression is a section of a line bundle on $\mathbf{E}_\bfT(X)$. The overlap is obtained by specializing $s=s^{(p_2)}(x,\hbar)$:
\begin{equation}
\label{JOverlap}
\langle \sB_{\bL(p_2),p_2}| \cJ(0, m\to \infty_{\mathfrak{C}})|\sN_{p_1}\rangle = \mathbf{S}_{p_1}(s^{(p_2)}(x,\hbar),x,\hbar,z).
\end{equation}
\paragraph{Ambiguity.} Note that our approach contains a fundamental ambiguity related to the choice of boundary theory $\Upsilon_{p_1}$, which is not unique. We can easily cook up a boundary 2d $(0,2)$ theory that is completely anomaly-free, and whose elliptic genus is a nontrivial meromorphic (fully elliptic) function of $(s,x,\hbar,z)$, which furthermore evaluates to $1$ after the specialization $s=s^{(p_1)}(x,\hbar)$. Adding such a theory to $\Upsilon_{p_1}$ is not prohibited by anything, and results in the ambiguity. It does not affect the line bundle that $\mathbf{S}_{p_1}(s,x,\hbar,z)$ is valued in, but simply multiples the latter by a meromorphic section of the trivial line bundle. It can be viewed as the normalization ambiguity of the vacuum states. One would need to impose some further conditions to remove it, e.g., that $|\sN_{p_1}\rangle$ is cohomologous to the normalized vacuum.

\subsection{Examples}
Let us evaluate $\mathbf{S}_p(s,x,\hbar,z)$ explicitly in several examples and compare it to the stable envelopes. We find that they are proportional, with $s$ playing the role of elliptic Chern roots, $(x,\hbar)$ -- equivariant, and $z$ -- K\"ahler parameters. The proportionality coefficient is such that $\mathbf{S}_p$ is better interpreted as the pole-subtraction matrix of \cite{Aganagic:2016jmx}.

\subsubsection{$T^* \C P^{n-1}$}
The most basic example is the $U(1)$ gauge theory with $n$ charge-1 hypermultiplets denoted $(I_i, J_i)$, $i=1..n$. The Higgs branch for a positive FI parameter $\zeta_\R>0$ is $X=T^* \C P^{n-1}$, with the base parameterized by $I_i$ and the fibers -- by $J_i$.

The flavor symmetry group is $PSU(n)$. We define the equivariant parameters $x_1,\dots, x_n$ and the real masses $m_1,\dots,m_n$, subject to
\begin{align}
x_1 x_2\dots x_n&=1,\cr
m_1 + m_2+\dots + m_n&=0,
\end{align}
such that the chiral $I_i$ has the equivariant parmeter $x_i \hbar^{1/2}$ and the real mass $m_i$. In addition to the choice of chamber $\zeta_\R>0$ in the real FI space, choose the following chamber in the real mass space:
\begin{equation}
\label{CPnChamber}
\mathfrak{C}=\{m_1> m_2>\dots > m_n\}.
\end{equation}
Other chambers can be accessed by the action of the permutation group $S_n$, the Weyl group of $PSU(n)$.

There are precisely $n$ fixed points corresponding to the massive vacua, all of which sit in the base of $T^*\C P^{n-1}$:
\begin{equation}
p^{\rm th} \text{ fixed point:}\qquad I_i = \delta_{i,p}\sqrt{\zeta_\R},\quad J_i=0,\quad \sigma = -m_p.
\end{equation}
In this vacuum, $(I_p, J_p)$ has no real mass, $(I_i,J_i)$ with $i<p$ receive large positive real masses $m_i-m_p$, and those with $i>p$ receive large negative real masses. Slightly abusing the notations, we can write:
\begin{align}
\cR_0(p) &\text{ is spanned by } I_p,\cr
\cR_+(p) &\text{ is spanned by } I_i, i<p,\cr
\cR_-(p) &\text{ is spanned by } I_i, i>p.
\end{align}
Furthermore, in the splitting $\cR_0(p)=\cR_0^D(p)\oplus \cR_0^N(p)$, $\cR_0^N(p)$ is empty, so $\cR_0^D(p)=\cR_0(p)$ (because $I_p$ gets a vev and $J_p$ does not).

We see that $G_p=G=U(1)$, so the theory $\cT^{\mathfrak{C}}_{p_1}$ corresponding to the vacuum $p_1$ is just a $U(1)$ gauge theory with one hypermultiplet $(I_{p_1}, J_{p_1})$. The boundary conditions between $\cT$ at $y>0$ and $\cT^{\mathfrak{C}}_{p_1}$ at $y<0$ are
\begin{align}
I_i\big|= \partial_y J_i\big|&=0, \, i<p_1,\cr
J_i\big|= \partial_y I_i\big|&=0, \, i>p_1.\cr
\end{align}

To define the future boundary conditions, we pick a polarization. It is most convenient to choose a polarization on $T^* \C P^{n-1}$ that corresponds to leaves along the base (the pull-back of the tangent bundle under the projection $T^*\C P^{n-1} \to \C P^{n-1}$). It lifts to the following linear polarization on the hypermultipelts:
\begin{equation}
\bL \text{ is spanned by } I_i, i=1..n,\qquad \bL^\perp \text{ is spanned by } J_i, i=1..n.
\end{equation}
Thus the $\sB_{\bL,p_2}$ boundary conditions imposed at the future boundary (where we pick the vacuum $p_2$) are simply the Dirichlet boundary conditions on the hypers:\footnote{The precise boundary value of $I_{p_2}$ is not important as it undergoes a non-trivial RG flow. In any case, it never appears in our formulas.}
\begin{equation}
I_i\big| = \delta_{i,p_2}\sqrt{\zeta_\R},\quad \partial_y J_i\big|=0,
\end{equation}
and the Dirichlet boundary conditions on the vectors, with the boundary flat connection
\begin{equation}
s = s^{(p_2)}(x,\hbar) = x_{p_2}^{-1}\hbar^{-1/2}.
\end{equation}
The latter equation follows from the $I_{p_2}$ having the weight $s x_{p_2} \hbar^{1/2}$.

The boundary conditions $\sN_{p_1}$ at the past boundary are:
\begin{equation}
\partial_y I_{p_1}\big| = J_{p_1}\big|=0,\quad \text{and (2,2) Neumann boundary conditions on the vector multipelt}.
\end{equation}
Additionally, we have to choose a boundary theory $\Upsilon_{p_1}$ there. The boundary chiral anomaly of $\sN_{p_1}$ is easily found to be
\begin{equation}
\left[(2p_1-2-n)(\mathbf{f}_\hbar-\mathbf{r})-2\mathbf{f}_z\right]\mathbf{f}.
\end{equation}
This tells us, according to our rules, that the anomaly of $\Upsilon_{p_1}$ must be:
\begin{equation}
P[\Upsilon_{p_1}] = 2\left(\mathbf{f} + \mathbf{f}_{p_1} + \frac12 \mathbf{f}_\hbar - \frac12 \mathbf{r}\right)\left(\mathbf{f}_z- \left(p_1-1-\frac{n}{2}\right)(\mathbf{f}_\hbar - \mathbf{r})\right),
\end{equation}
where $\mathbf{f}_{p_1}$ is the field strength for the flavor symmetry $U(1)\subset \mathbf{A}$ whose fugacity is $x_{p_1}$. We need to construct a 2d system with such an anomaly, which also has the property that its torus partition function (elliptic genus) is $1$ at $s=s^{(p_1)}(x,\hbar)$. It is not hard to just write the elliptic genus which satisfies such conditions:
\begin{equation}
\mathbb{W}[\Upsilon_{p_1}] = \frac{\vartheta(s x_{p_1}\hbar^{1/2} \times z \hbar^{-(p_1-n/2)})\vartheta(\hbar^{-1})}{\vartheta(s x_{p_1}\hbar^{1/2}\times \hbar^{-1})\vartheta(z \hbar^{-(p_1-n/2)})},
\end{equation}
where we organized factors in a way that makes anomalies manifest. This elliptic genus can be realized via the following set of multiplets:
\begin{center}
\begin{tabular}{|c|c|c|c|c|}
\hline
Multiplet & weight under $U(1)\times SU(n)\times U(1)_\hbar\times U(1)_{\rm top}$\\
\hline
Fermi & $s x_{p_1} \hbar^{\frac12(n+1)-p_1} z$ \\
Fermi & $\hbar^{-1}$ \\
Chiral & $sx_{p_1}\hbar^{-1/2}$\\
Chiral & $\hbar^{\frac12 n -p_1} z$\\ 
\hline
\end{tabular}
\end{center}
where instead of explicitly writing out charges of each multiplet, we wrote the corresponding weight (fugacity).

It is straightforward to compute the remaining ingredients of the interval partition function. Since the gauge group is abelian, there is no $\mathbb{V}_{p_1}$. The factor $\mathbb{M}_{p_1,p_2}$ only receives contributions from $(I_i, J_i)$, $i<p_1$, since they obey identical boundary conditions on the two ends. We thus find:
\begin{equation}
\mathbf{S}_{p_1}(s,x,\hbar,z) = \frac{\vartheta(s x_{p_1}\hbar^{1/2} z \hbar^{-(p_1-n/2)})\vartheta(\hbar^{-1})}{\vartheta(s x_{p_1}\hbar^{-1/2})\vartheta(z \hbar^{-(p_1-n/2)})}  \prod_{i<p_1} \frac{\vartheta(s x_i\hbar^{1/2})}{\vartheta(s x_i\hbar^{-1/2})}.
\end{equation}
To make contact with the math literature, we have to redefine some variables. First, it makes sense to replace:
\begin{equation}
s \hbar^{1/2} \mapsto s,
\end{equation}
such that at the $p$'th vacuum one has $s x_p=1$, which is more common in the literature on this subject \cite{Aganagic:2016jmx,Smirnov:2018drm}. Second, let us rescale:\footnote{This, most likely, has to do with the fact that $z$ counts BPS monopoles, and, since monopoles transform under $SU(2)_C$, they have non-zero $U(1)_\hbar$ charge. Thus the natural expansion parameter, in our conventions, is not $z$ but rather $z\hbar^{-n/2}$.}
\begin{equation}
z\hbar^{-n/2}\mapsto z.
\end{equation}
This gives:
\begin{equation}
\label{CPnStilde}
\tilde{\mathbf{S}}_{p_1}(s,x,\hbar,z) = \frac{\vartheta(\hbar^{-1})}{\prod_{i=1}^n\vartheta(s x_i\hbar^{-1})} \times \underbrace{\prod_{i<p_1} \vartheta(s x_i) \frac{\vartheta(s x_{p_1}z \hbar^{-(p_1-n)})}{\vartheta(z \hbar^{-(p_1-n)})} \prod_{i>p_1} \vartheta(s x_i\hbar^{-1})}_{{\rm Stab}(p_1)},
\end{equation}
where in the second factor, upon comparison with \cite{Aganagic:2016jmx}, we clearly recognize the elliptic stable envelope of a point $p_1$ (after an insignificant replacement $\hbar\mapsto\hbar^{-1}$, due to an unfortunate mismatch of conventions). The first factor in \eqref{CPnStilde}, which in the language of \cite{Aganagic:2016jmx} would be written as a section of\footnote{Here $T^{1/2}X$ is the polarization of $X$ that we chose, $(T^{1/2}X)^\vee$ is the dual polarization, $\hbar$ is the trivial line bundle acted on by $U(1)_\hbar$, and $\Theta(\cdot)$ is the elliptic Thom class.} $\Theta((T^{1/2}X)^\vee - \hbar)^{-1}$, makes $\tilde{\mathbf{S}}_{p_1}$ look more like the pole-subtraction matrix $\mathfrak{P}_{\mathfrak{C}}$ (compare with the Section 5.4.1 of \cite{Aganagic:2017smx}). The pole-subtraction matrix is indeed just a differently normalized elliptic stable envelope. The reason we find such a normalization here is that our Janus interfaces can be used to construct a natural interface between the Coulomb and Higgs phases, as we will explore more fully in the upcoming paper.

Finally, the matrix elements of the interface are given by \eqref{JOverlap}, and they are
\begin{equation}
\mathbf{S}_{p_1}(s^{(p_2)}(x,\hbar),x,\hbar,z) \neq 0 \Leftrightarrow p_1 \leq p_2.
\end{equation}
Components with $p_2<p_1$ vanish because $\vartheta(1)=0$. Hence the transition between vacua across the interface is possible if $p_1 \leq p_2$. We can write it as the following ordering of massive vacua:
\begin{equation}
\label{order}
1 \to 2 \to \dots \to n,
\end{equation}
where arrows denote possible gradient flows. This is the ordering of fixed points corresponding to the choice of chamber in \eqref{CPnChamber}.

Note that our conventions differ from \cite{Aganagic:2016jmx} in one more little detail: our stable envelope is supported on the full repellent $\bar{\rm Rep}(p_1)$ of a fixed point, while theirs is supported on the full attractor $\bar{\rm Attr}(p_1)$. Thus, our answer for the chamber $\mathfrak{C}$ equals their answer for $-\mathfrak{C}$.

We studied $\cJ(0,m\to\infty_{\mathfrak{C}})|p\rangle$ (or rather $\cJ(0,m\to\infty_{\mathfrak{C}})|\sN_p\rangle$), in which case we naturally obtain $\bar{\rm Rep}(p_1)$. We could of course consider a reflected configuration, when $\cT$ occupies the $y<0$ side, and $\cT^{\mathfrak{C}}_p$ -- the $y>0$ side of the interface, which would compute $\langle p|\cJ(m\to\infty_{\mathfrak{C}},0)$. Such a setup would produce stable envelopes supported on the attractor, like in \cite{Aganagic:2016jmx}. The choice made in the current text was motivated by the intention to contrast/compare the two approaches.

\subsubsection{$A_{n-1}$ ALE space}
In the $T^*\C P^{n-1}$ example of the previous section, we were able to choose a linear polarization $\bL$ on the hypermultipelts, such that at all fixed points, only the chirals $I\in \bL$ were getting vevs. Such a global $\bL$ does not exist in general, and the simplest example to illustrate this is the $A_N$ ALE space, which we will simply denote $A_N$. The gauge theory that has $A_N$ for its Higgs branch is the circular abelian quiver with $N+1$ $U(1)$ gauge nodes connected to each other by the bi-fundamental hypers $H_{i,i+1}$. In fact, this theory is 3d mirror to the SQED$_n$ considered in the previous subsection, with $N=n-1$, which makes it even more appropriate to study now.

In the circular quiver, the diagonal $U(1)$ vector multipelt decouples and is usually remover from the theory, so it is better to consider an equivalent linear quiver, which is more transparent to study:
\begin{center}
	\includegraphics[scale=0.7]{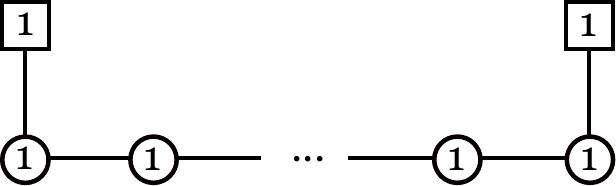}
\end{center}
There are $N$ gauge nodes, and this is a 3d $\cN=4$ quiver, meaning that each (unoriented) line denotes a bifundamental hypermultiplet. Let us denote the hypermultiplets by
\begin{equation}
(I_{i,i+1}, J_{i,i+1}),\ i=0\dots N,
\end{equation}
where each $I_{i,i+1}$ has charges $(1,-1)$ with respect to the $i$'th and $(i+1)$'th $U(1)$ factor. In this notation, the 0'th and the $(N+1)$'th copies of $U(1)$ are not gauged, while those labeled by $1,2,\dots,N$ are gauged. Despite appearance of two $U(1)$ flavor nodes, the flavor symmetry group here is
\begin{equation}
G_H=U(1)_f,
\end{equation}
and correspondingly, there is only one real mass $m$, which has two chambers, $m>0$ and $m<0$ (like the FI parameter in the mirror SQED). We choose
\begin{equation}
\mathfrak{C}=\{m>0\}.
\end{equation}
We incorporate this mass in the Lagrangian by giving it to the left-most hyper:
\begin{equation}
(I_{0,1},J_{0,1})\quad \text{are given real masses } (m,-m).
\end{equation}
We can identify it with the real scalar in the 0'th, i.e. non-dynamical, vector multiplet:
\begin{equation}
\sigma_0=m.
\end{equation}
There are $N$ real FI parameters $(\zeta_1,\dots,\zeta_N)$ corresponding to the gauge nodes. One usually introduces parameters $\omega_0,\dots,\omega_N$, such that
\begin{equation}
\zeta_i=\omega_{i-1}-\omega_i,\quad \text{and}\quad \sum_{i=0}^N\omega_i=0.
\end{equation}
These $\omega_i$'s are mapped to masses in SQED$_{N+1}$ under the mirror symmetry. The FI chambers correspond to different orderings of the $\omega_i$'s, and we choose to work in the chamber:
\begin{equation}
\omega_0>\omega_1>\dots>\omega_N,
\end{equation}
or in other words, all the FI parameters are positive $\zeta_i>0$. The Higgs branch is described by $XY=Z^{N+1}$, with
\begin{equation}
X=\prod_{i=0}^N I_{i,i+1},\quad Y=\prod_{i=0}^N J_{i,i+1},\quad Z=I_{i,i+1}J_{i,i+1},
\end{equation}
where in the last equation $i$ is any number from $0$ to $N$, since the complex moment map constraints ensure
\begin{equation}
\label{complMuAN}
I_{0,1}J_{0,1}=I_{1,2}J_{1,2}=\dots=I_{N,N+1}J_{N,N+1}.
\end{equation}
The singularity of $XY=Z^{N+1}$ is resolved by the FI parameters, with the real moment map constraints given by
\begin{equation}
\label{realMuAN}
|I_{i,i+1}|^2 - |J_{i,i+1}|^2 - |I_{i-1,i}|^2 + |J_{i-1,i}|^2 = \zeta_i,\ i=1\dots N.
\end{equation}
There are $N+1$ massive vacua, i.e. fixed points of $U(1)_f$, which we label by $k=0,\dots,N$. In the $k$'th massive vacuum, the dynamical real scalars in the vector multiplet take values:
\begin{align}
\sigma_1 = \sigma_2=\dots\sigma_k&=m,\cr
\sigma_{k+1}=\sigma_{k+2}=\dots\sigma_N&=0,
\end{align}
and in the $k=0$ vacuum, they simply all vanish. The effective real masses of all the hypermultiplets vanish, except for the $k$'th hyper $(I_{k,k+1},J_{k,k+1})$, which obtains effective real mass
\begin{equation}
(m_{\rm eff}, -m_{\rm eff}),\qquad m_{\rm eff}=\sigma_k - \sigma_{k+1}=m.
\end{equation}
The $k$'th hyper thus has zero expectation value. The vevs of others are fixed by the equations \eqref{complMuAN} and \eqref{realMuAN}, and we find:
\begin{align}
\label{VacVevAN}
\text{For }i<k:\quad &I_{i,i+1}=0,\ |J_{i,i+1}|=\sqrt{\zeta_{i+1}+\zeta_{i+2}+\dots +\zeta_k},\cr
\text{For }i>k:\quad &J_{i,i+1}=0,\ |I_{i,i+1}|=\sqrt{\zeta_{k+1}+\zeta_{k+2}+\dots +\zeta_i}.
\end{align}

It is clear that the equivariant parameters basically repeat this pattern, up to a modification by $\hbar$. Namely, there is one flavor parameter $x$ (in addition to $\hbar$), which is assigned to the $0$'th hypermultiplet. There are $N$ gauge equivariant parameters (Chern roots) $s_1,\dots,s_N$, and at the $k$'th vacuum they adjust to:
\begin{align}
\label{skAN}
s_i^{(k)}&=x\hbar^{-i/2},\ i=1\dots k\cr s_{i}^{(k)}&=\hbar^{-\frac{N+1-i}{2}},\ i=k+1\dots N.
\end{align}

As we send $m\to+\infty$, the $k$'th hyper freezes and disappears from the spectrum. What remains is our theory $\cT^{\mathfrak{C}}_k$, which is now seen to be described by a disconnected pair of quivers:
\begin{center}
	\includegraphics[scale=0.7]{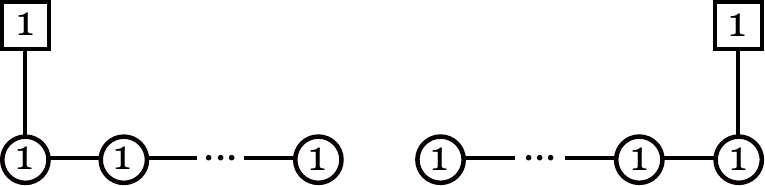}
\end{center}
where the left quiver has $k$ gauge nodes, and the right one has $N-k$ gauge nodes. Each of these two theories has no flavor symmetries (as expected) and only one isolated vacuum described exactly as before (which is automatically massive), confirming that $\cT^{\mathfrak{C}}_k$ has only one massive vacuum.

We see from \eqref{VacVevAN} that which chiral in a pair $(I_{i,i+1}, J_{i,i+1})$ receives a vev depends on the vacuum $k$. Thus there is no one single polarization $\bL$ such that only chirals in $\bL$ receive vevs. We still have to choose polarization, however, because $(I_{k,k+1}, J_{k,k+1})$ does not have a vev in the vacuum $k$. Let us pick polarization spanned by $I_{i,i+1}$, so in the vacuum $k$, $I_{k,k+1}$ is given Dirichlet boundary conditions.

Now we can define $\sB_{\bL,p}$ boundary conditions. On the hypers we impose:
\begin{align}
I_{p,p+1}\big|= \partial_y J_{p,p+1}\big| &=0,\cr
\text{For } i<p:\quad \partial_y I_{i,i+1}\big|&=0,\quad J_{i,i+1}\big| = c_i,\cr
\text{For } i>p:\quad \partial_y J_{i,i+1}\big|&=0,\quad I_{i,i+1}\big|=c_i,
\end{align}
where $c_i$ are the vevs as in \eqref{VacVevAN}. On the vectors we impose the usual $(2,2)$ Dirichlet boundary conditions with the boundary flat connection $s=s^{(p)}(x,\hbar)$.

If we have $\cT$ in the $y>0$ half-space and $\cT^{\mathfrak{C}}_k$ in the $y<0$, then the hyper $(I_{k,k+1}, J_{k,k+1})$ only lives on the $y>0$ side. For $m\to+\infty$, at the interface it obeys:
\begin{equation}
I_{k,k+1}\big| = \partial_y J_{k,k+1}\big|=0.
\end{equation}

We also have to describe $\sN_k$ boundary conditions in the $\cT^{\mathfrak{C}}_k$ theory defined by the disconnected quiver shown earlier. The gauge fields obey $(2,2)$ Neumann boundary conditions. Hypermultiplets, as it follows from \eqref{VacVevAN}, obey:
\begin{align}
\text{For }i<k:\quad &I_{i,i+1}\big|=0,\ \partial_y J_{i,i+1}\big|=0,\cr
\text{For }i>k:\quad &J_{i,i+1}\big|=0,\ \partial_y I_{i,i+1}\big|=0.
\end{align}
Finally, we need to include the boundary theory $\Upsilon_k$. For that, we have to compute the boundary chiral anomaly of $\sN_k$. First we note that $\cR_-(k)$ is 0, $\cR_+(k)$ only contains $I_{k,k+1}$, and $\cR_0^D(k)$ spans $\{I_{k+1,k+2}, I_{k+2, k+3},\dots,I_{N,N+1}\}$, while $\cR_0^N(k)$ spans $\{I_{0,1}, I_{1,2},\dots, I_{k-1,k}\}$. With the help of this data, we find the boundary chiral anomaly of $\sN_k$ to be:
\begin{equation}
-2\left[\sum_{i=1}^N \mathbf{f}_i \mathbf{f}_z^{(i)} + (\mathbf{f}_\hbar - \mathbf{r})\mathbf{f}_{k+1} \right],
\end{equation}
where $\mathbf{f}_i$ denote the gauge field strengths, and $\mathbf{f}_z^{(i)}$ -- the corresponding topological symmetries. The last term here must be dropped for $k=N$, since $\mathbf{f}_{N+1}$ does not correspond to the gauge symmetry.

From equations \eqref{skAN}, we know that $\mathbf{f}_i - \mathbf{f}_x + \frac{i}{2}\mathbf{f}_\hbar$ for $i\leq k$ and $\mathbf{f}_i + \frac{N+1-i}{2}\mathbf{f}_\hbar$ for $i>k$ are set to zero in the $k$'th vacuum. We thus find that the anomaly polynomial of $\Upsilon_k$ must be:
\begin{equation}
P[\Upsilon_k] = 2\left[\sum_{j=1}^k \left(\mathbf{f}_j- \mathbf{f}_x + \frac{j}{2}\mathbf{f}_\hbar\right) \mathbf{f}_z^{(j)} + \sum_{j=k+1}^N \left(\mathbf{f}_j + \frac{N+1-j}{2}\mathbf{f}_\hbar\right) \mathbf{f}_z^{(j)} + (\mathbf{f}_\hbar - \mathbf{r})\left(\mathbf{f}_{k+1} + \frac{N-k}{2}\mathbf{f}_\hbar\right) \right],
\end{equation}
where the last term should also be dropped for $k=N$. This anomaly polynomial can be engineered by $N$ copies of the system, which, like in the $T^*\C P^{n-1}$ case, contains two Fermi and two chiral multiplets in each copy. We can just directly write (a non-unique) elliptic genus corresponding to the anomaly polynomial $P[\Upsilon_k]$:
\begin{align}
\mathbb{W}[\Upsilon_k] &= \frac{\vartheta(s_{k+1} \hbar^{(N-k)/2+1} z_{k+1}\xi_{k+1})\vartheta(\xi_{k+1})}{\vartheta(s_{k+1} \hbar^{(N-k)/2}\xi_{k+1})\vartheta(z_{k+1}\hbar \xi_{k+1})}\cr
&\times\prod_{j=1}^k \frac{\vartheta(s_j x^{-1} \hbar^{j/2} z_j\xi_j)\vartheta(\xi_j)}{\vartheta(s_j x^{-1}\hbar^{j/2}\xi_j)\vartheta(z_j\xi_j)} \prod_{j=k+2}^N \frac{\vartheta(s_j \hbar^{(N+1-j)/2} z_j\xi_j)\vartheta(\xi_j)}{\vartheta(s_j \hbar^{(N+1-j)/2}\xi_j)\vartheta(z_j\xi_j)},
\end{align}
and it is straightforward, though tedious, to describe charges of the corresponding multiplets. Notice that we have incorporated undefined variables $\xi_1,\dots,\xi_N$, which can be some products of powers of the fugacities $(s,x,\hbar,z)$. Their presence does not change the anomaly polynomial or spoil the fact that at $s=s^{(k)}(x,\hbar)$, $\mathbb{W}[\Upsilon_k]$ evaluates to $1$. We just have to make sure that we do not introduce unwanted poles or zeros, for example $\xi_j=1$ would be a bad choice. This is of course part of the previously mentioned ambiguity of multiplication by a fully elliptic meromorphic function.

Let us look at the remaining ingredients. Again, there is no $\mathbb{V}_k$ because the gauge group is abelian. The factor $\mathbb{M}_{k,p}$ receives contributions from the hypermultiplets. At the $\sB_{\bL,p}$ boundary (say at $y=y_+>0$), $(I_{i,i+1},J_{i,i+1})$ obey (D,N) for $i\geq p$ and (N,D) for $i<p$. At the interface (say at $y=0$), $(I_{k,k+1},J_{k,k+1})$ obey $(D,N)$, and at the opposite end (say at $y=y_-<0$), $(I_{i,i+1},J_{i,i+1})$ obey (D,N) for $i< k$ and (N,D) for $i>k$. We see that when $p>k$, only the multiplets $(I_{i,i+1},J_{i,i+1})$ with $k<i<p$ have interval zero modes, while for $p\leq k$, only those with $p\leq i\leq k$ have them. We can thus write the elliptic genus of zero modes:
\begin{align}
\label{MnaiveAN}
p>k:\quad \mathbb{M}_{k,p} &= \prod_{k<i<p} \frac{\vartheta(s_{i+1}s_i^{-1}\hbar^{1/2})}{\vartheta(s_i s_{i+1}^{-1}\hbar^{1/2})},\cr
p\leq k:\quad \mathbb{M}_{k,p} &= \prod_{p\leq i\leq k} \frac{\vartheta(s_{i}s_{i+1}^{-1}\hbar^{1/2})}{\vartheta(s_{i+1}s_{i}^{-1}\hbar^{1/2})}.
\end{align}
These expressions are to be evaluated at $s=s^{(p)}(x,\hbar)$, as usual. One can notice that this gives zero, unless $p=k$ or $p=k+1$. This is not the expected answer: rather, the vacuum $k$ should be able to flow to any vacuum $p\geq k$.

This can be seen, first of all, from the gradient flow equations. One can see that for $m>0$, the flow goes to the right in the quiver, and the vacua are ordered according to:
\begin{equation}
\label{VacOrderAN}
0\to1\to2\to\dots\to N.
\end{equation}
An illustration of this is given on Figure \ref{fig:a5}.
\begin{figure}[h]
	\centering
	\includegraphics[scale=0.7]{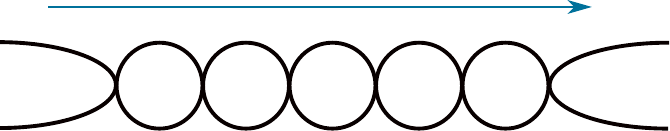}
	\caption{Schematic illustration of the $A_5$ space. The arrow denotes direction of the ``wind'', i.e., the gradient flow. The fixed points are where the segments touch each other.}\label{fig:a5}
\end{figure}\\

In our effective approach, the direction of the flow is determined by the boundary conditions on the hypers that terminate at the $\cT|\cT^{\mathfrak{C}}_k$ interface. Indeed, these boundary conditions are chosen in accordance with the direction of the flow on the massive side. In the current example, only $(I_{k,k+1},J_{k,k+1})$ ends there and obeys $(D,N)$ boundary conditions. If $p<k$, then $(I_{k,k+1},J_{k,k+1})$ also obeys the $(D,N)$ boundary conditions on the other end, and furthermore $I_{k,k+1}$ is given a boundary vev there, which implies that the partition function of this multiplet is zero.\footnote{BPS, or gradient flow, equation does not have a solution that connects $I_{k,k+1}\big|_{y=0}=0$ to $I_{k,k+1}\big|_{y=y_+}=c_k$ in finite time. Even if we could find such a solution somewhere in field space (say at infinite $\sigma_k$), there would be a fermion zero mode that at $s=s^{(p)}$ does not have a fugacity and forces the path interval to vanish.} For $p\geq k$ this does not happen, so one naturally expects to have transitions from $k$ to all possible vacua on the right in \eqref{VacOrderAN}.

The fact that the answer \eqref{MnaiveAN} vanishes at $s^{(p)}(x,\hbar)$ for $p\geq k+2$ does therefore look like a problem. However, we can use the ambiguity in $\mathbb{W}[\Upsilon_k]$ to cancel the unwanted zeroes in $\mathbb{M}_{k,p}$. For example, we may pick:
\begin{equation}
\xi_j=s_{j-1}^{-1}\hbar^{(j-N)/2},\quad \text{For } j\geq k+2,
\end{equation}
while $\xi_j=\hbar^{-1}$ for $j\leq k+1$. This gives:
\begin{align}
\mathbb{W}[\Upsilon_k] &= \frac{\vartheta(s_{k+1} \hbar^{(N-k)/2} z_{k+1})\vartheta(\hbar^{-1})}{\vartheta(s_{k+1} \hbar^{(N-k)/2-1})\vartheta(z_{k+1})}\cr
&\times\prod_{j=1}^k \frac{\vartheta(s_j x^{-1} \hbar^{j/2-1} z_j)\vartheta(\hbar^{-1})}{\vartheta(s_j x^{-1}\hbar^{j/2-1})\vartheta(z_j\hbar^{-1})} \prod_{j=k+2}^N \frac{\vartheta(s_j s_{j-1}^{-1}\hbar^{1/2} z_j)\vartheta(s_{j-1}^{-1}\hbar^{(j-N)/2})}{\vartheta(s_j s_{j-1}^{-1}\hbar^{1/2})\vartheta(z_j s_{j-1}^{-1}\hbar^{(j-N)/2})}.
\end{align}
We thus find:
\begin{align}
\mathbf{S}_k(s^{(k)}(x,\hbar),x,\hbar,z)&=\mathbb{W}[\Upsilon_k]\big|_{s=s^{(k)}(x,\hbar)}=1,\cr
\mathbf{S}_k(s^{(k+1)}(x,\hbar),x,\hbar,z)&=\mathbb{W}[\Upsilon_k]\big|_{s=s^{(k+1)}(x,\hbar)}=\frac{\vartheta(x z_{k+1}\hbar^{(N-1)/2-k})\vartheta(\hbar^{-1})}{\vartheta(x \hbar^{(N-1)/2-k-1})\vartheta(z_{k+1})},\cr
\end{align}
while for $p\geq k+2$:
\begin{align}
\mathbf{S}_k(s^{(p)},x,\hbar,z)&=\mathbb{W}[\Upsilon_k]\mathbb{M}_{k,p}\big|_{s=s^{(p)}(x,\hbar)}\cr
&= \frac{\vartheta(x z_{k+1}\hbar^{(N-1)/2-k})\vartheta(\hbar^{-1})}{\vartheta(x \hbar^{(N-1)/2-k-1})\vartheta(z_{k+1})}\prod_{j=k+2}^p \frac{\vartheta(z_j)\vartheta(x^{-1}\hbar^{j-(N+1)/2})}{\vartheta(\hbar)\vartheta(z_j x^{-1}\hbar^{j-(N+1)/2})}.
\end{align}
The normalization obtained here is different from the normalization of stable envelopes in the mathematical literature \cite{Maulik:2012wi,Aganagic:2016jmx}. Namely, their normalization condition is that ${\rm Stab}(p)$ restricted to ${\rm Attr}(p)$ (not the full attractor!) is equal to $i_* j^*$, where $j^*$ is the pullback from $p$ to ${\rm Attr}(p)$, and $i_*$ is the pushforward from ${\rm Attr}(p)$ into the ambient space $X$. If one restricts this back to the point $p$, one simply obtains the Euler class of the repelling subspace $TX_{>}$, which in the elliptic case is $\Theta(TX_>)$. Since in our treatment attractors and repellants are swapped, the proper normalization of the elliptic stable envelope (in the current conventions) involves the attractor, which corresponds to the $I_{k,k+1}$ direction at the fixed point $k$:
\begin{equation}
{\rm Stab}(k)\big|_k = \Theta(T_pX_{<})=\vartheta(s_k s_{k+1}^{-1}\hbar^{1/2})\big|_{s=s^{(k)}}=\vartheta(x\hbar^{(N+1)/2-k})
\end{equation}
It would be interesting to compare our formulas to known results in the literature, e.g., \cite{dinkins2021elliptic}.

\subsubsection{$T^* {\rm Gr}(N,L)$}

This space is realized as the Higgs branch of the $U(N)$ gauge theory with $L$ fundamental hypermultiplets $(I_i^{a},J^i_{a})$, $a = 1, \ldots , N$, $i = 1, \ldots , L$, with the flavor group $U(1)_{\hbar} \times PSU(L)$. The topological symmetry is just $\mathbf{A}'=U(1)_{\rm top}$.

The $F$-term equation is
\be
\sum_{i=1}^{L} I_{i}^{a} J^{i}_{b} = 0\, , \ a,b = 1, \ldots , N,
\label{eq:ftgr}
\ee
while the $D$-term equation reads:
\be
\sum_{i=1}^{L} I_{i}^{a} {\bar I}_{i}^{b} - {\bar J}^{i}_a J^{i}_{b} = {\zeta} {\delta}_{b}^{a} \, , \ \ a,b = 1, \ldots , N.
\label{eq:dtgr}
\ee
As in the previous example, we have the equivariant parameters $x_{1}, \ldots , x_{L}$ obeying 
\be
\prod_{i} x_{i} = 1\ , 
\ee
and $\hbar$, as well as $L$ real masses $m_{i}$, obeying
\be
m_{1} + \ldots + m_{L} = 0,
\ee
and, when generic enough, breaking the flavor group to its maximal torus $\mathbf{A}$.

The $\mathbf{A}$-fixed points $p \in T^{*}Gr(N,L)$ are the solutions to \eqref{eq:ftgr} and \eqref{eq:dtgr}, which, additionally, solve:
\be
\left( m_{i} - {\sigma}_{a} \right) I_{i}^{a} = 0\, , \ \left( {\sigma}_{a} - m_{i} \right) J^{i}_{a} = 0 \, , 
\label{eq:fixtgr}
\ee
for some ${\sigma}^{(p)} = {\rm diag} \left( {\sigma}_{1}, \ldots , {\sigma}_{N} \right) \in \mathfrak{u}(N)$. 

Let us denote the color and flavor spaces ${\C}^{N}$, ${\C}^{L}$  by ${\bf N}$ and ${\bf L}$, respectively. Then $\Vert I_{i}^{a} \Vert$ is a matrix of an operator $I: {\bf L} \to {\bf N}$, while $\Vert J^{i}_{a} \Vert$ is a matrix of an operator $J: {\bf N} \to {\bf L}$. 

For masses, we again choose the chamber:
\begin{equation}
\mathfrak{C}=\{m_1>m_2>\dots>m_L\}.
\end{equation}
The FI parameters are taken from the chamber
$\mathfrak{C}'=\{\zeta > 0\}$ (for $\zeta < 0$ exchange $I$ with $J^\dagger$), so $I$ has to have a rank $N$ by virtue of \eqref{eq:dtgr}, implying
\be
J^{i}_{a} = 0\, , \ I_{i}^{a} = \sqrt{\zeta} {\delta}^{l(a)}_{i}\, , \ {\sigma}_{a} = m_{l(a)} 
\ee
for some injective map $l: \{ 1, \ldots, N \} \to \{ 1, \ldots , L \}$, modulo the permutation group $S_N$ (the Weyl group of $U(N)$) acting on the domain. In other words, two such maps that differ by a permutation in the domain give gauge-equivalent vacua, and in total we obtain $L\choose N$ isolated vacua.
%\be
%l(i) = l (j) \Longrightarrow i = j
%\ee

The choice of the function $l$ modulo $S_N$ determines the fixed point $p_{l} \in T^{*}Gr(N,L)^{\mathbf A}$. Now let us look at the tangent space and
its transformations under $\mathbf{A}$, as well as the gauge group $G_{p_{l}} \subset U(N)$ left unbroken by $\sigma^{(p_l)}$. The latter consists of gauge transformations
$g \in U(N)$, such that $g^{-1} {\sigma}^{(p_l)} g = {\sigma}^{(p_l)}$. For generic $m_i$'s, meaning $m_{i} \neq m_{i}$
for $i \neq j$ (which is the case inside the chamber $\mathfrak{C}$,) this means $g =  {\rm diag} \left( e^{{\ii}{\varphi}_{a}} \right)_{a=1}^{N} \in U(1)^{N} \subset U(N)$, i.e.
the maximal torus of the gauge group. So in this case the gauge group of the theory $\cT^{\mathfrak{C}}_{p_l}$ coincides with the maximal torus:
\begin{equation}
G_{p_l}=U(1)^N = \mathbf{H}.
\end{equation}
The roots $e_a-e_b$, $a\neq b$ of $U(N)$ receive large masses $m_{l(a)}-m_{l(b)}$, so the W-boson multiplets are integrated out. Given the ordering in $\mathfrak{C}$, the positive mass roots are:
\begin{equation}
\Delta_+(p_l)=\{e_a-e_b: l(a)<l(b)\}.
\end{equation}

The tangent space $T_{p_{l}}T^{*}Gr(N,L)$ %represents the relatively light fields, still charged under $G_{p_{l}}$. It 
is parametrized by the fluctuations ${\delta}I, {\delta}J$, obeying the linearized Eqs. \eqref{eq:ftgr}, \eqref{eq:dtgr}, modulo the linearized gauge transformations. In the gauge ${\delta}I I^{\dagger}  - J^{\dagger} {\delta}J - I {\delta} I^{\dagger}  + {\delta} J^{\dagger} J = 0$, these are the arbitrary fluctations ${\delta}I_{i}^{a}, {\delta}J^{i}_{a}$, with $i \in \{1, \ldots, L\} \backslash \{ l(1), \ldots , l(N) \}$, in other words:
\be
{\delta}I_{l(b)}^{a}  = 0\, , \ {\delta}J_{a}^{l(b)} = 0\, , \ a,b = 1, \ldots , N.
\ee
The corresponding equivariant weights of ${\delta}I_{i}^{a}, {\delta}J^{i}_{a}$ are $x_{i}x_{l(a)}^{-1}$, $x_{i}^{-1} x_{l(a)} \hbar$, as follows from the flat gauge field in the vacuum $p_l$:
\begin{equation}
\label{skGr}
s^{(p_l)}_a = x_{l(a)}^{-1}\hbar^{-1/2}.
\end{equation}

The real mass of the hypermultiplet $\left( I_{i}^{a}, J^{i}_{a} \right)$ in the vacuum $l$ is equal to $m_{i} - m_{l(a)}$. It is nonzero and large if $i\neq l(a)$, hence such hypermultiplets are integrated out. In the chamber $\mathfrak{C}$, the mass is positive for $i<l(a)$, so the positive and negative weight subspaces are:
\begin{align}
\cR_+(p_l) &=\text{ span of the weights of } I^a_i,\ i< l(a),\cr
\cR_-(p_l) &=\text{ span of the weights of } I^a_i,\ i> l(a).
\end{align}
The remaining hypermultiplets form the matter content of $\cT^{\mathfrak{C}}_{p_l}$:
\begin{equation}
\cR_0(p_l) = \text{ span of the weights of } I^a_{l(a)},\ a=1\dots N.
\end{equation}
Furthermore, because only $I^a_i$'s get vevs, $\cR_0^D(p_l)=\cR_0(p_l)$, and $\cR_0^N(p_l)=0$.

We thus see that $\cT^{\mathfrak{C}}_{p_l}$ is a tensor product of $N$ copies of the SQED$_1$ ($U(1)$ gauge theory with one hyper). All of its $N$ FI parameters are equal to $\zeta>0$, and there is only one vacuum. To avoid clumsy notations in the following, let us denote $\cT^{\mathfrak{C}}_{p_l}$ as $\cT^{\mathfrak{C}}_l$, and in general simply label fixed points $p_l$ as $l$.

To define the $\sB_{\bL,p_l}\equiv \sB_{\bL,l}$ boundary conditions, it is most convenient to choose the global linear polarization $\bL$ spanned by the $I^a_i$'s. Then the $\sB_{\bL,l}$ is
\begin{align}
I^a_i\big| = \sqrt{\zeta} \delta^{l(a)}_i,\quad \partial_y J_a^i\big| = 0,
\end{align}
accompanied by the usual Dirichlet boundary conditions on the vector multiplets, with the boundary flat connection \eqref{skGr}.

Consider the interface $\cT|\cT^{\mathfrak{C}}_{l}$, such that on the massive side (which is $y<0$) we pick the vacuum $l$. The boundary conditions on the $\cR_+(l)\oplus \cR_-(l)$-valued hypermultiplets (living on the $y>0$ side) at the interface are:
\begin{align}
I^a_i\big| = \partial_y J_a^i\big|  =0,\quad \text{for } i<l(a),\cr
J_a^i\big| = \partial_u I^a_i\big| =0,\quad \text{for }i>l(a).
\end{align}
The W-boson vector multiplets also terminate at $y=0$, and obey the Dirichlet and Neumann boundary conditions for the roots in $\Delta_+(l)$ and $-\Delta_+(l)$, respectively.

The $\sN_{l}$ boundary conditions in the theory $\cT^{\mathfrak{C}}_{l}$ are imposed at $y=y_-<0$. All the vector multiplets are given Neumann boundary conditions, and the $\cR_0(l)$-valued hypermultiplets obey:
\begin{equation}
J_a^{l(a)}\big|=\partial_y I^a_{l(a)}\big| = 0.
\end{equation}
The boundary chiral anomaly of $\sN_{l}$ is:
\begin{equation}
\label{chiralGr}
-2\mathbf{f}_z\Tr \mathbf{f}+2\mathbf{f}_\hbar \sum_{\alpha\in\Delta_+(l)}\langle\alpha,\mathbf{f}\rangle +(\mathbf{f}_\hbar-\mathbf{r})\left[ \Tr_{\cR_+(l)}-\Tr_{\cR_-(l)}-\Tr_{\cR_0^D(l)} \right]\mathbf{f}.
\end{equation}
Note that the sum over positive roots evaluates to:
\begin{equation}
\sum_{\alpha\in\Delta_+(l)}\alpha = 2\rho_{l},
\end{equation}
where $\rho_{l}$ is a permutation of the standard Weyl vector $\rho$,
\begin{equation}
2\rho=(N-1,N-3,\dots,1-N).
\end{equation}
Since we are going to Weyl-average in the end anyways, we may assume without loss of generality that $l(1)<l(2)<\dots <l(N)$, in which case $\rho_{l}=\rho$. Then \eqref{chiralGr} evaluates to:
\begin{equation}
\sum_{a=1}^N\left[-2\mathbf{f}_z \mathbf{f}_{aa} + 4\rho_a \mathbf{f}_\hbar \mathbf{f}_{aa} + (2l(a)-2-L)(\mathbf{f}_\hbar - \mathbf{r})\mathbf{f}_{aa}  \right],
\end{equation}
where $\mathbf{f}_{aa}$ are the gauge fields corresponding to the maximal torus $\mathbf{H}=G_{l}$. To cancel this anomaly, we add the boundary theory $\Upsilon_{l}$ with the anomaly polynomial:
\begin{equation}
P[\Upsilon_{l}]=2\sum_{a=1}^N \left(\mathbf{f}_{aa} + (\mathbf{f}_x)_{l_1(a)} + \frac12 \mathbf{f}_\hbar - \frac12\mathbf{r}\right)\left(\mathbf{f}_z + \left(\frac{L}2+1-l(a)-2\rho_a\right)(\mathbf{f}_\hbar-\mathbf{r})-2\rho_a\mathbf{r} \right).
\end{equation}
We again build $\Upsilon_{l}$ from $N$ systems of two Fermi and two chiral multiplets, whose charges can be read off from their elliptic genus:
\begin{align}
\mathbb{W}[\Upsilon_{l}]=\prod_{a=1}^N \frac{\vartheta(s_a x_{l(a)}\hbar^{1/2}\times z \hbar^{\frac{L}{2}-l(a)-2\rho_a})\vartheta(\hbar^{-1})}{\vartheta(s_a x_{l(a)}\hbar^{1/2}\times \hbar^{-1})\vartheta(z \hbar^{\frac{L}{2}-l(a)-2\rho_a})}.
\end{align}
Now we compute other contributions to the interval index. In this example, there is a non-trivial effect from the W-boson multiplets expressed as a product over $\alpha\in\Delta_+(l)$ \eqref{WbosEll}:
\begin{equation}
\mathbb{V}_{l} = \prod_{\substack{a,\ b\\ l(a)<l(b)}} \frac{\vartheta(s_b s_a^{-1}\hbar)}{\vartheta(s_a s_b^{-1})}=\prod_{a<b} \frac{\vartheta(s_b s_a^{-1}\hbar)}{\vartheta(s_a s_b^{-1})},
\end{equation}
where we used the ordering $l(1)<l(2)<\dots<l(N)$. Finally, the matter index $\mathbb{M}_l$ receives contributions from $(I^a_i, J_a^i)$ with $i<l(a)$, as these are the only hypers that have the interval zero modes (one Fermi multiplet for $I^a_i$ and one chiral for $J_a^i$). Their effect is captured by:
\begin{equation}
\mathbb{M}_{l}=\prod_{\substack{i,\ a\\ i<l(a)}}\frac{\vartheta(s_a x_i\hbar^{1/2})}{\vartheta(s_a x_i\hbar^{-1/2})}.
\end{equation}
Then the final answer is
\begin{equation}
\mathbf{S}_{l}(s,x,\hbar,z) = {\rm Symm}_{S_N} \left(\mathbb{V}_{l}\mathbb{M}_{l}\mathbb{W}[\Upsilon_{l}]\right),
\end{equation}
where the symmetrization is performed over the gauge variables $s_a$ (elliptic Chern roots).

To compare this to the answer for ${\rm Stab}(l)$ given in \cite{Aganagic:2016jmx}, let us introduce
\begin{equation}
f_k(S,Z)=\prod_{i<k}\vartheta(S x_i) \frac{\vartheta(S x_k Z \hbar^{L-k})}{\vartheta(Z \hbar^{L-k})}\prod_{i>k}\vartheta(S x_i \hbar^{-1}),
\end{equation}
and rewrite $\mathbf{S}_{l}$ as
\begin{align}
\label{SlGr}
\mathbf{S}_{l} = \frac{\prod_{a\neq b}\vartheta(s_b s_a^{-1}\hbar)}{\prod_{i, a}\vartheta(s_a x_i\hbar^{-1/2})} \times {\rm Symm}_{S_N} \frac{\prod_{a=1}^N f_{l(a)}(s_a\hbar^{1/2}, z\hbar^{-\frac{L}{2}-2\rho_a})}{\prod_{a<b}\vartheta(s_a s_b^{-1})\vartheta(s_a s_b^{-1}\hbar)}.
\end{align}
The second factor exactly agrees with the elliptic stable envelope ${\rm Stab}(l)$ from \cite{Aganagic:2016jmx}, once we perform the same rescalings as in the $T^*\C P^{n-1}$ case:
\begin{equation}
s_a\hbar^{1/2}\mapsto s_a,\quad z\hbar^{-\frac{L}{2}}\mapsto z,
\end{equation}
and then replace $\hbar$ by $\hbar^{-1}$.

The first factor in \eqref{SlGr} is a Weyl-invariant combination that must be attributed to the difference in normalizations of the pole-subtraction matrix $\mathfrak{P}_{\mathfrak{C}}$ and ${\rm Stab}(l)$. To find the overlap \eqref{matr_el_to_compute}, i.e., the matrix element of the interface between the vacuum $l$ on the massive side and $l'$ on the massless, we have to substitute $s=s^{(l')}$ from \eqref{skGr} into \eqref{SlGr}.

\subsubsection{Hilbert scheme of points}

In this section we use the notations ${\bf W} \cong {\C}^{1}, {\bf N} \cong {\C}^{N}$ for the flavor and color spaces, respectively. The Hilbert scheme $M_{N} \equiv Hilb^{[N]}({\C}^{2})$ of $N$ points on ${\C}^2$ is the 
the Higgs branch of the $U(N)$ gauge theory with $1$ fundamental hypermultiplet $(I,J)$, $I: {\bf W} \to {\bf N}$, $J: {\bf N} \to {\bf W}$, and one adjoint hypermultiplet $(B_1, B_2)$, $B_{1,2} \in {\rm End}({\bf N})$:
\begin{center}
    \includegraphics[scale=1.0]{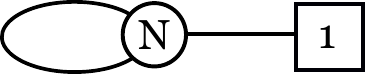}
\end{center}
The $\cN=4$ flavor group is $\mathbf{A}=U(1)$ acting on the adjoint hypermultiplet only, so the $\cN=2$ flavor group is $U(1)_\hbar \times U(1)$. The $F$-term equation reads
\be
IJ + [B_1, B_2] = 0\, , 
\label{eq:fhs}
\ee
while the $D$-term equation reads:
\be
I I^{\dagger} - J^{\dagger}J + [B_{1}, B_{1}^{\dagger}] + [ B_{2}, B_{2}^{\dagger}]\  = \ {\zeta} {\bf 1}_{\bf N} \, , 
\label{eq:dhs}
\ee
Unlike the previous example, we only have two equivariant parameters $x$ 
and $\hbar$, and just one mass $m$. It is well-known \cite{Nakajima:2014} that for $\zeta > 0$ the Eqs. \eqref{eq:fhs}, \eqref{eq:dhs} imply $J = 0$. We henceforth work in the chambers:
\begin{equation}
    \mathfrak{C}=\{m>0\},\qquad \mathfrak{C}'=\{\zeta>0\}.
\end{equation}

The $A$-fixed points $p_{\lambda} \in M_{N}^{A}$ are in one-to-one correspondence with the partitions $\lambda$, of size $|{\lambda}|= N$, i.e.
\be
{\lambda} \ = \ \left[ {\lambda}_{1} \geq {\lambda}_{2} \geq \ldots \geq {\lambda}_{{\ell}({\lambda})} \right] \ , 
\ee
with
\be
N = {\lambda}_{1} + {\lambda}_{2} + \ldots + {\lambda}_{{\ell}({\lambda})}\ . 
\ee
We shall also use the dual partition $\lambda^t$, with ${\lambda}^{t}_{j} = \# \{ \, i \, | \, 1 \leq i \, , \ {\lambda}_{i} \geq j \, \}$. 
The fixed point equations read, 
\be
m B_{1} = [ {\sigma}_{\lambda} , B_{1}]\, , - mB_{2} = [ {\sigma}_{\lambda}, B_{2}]\, , \ {\sigma}_{\lambda} I = 0\ ,  \ J {\sigma}_{\lambda} = 0
\label{eq:fixhs}
\ee
We cannot solve \eqref{eq:fhs}, \eqref{eq:dhs}, \eqref{eq:fixhs} modulo $U(N)$, but we can solve \eqref{eq:fhs} and \eqref{eq:fixhs} modulo $GL(N)$, and here is the result:
there is an orthonormal basis ${\bf e}_{(i,j)}$ of ${\bf N}$, with $(i,j) \in {\lambda}$, i.e. $1 \leq i \leq {\ell}({\lambda})$, $1 \leq j \leq {\lambda}_{i}$, in which
\be
B_{1} {\bf e}_{(i,j)} \ = \ \sqrt{\zeta \frac{h_{i+1,j}}{h_{i,j}}} {\bf e}_{(i+1, j)}\, , \ B_{2} {\bf e}_{(i,j)} \ = \ \sqrt{\zeta \frac{h_{i,j+1}}{h_{i,j}}} {\bf e}_{(i, j+1)} \, , \ I ({\bf W})\ = \ \sqrt{N {\zeta}}\, {\bf e}_{(1,1)}
\label{eq:part}
\ee
where the real numbers $h_{i,j}$ obey \cite{Braden:1999zp} the system of quadratic equations:
\be
\frac{h_{i,j}}{h_{i-1,j}} - \frac{h_{i+1,j}}{h_{i,j}} + \frac{h_{i,j}}{h_{i,j-1}} - \frac{h_{i,j+1}}{h_{i,j}} + N  {\delta}_{i,1}{\delta}_{j,1} =  1\, , \ (i,j) \in {\lambda} 
\label{eq:stabhs}
\ee
supplemented with the boundary conditions $h_{i,j} = +{\infty}$ when either $i = 0$ or $j = 0$, $h_{1,1} = 1$, and $h_{i,j} = 0$
when $j > {\lambda}_{i}$, or $i > {\lambda}_{j}^{t}$. The equations \eqref{eq:stabhs} have a unique solution, the critical point of a convex Morse function of $h_{i,j}$'s. 

The scalar $\sigma$ in the vector multiplet takes the value 
\be
{\sigma}^{(p_{\lambda})} \ =\  \sum_{(i,j) \in {\lambda}}\  m (i-j)\,  {\bf e}_{i,j} {\bf e}_{i,j}^{\dagger}
\label{eq:sighs}
\ee
The tangent space $T_{p_{\lambda}} M_{N}$ has a basis $\left( {\delta}_{i,j}^{\pm}  B_{1}, {\delta}_{i,j}^{\pm}  B_{2}, {\delta}_{i,j}^{\pm}  I, {\delta}_{i,j}^{\pm}  J \right)$, where $(i,j) \in \lambda$ with the corresponding hypermultiplets
having the mass $m \left( {\lambda}_{i} - j + {\lambda}_{j}^{t} - i +1 \right) $ (the hook formula).
The corresponding equivariant weights are (the arm-leg formula)
\be
x^{\pm \left( {\lambda}_{i} - j + {\lambda}_{j}^{t} - i +1 \right)} {\hbar}^{
\frac 12 \pm \frac{{\lambda}_{i} - j-{\lambda}_{j}^{t} + i}{2}}\, , \
\label{eq:twhs}
\ee
which are computed from the character 
\be
{\chi}_{T_{p_{\lambda}}} (q_{1}, q_{2}) = K^{*} + q_{1}q_{2} K - (1-q_{1})(1-q_{2}) K K^*
\label{eq:charths}
\ee
with
\be
K (q_{1},q_{2}) = \sum_{(i,j) \in {\lambda}} q_{1}^{i-1} q_{2}^{j-1} \, , \ K^{*}(q_{1},q_{2}) = K(q_{1}^{-1}, q_{2}^{-1})
\ee
using the representation of the 
tangent space as a cohomology of a three-term equivariant complex \cite{Nakajima:2014}, or by direct examination of the linearized equations and symmetries. 
The unbroken gauge group $G_{p_{\lambda}} \subset U(N)$ is the centralizer of ${\sigma}^{(p_{\lambda})}$ in \eqref{eq:sighs}, i.e.
\be
\label{HSGp}
G_{p_{\lambda}} \ = \ \prod_{h=-\infty}^{\infty} \, U(n_{h, \lambda})
\ee
where
\be
n_{h,\lambda} = \# {\cI}_{h, \lambda}\, , \qquad {\cI}_{h, \lambda} = \left\{\,  i \, | \,  i - {\lambda}_{i} \leq h \leq i-1 \, \right\} 
\ee
The ranks $n_{h,\lambda}$ can be read off the reduced character
\be
{\kappa}(x) = K(x, x^{-1}) = \sum_{h} n_{h, \lambda} x^{h}
\label{eq:kapnh}
\ee
It is easy to see that ${\cI}_{h,\lambda} \subset {\cI}_{h\mp 1,\lambda}$ for $\pm h \geq 1$, hence the dimensions
$n_{h,\lambda}$ are non-increasing in both positive and negative $h$ directions.

It is also easy to see from \eqref{eq:twhs} that the reduced character of the tangent space
\be
{\tau}_{\lambda}(x) = {\chi}_{T_{p_{\lambda}}} (x, x^{-1}) \ =\  \sum_{{\square}\in {\lambda}} \, x^{h_{\square}} + x^{-h_{\square}}
\ee
does not have a constant term: $[x^{0}] {\tau}_{\lambda}(x) = 0$. Therefore (cf. \eqref{eq:charths}):
\be
\left( {\kappa}(x) - {\kappa}(x){\kappa}(x^{-1})(1-x) \right)_{0} = - \left( {\kappa}(x^{-1}) - 
{\kappa}(x){\kappa}(x^{-1})(1-x^{-1}) \right)_{0} = 0
\ee
Now plug into this the relation \eqref{eq:kapnh} to conclude
\be
\label{HSidentity}
n_{0} + \sum_{h} n_{h} n_{h+1}  = \sum_{h} n_{h}^{2}
\ee

In addition to the  vector multiplets for the gauge group \eqref{HSGp}, the light degrees of freedom associated to the vacuum $\lambda$ include: A bunch of bi-fundamental hypermultiplets originating from the massless components of $B_1$ and $B_2$, and a fundamental hypermultiplet for the group $U(n_0)$ originating from the $(i,i)\in\lambda$ componenets of $(I,J)$ (because $(I,J)$ only get masses from the vev of $\sigma^{(p_\lambda)}$, whose $(i,i)$ components vanish). The massless components of $B_1$ and $B_2$ only couple the adjacent blocks of the block-diagonal matrices in $\prod_h U(n_h)\subset U(N)$, and together, the light fields assemble into the quiver that defines $\cT^{\mathfrak{C}}_\lambda$:
\begin{center}
    \includegraphics[scale=1.0]{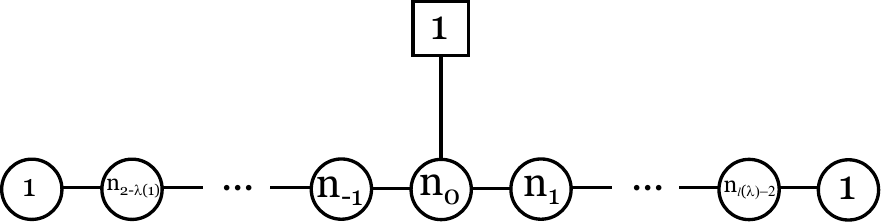}
\end{center}
where we have indicated that the very first and last non-zero $n_h\equiv n_{h,\lambda}$ are both $1$. This theory has $\ell(\lambda)+\lambda_1-1$ FI parameters and the same number of topological symmetries, but due to its origin as a light subsector in the ADHM quiver, all FI parameters should be equal to $\zeta$, and all the topological symmetries should have the same fugacity $z$. By construction, there is only one Higgs branch vacuum, and the equality \eqref{HSidentity} confirms that the dimensionality counting is in agreement with that.

As per usual story, the massive matter is organized into the spaces $\cR_\pm(\lambda)$ depending on the sign of real mass, and given Dirichlet/Neumann boundary conditions at the $\cT|\cT^{\mathfrak{C}}_\lambda$ interface. Defining the $\sB_{\bL,\lambda}$ boundary conditions follows the standard route, except the scenario mentioned in the footnote \ref{nonVevR0} takes place in this case, with some of the massless hypermultiplets in the ADHM quiver remaining at zero vev. This is easiest to see for $I$: the whole $n_0$-component subvector of $I$ (fundamental under the $U(n_0)$) remains massless, but in the description of fixed points given above, only the $(1,1)\in\lambda$ component of $I$ is non-zero. The same occurs for $B_{1,2}$. This is not really a problem, and it is still possible to choose boundary conditions for such components consistent with the requirement that the $U(N)$ symmetry is fully broken at the $\sB_{\bL,\lambda}$ boundary.

The definition of $\sN_\lambda$ boundary conditions becomes more subtle. If we follow the standard route here, we will end up in the situation mentioned in the Remark towards the end of Section \ref{sec:bndry_vacua}. Namely, the $N_p$ boundary conditions appear to explicitly break the gauge symmetry, even thought the Neumann boundary conditions on the gauge fields require its presence. As explained there, the way out is to impose the $(0,2)$ Neumann boundary conditions on all the chiral multiplets that constitute the matter content of our quiver theory $\cT^{\mathfrak{C}}_\lambda$. This, on the other hand, produces the boundary gauge anomaly, which has to be canceled by the boundary theory $\Upsilon_\lambda$. As usual, we have to make sure that $\langle \sN_\lambda|\sD_\lambda\rangle=1$, which does not entirely fix the normalization, and leaves a meromorphic elliptic ambiguity.

We thus see that our method also works straightforwardly, with very minor adjustments, in the case of Hilbert scheme of points. This is unlike the abelianization approach \cite{Aganagic:2016jmx}, which becomes significantly more involved, since the abelianization of $Hilb^{[N]}({\C}^{2})$ is a hypertoric variety with non-isolated $\mathbf{A}$-fixed points. These difficulties were overcome by Smirnov in \cite{Smirnov:2018drm}. We will complete the computations for $Hilb^{[N]}({\C}^{2})$ elsewhere, but to make them really useful, we would have to come up with a way to fix the elliptic ambiguity in the definition of $|\sN_\lambda\rangle$. Only then it would be possible to compare results with the sum over trees in \cite{Smirnov:2018drm}.

%\section{Chamber R-matrices and Higgsing}
%\section{Yangian interfaces: the two perspectives}
%\section{Bethe/gauge correspondence revisited}

\section{Outlook}

In this paper we gave a physical interpretation of the stable envelope ${\rm Stab}_{\mathfrak{C}}(X)$ constructions of \cite{Maulik:2012wi,Aganagic:2016jmx}. 
We realized them as the interfaces in softly broken ${\cN}=4$ 3d quiver gauge theories, which have $X$ as their Higgs branches, which preserve a fraction of supersymmetry, while having some mass/FI parameters varying in time. One can also consider the domain walls, where the masses vary along some spatial direction. These are more familiar as the supersymmetric boundary conditions in the folded theories. The advantage of time varying masses is that they act as operators in the space of states. The supersymmetry preserving interfaces define operators acting on the space of supersymmetric vacua. In this paper we are only interested in the cohomology of one of the supercharges $\cQ$. So, at times, we take the limit, in which the masses change in a quench-like manner, which we relate to Dirichlet and Neumann boundary conditions on the supermultiplets at the interface wall. In string theory brane realizations, these limits correspond to taking some of the nearly parallel flavor branes and rotating them to nearly $90$ degrees, obtaining essentially the intersecting branes configuration, with one stack of branes producing interfaces on the worldvolumes of another stack. Some representatives of the $\cQ$-cohomology might be more convenient to work with then others. In this respect it would be interesting to generalize the recent work \cite{Cuomo:2021rkm} on the renormalization group flows on line defects in conformal field theories
to the case of supersymmetric interfaces (our theories flow to superconformal fixed points both for vanishing and infinite masses). Additionally, it would be interesting to put the study of supersymmetric interfaces and boundary conditions in perspective of the paper \cite{Thorngren:2020yht}, and understand whether there can exist any subtle obstructions for the SUSY at the boundary.

The equivariant cohomology ($K$-theory, elliptic cohomology) of the fixed point set $X^{A}$ in \cite{Maulik:2012wi,Aganagic:2016jmx} corresponds to the space of supersymmetric vacua of the theory with real masses, which asymptote to infinity along a ray within a chamber $\mathfrak{C}$
in ${\rm Lie}(\mathbf{A})$. The equivariance is achieved by turning on a flat flavor (including $U(1)_{\hbar})$ connection along the two-torus in the case of a compactified 3d theory, or the appropriate combination of a flat connection and a twisted mass in 2d or 1d reductions. The equivariant cohomology ($K$-theory, elliptic cohomology) of $X$ is identified with the space of supersymmetric vacua of the theory with vanishing real masses, while keeping the same flat flavor (including $U(1)_{\hbar})$ connection along the two-torus in the case of a compactified 3d theory, or the appropriate combination of a flat connection and a twisted mass in 2d or 1d reductions. The stable envelope map is then the operator induced by the supersymmetric interface on the space of supersymmetric vacua. Mathematically, these are the harmonic representatives of $\cQ$-cohomology, for some of the nilpotent
(up to the global symmetry corresponding to the equivariance) supercharges $\cQ$. 

In the companion papers we shall be exploring the backgrounds changing the nature of $\cQ$, from the $A$-model supercharge to our supercharge, and their lifts to three dimensions, as well as their $\Omega$-deformed versions. Namely, we will study the cigar background that is the 3d uplift of the Gomis-Lee \cite{Gomis:2012wy} squashed sphere background, which thus preserves two supercharges. This 3d background is the squashed version of the usual 3d $\cN=2$ background used in the study of 3d index \cite{Imamura:2011su} and half-index \cite{Gadde:2013wq}. One can prove using the methods of \cite{Closset:2012ru,Closset:2013vra,Closset:2014uda} that the squashing parameter is a trivial deformation of the transversally holomorphic foliation (THF), thereby allowing to take the infinite squashing limit without affecting the BPS observables. Taking this limit connects the half-index to the topologically twisted half-index, known in the literature under a variety of names \cite{Shadchin:2006yz,Dimofte:2010tz,Bonelli:2011fq,Bonelli:2013mma,Beem:2012mb,Okounkov:2015spn,Gukov:2017kmk}. The tradition seems to be that each author comes up with a new name for this object, which we thus unconventionally called the ``topologically twisted index'' above (in the part II, however, we will mention all the known names). An interesting relation between this object and our interfaces is that one can change parameters of the theory on the cigar by acting with the Janus interface on its boundary, see Figure \ref{fig:act_on_cigar}.

\begin{figure}[h]\label{fig:act_on_cigar}
	\centering
	\includegraphics[scale=0.8]{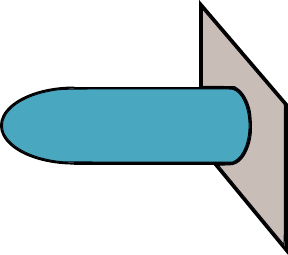}
	\caption{Acting with an interface on the cigar partition function.}
\end{figure}
Thus, our interfaces encode the behavior of the cigar partition function under the transition between the Higgs and Coulomb phases (the duality interface), or in the massive case, under the wall-crossing between the chambers $\mathfrak{C}_1$ and $\mathfrak{C}_2$ (the chamber R-matrices \cite{Maulik:2012wi,Aganagic:2016jmx}, to be discussed in part III).

We shall also consider the generalizations of our interfaces where some of the masses are replaced by the flavor flat connection. It is in this compactified (or T-dualized) setup, that we shall find the connection between the twisted transfer matrices and the local operators of the twisted chiral ring, earlier observed as the Bethe/gauge correspondence. We shall also connect our interfaces to the arrangements of line operators in the four dimensional Chern-Simons theory, in the case of a finite ADE quiver gauge theory. Of course, our constructions work for more general quiver theories, for which the corresponding 4d Chern-Simons is not fully understood (but string theory considerations suggest a higher dimensional theory \cite{NikHamburg:2017}).

Another extension of our work is the study of the analogous interfaces in theories, which start their life as higher dimensional ones, 5d/4d/3d, subject to an $\Omega$-deformation in two noncompact directions $D^2$ (which could be taken to have a cigar geometry). These theories are not sigma models with some finite dimensional target space $X$, rather they share many common features with a would-be sigma model with an infinite-dimensional target space. Thus, the quiver ${\cN}=2$ theories in four dimensions and their five dimensional uplifts would be related to the spin chains with infinite dimensional representations of the Lie algebra of the MacKay dual group at the spin sites \cite{Nekrasov:2012xe,Nekrasov:2013xda,Lee:2020hfu,Jeong:2021bbh}. The r{\^o}le of the $U(1)_{\hbar}$ symmetry is played by the rotations of $D^2$, so the Planck constant of the quiver spin chain is the $\Omega$-deformation parameter, in agreement with  \cite{Nekrasov:2009rc}. The novelty of this class of theories compared to the gauged linear sigma models considered in this paper is the complex spin of the associated representations of quantum algebras. In the $A_1$ case, considered in detail in the recent papers \cite{Lee:2020hfu,Jeong:2021bbh}, the $2N$ masses of quarks in the ${\cN}=2$ $SU(N)$ gauge theory in four dimensions determine the $N$ inhomogeneities, and $N$ (complex) spins of the $\mathfrak{sl}_{2}$ XXX-spin chain, governing the BPS sector of the $\Omega$-deformed theory.\footnote{Transfer matrices labeled by the complex spins of $\mathfrak{sl}_2$, but in the $XXX$ spin chain with the usual spin $1/2$ at each site, have also appeared in \cite{Dedushenko:2020yzd}, in a seemingly quite different context, but in exactly the same $SU(N)$, $N_f=2N$ gauge theory, this time with a codimension-one defect inserted. In that case, it is also true that inhomogeneities and the complex spins of representations of the auxiliary space are determined by the $2N$ masses.  The physical space of the spin chain realized in gauge theory can also be the usual Verma modules of $\mathfrak{sl}_2$, as in \cite{Dorey:2011pa}, or the unbounded modules found in \cite{Lee:2020hfu,Jeong:2021bbh}, which depend on the additional $N$ Coulomb moduli.} To compare to the construction of the present paper, we should start with the 3d reduction of that theory, and use the $m_{\R}$ component of the mass vector ${\vec m} = (m_{\R} ,m_{\C}, {\bar m}_{\C} )$ to define the Janus interfaces, while keeping the ``equivariant'' masses $m_{\C}$ constant across the defect. The 4d theory would correspond to the models with twisted transfer matrix (the twist determined by the complexified gauge coupling), to be analyzed in the part III of our series.  

We should also point out that some of our models can be mirror mapped to the ${\cN}=2$ Landau-Ginzburg theories in two dimensions. 
The interfaces in these theories and the $\infty$-structures they define were extensively studied in \cite{Gaiotto:2015zna,Gaiotto:2015aoa}. It seems interesting to find the corresponding quantum algebra generalizing the Yangians and quantum loop algebras (for 3d uplifts of those models) for Landau-Ginzburg theories. Conversely, the Yangian might
uplift to some $L_{\infty}$-structure, with the inclusion of junctions of Janus interfaces of our paper.

\appendix
\section{3d $\cN=2$ conventions}\label{app:conv}
\subsection{Flat space}\label{app:FlatConv}
We choose the Euclidean gamma matrices to be given by Pauli matrices $\tau_\mu$:
\begin{equation}
\gamma_\mu = \tau_\mu,\quad \mu=1,2,3.
\end{equation}
Spinors are by default assumed to have lower indices, which can be raised according to
\begin{equation}
\psi^\alpha= \varepsilon^{\alpha\beta}\psi_\beta,
\end{equation}
where $\varepsilon^{12}=\varepsilon_{21}=1$. As usual, the contractions work as
\begin{equation}
\psi\chi = \psi^\alpha\chi_\alpha.
\end{equation}
A 3d $\cN=2$ gauge theory is built from two types of multipelts: 1) A gauge or vector multiplet $V=(A_\mu,\sigma,\lambda,\bar\lambda,D)$, where $\sigma$ is a real scalar, $(\lambda,\bar\lambda)$ is a Dirac spinor, and $D$ is a real auxiliary field; 2) A chiral multiplet $\Phi=(\phi,\psi,\bar\psi,F)$, where $\phi$ is a complex scalar, $(\psi,\bar\psi)$ a Dirac spinor, and $F$ a complex auxiliary field. Note that $\Phi$ and $\phi$ are just generic names in this appendix, which should not be confused with the specific adjoint-valued chiral $\Phi$ that enters the construction of 3d $\cN=4$ theories in the main text.

In flat Euclidean space, the SUSY transformations are parameterized by a pair
\begin{equation}
\epsilon=\left(\begin{matrix}
\epsilon_1\\ \epsilon_2
\end{matrix} \right),\quad \bar\epsilon=\left(\begin{matrix}
\bar\epsilon_1\\ \bar\epsilon_2
\end{matrix} \right),
\end{equation}
which contain four independent infinitesimal parameters. Upon Wick rotation to Minkowski signature, $\epsilon$ and $\bar\epsilon$ should be thought of as complex spinors related to each other by a proper conjugation operation. We find it more convenient to treat $\epsilon$, $\bar\epsilon$ as commuting spinors. The SUSY variations of the vector multiplet are given by the following equations:
\begin{align}
\delta A_\mu &\ = \ \frac{\ii}{2} (\bar\epsilon \gamma_\mu\lambda + \epsilon\gamma_\mu \bar\lambda)\cr
\delta \sigma &\ = \ \frac12 (\bar\epsilon\lambda - \epsilon\bar\lambda)\cr
\delta\lambda&\ = \ -\frac12 F_{\mu\nu} \gamma^{\mu\nu}\epsilon - D\epsilon + {\ii}D_\mu\sigma \gamma^\mu\epsilon\cr
\delta\bar\lambda&\ = \ -\frac12 F_{\mu\nu}\gamma^{\mu\nu}\bar\epsilon + D \bar\epsilon -{\ii}D_\mu \sigma \gamma^\mu \bar\epsilon\cr
\delta D&\ =\  \frac{\ii}{2} \left( - {\bar\epsilon} \gamma^\mu D_\mu\lambda +  \epsilon \gamma^\mu D_\mu\bar\lambda + \bar\epsilon [\lambda,\sigma] + \epsilon[\bar\lambda,\sigma] \right) \ , \cr
\end{align}
and for the chiral multiplet, they are:
\begin{align}
\delta \phi &\ =\ \bar\epsilon \psi\cr
\delta \bar\phi&\ =\ \epsilon \bar\psi\cr
\delta\psi&\ =\ {\ii} \gamma^\mu \epsilon D_\mu\phi + {\ii}\epsilon\sigma\phi + \bar{\epsilon} F\cr
\delta\bar\psi&\ = \ {\ii} \gamma^\mu \bar\epsilon D_\mu\bar\phi + {\ii}\bar\phi \sigma\bar\epsilon + \bar{F}\epsilon\cr
\delta F&\ = \ {\ii}{\epsilon} \left(  \gamma^\mu D_\mu\psi - \sigma\psi - \lambda\phi \right)\cr
\delta \bar{F}&\ =\ {\ii}{\bar\epsilon} \left( \gamma^\mu D_\mu\bar\psi- \bar\psi\sigma+ \bar\phi\bar\lambda \right)\ .
\end{align}
The invariant actions are
\begin{align}
\cL_{\rm v}&\ =\ {\Tr}\Big[ \frac12 F^{\mu\nu}F_{\mu\nu} + D^\mu\sigma D_\mu\sigma+ D^2 -\epsilon^{\mu\nu\rho}F_{\mu\nu}D_\rho\sigma\Big]\ + \ {\ii}{\Tr}\Big[ \lambda{\gamma}^{\mu} D_{\mu}{\bar\lambda} +  {\lambda} [ {\bar\lambda},{\sigma}] \Big],\\
\cL_{\rm c}&\ =\ - {\bar\phi} D^\mu D_\mu\phi +\bar\phi \sigma^2 \phi + {\ii} \bar\phi D\phi + \bar{F} F -{\ii}\bar\psi\gamma^\mu D_\mu \psi + {\ii}\bar\psi\sigma\psi  +{\ii} \bar\psi\lambda\phi -{\ii}\bar\phi\bar\lambda\psi.
\end{align}
Additionally, one can have an FI term ${\ii}\Tr(\zeta D)$, and a superpotential term:
\begin{equation}
\cL_W = F \frac{\partial W(\phi)}{\partial \phi} - \bar{F} \frac{\partial \bar{W}(\bar{\phi})}{\partial \bar{\phi}} - \frac{\partial^2 W(\phi)}{\partial\phi \partial\phi}\psi\psi + \frac{\partial^2 \bar{W}(\bar{\phi})}{\partial\bar{\phi} \partial\bar{\phi}}\bar\psi\bar\psi,
\end{equation}
where we omitted the obvious index contractions.

\subsection{Localizing deformations in flat space}\label{app:loc_def_flat}
For references, we also list the flat space localization equations here. They follow from the various $\cQ$-exact deformations used in the literature, which we skip for brevity. % They can either be obtained from the flat space SUSY, or by simply taking the $\theta=\pi/2$ specialization and $f(\theta)\to\infty$ limit of the above equations on $S^1\times S^2_b$. in the process we also have to remember to replace $D_\theta \mapsto f(\theta)D_y$.

\paragraph{The Coulomb branch localization.} The BPS equations for $\cQ$ in the Coulomb branch localization scheme \cite{Imamura:2011su} look as follows. The 3d $\cN=2$ vector multiplet equations are:
\begin{align}
\label{CBv}
D\ &\ =\ 0,\quad D_\mu\sigma=0,\quad F_{\mu\nu}=0.
\end{align}
The 3d $\cN=2$ chiral multiplet equations depend on whether we include the superpotential into the localizing deformation. If we do not, we get:
\begin{align}
\label{CBm1}
F=\bar{F}&=0,\quad D_\alpha \phi = D_\varphi\phi =0,\cr
(D_y + \hat{\sigma})\phi&=0,
\end{align}
where as before, $\hat{\sigma}=\sigma+m$. If we include superpotential in the localization, we find instead:
\begin{align}
\label{CBm2}
F=e^{-i\nu}\bar{\partial W},\quad \bar{F}&=e^{{\ii}\nu}\partial W,\cr
\quad D_\alpha \phi = D_\varphi\phi &=0,\cr
(D_y + \hat{\sigma})\phi-{\ii} e^{-i\nu}\bar{\partial W}&=0.
\end{align}
In the $\cN=4$ case, it is possible to further simplify this, if we fix vacua at $y=\pm\infty$ (or the appropriate boundary conditions, if we are on a half-line or on the interval). In this case both $\partial W$ and $W$ vanish at the two infinities, and we can argue that the equations become:
\begin{align}
\label{CBm3}
F=\bar{F}=\partial W&=\bar{\partial W}=0,\cr
\quad D_\alpha \phi = D_\varphi\phi &=0,\cr
(D_y + \hat{\sigma})\phi&=0.
\end{align}

\paragraph{The Higgs branch localization.} Working in the Higgs branch localization scheme \cite{Benini:2013yva} does not affect the matter BPS equations: there are still two options described above. The vector multipelt equations do modify. As usual, we use the appropriate Q-exact deformation, integrate out the auxiliary field $D$, and find:
\begin{align}
\label{HBv}
D_\varphi\sigma &= D_\alpha\sigma =0,\quad F_{\mu\nu}=0,\cr
H(\phi) &= D_y \sigma,
\end{align}
where one normally writes:
\begin{equation}
H(\phi)=e^2 \mu_\R.
\end{equation}

\subsection{Interval partition function}\label{app:interval}
In this paper, partition functions of 3d $\cN=2$ theories on $\bE_\tau\times I$ play some role, where $I=(0,y_0)$ is an interval. At each boundary, we either fix Dirichlet or Neumann boundary conditions for the gauge fields, and the hypermultiplets are given some $(2,2)$ boundary conditions determined by the complex Lagrangian $L$, as in Section \ref{sec:gauge_review}. The partition function can be computed using the localization.

\begin{itemize}
	\item Neumann-Neumann case. Interval partition functions with the Neumann boundary conditions on the vector multiplets at both ends were discussed in \cite{Sugiyama:2020uqh}. In this case, we are not allowed to give any boundary vevs to the hypers, as they are inconsistent with gauge invariance. It is most convenient to use the Coulomb branch localization scheme that does not involve the superpotential, i.e., the BPS equations are \eqref{CBv} and \eqref{CBm1}. The gauge field is flat on the localization locus, and the $A_y$ components can be gauged away, so one simply obtains a flat connection on $\bE_\tau$. For each multiplet, the 1-loop determinants of non-zero modes on the interval cancel out between bosons and fermions. Only the interval zero modes contribute, which is of course consistent with the answer being independent of the length $y_0$ of the interval.
	
	The problem thus effectively reduces to the 2d computation of \cite{Benini:2013nda,Benini:2013xpa}. Chiral multiplets with Dirichlet boundary conditions on the one end and Neumann on the other have no zero modes. A chiral multiplet valued in $\cR$ with Dirichlet on both ends leads to a 2d $(0,2)$ Fermi multiplet valued in $\bar\cR$. In the case of Neumann boundary conditions on both ends, the zero mode is described by a 2d $(0,2)$ chiral in $\cR$. One simply multiplies the 2d one-loop determinants of these zero modes, and includes the one-loop determinants of the boundary matter, if present. After that, one computes the Jeffrey-Kirwan residues of the resulting expression. The reader should consult original papers mentioend above for more details on this.
	
	\item Neumann-Dirichlet. If one of the ends (let us say it is $y=0$) supports Dirichlet boundary conditions for the gauge fields, the computation becomes somewhat simpler. At the same time, we are allowed to turn on non-trivial boundary vevs at the Dirichlet end, since this is not prohibited by gauge invariance. In general, we fix a boundary flat connection at the Dirichlet end:
	\begin{equation}
	A_\parallel\big|_{y=0} = s.
	\end{equation}
	By the localization equations \eqref{CBv}, the gauge field is flat everywhere, hence now we simply have a constant flat connection $s\in {\rm Hom}(\pi_1(\bE_\tau),G)/G$ everywhere on $\bE_\tau\times I$. In particular, there is no need to integrate over $s$ and/or take JK residues. If a chiral multiplet obeys Dirichlet boundary conditions at $y=0$, we can generalize it to
	\begin{equation}
	\phi\big|_{y=0}=c,
	\end{equation}
	for some constant $c$. If it obeys Neumann conditions at the other end,
	\begin{equation}
	\partial_y\phi\big|_{y=y_0}=0,
	\end{equation}
	then the BPS equation $(D_y+\hat{\sigma})\phi=0$ has a solution. No interval zero modes survives (this is always the case when we have Neumann and Dirichlet on the opposite ends), and this multiplet does not contribute to the one-loop determinant. On the other hand, if it obeys Dirichlet boundary conditions at the other end,
	\begin{equation}
	\phi\big|_{y=y_0}=0,
	\end{equation}
	then the equation $(D_y+\hat{\sigma})\phi=0$ has no solutions, as this is inconsistent with $\phi\big|_{y=0}\neq 0$. To be more precise, such solution could be found ``at infinity'', for infinite $\hat{\sigma}$, but the value of $\sigma$, and hence also $\hat{\sigma}=\sigma+m$, is fixed by the boundary conditions for the vector multiplet. Indeed, the $(2,2)$ Neumann boundary conditions fix the boundary value of $\sigma$, and the BPS equations \eqref{CBv} further imply that in the Coulomb branch localization, $\sigma$ is constant.
	
	Thus we see that when $\phi\big|_{y=0}\neq 0$ and $\phi\big|_{y=y_0}=0$, the BPS equations have no solutions. This implies that the interval partition function simply vanishes, meaning that the inner product of the regularized boundary states is zero. This can also be phrased as a spontaneous SUSY breaking: the Hilbert space on $S^1\times I$ with chosen boundary conditions at the endpoints has no SUSY vacuum.
	
	In case the answer is non-zero (i.e. BPS equations have solutions,) to compute the interval partition function, we simply collect contributions of those multiplets that have zero modes, and also include contributions of the possible boundary matter. The vector multiplet leaves no zero modes, and simply provides a background flat connection $s$. Each $\cR$-valued chiral obeying Dirichlet with zero boundary vevs,
	\begin{equation}
	\phi\big|_{y=0} = \phi\big|_{y=y_0}=0,
	\end{equation}
	contributes a $(0,2)$ Fermi multiplet in $\bar\cR$, and its one-loop determinant is taken as in 2d, see equations \eqref{oneloopdet} in the main text. If a chiral obeys Neumann boundary conditions on both ends, it contributes a $(0,2)$ chiral in $\cR$, and its one-loop determinant is again as in 2d, see \eqref{oneloopdet}. Including contributions of the boundary matter, we get schematically:
	\begin{equation}
	Z_{\bE_\tau\times I}=\prod_{(w,f)\in L_1\cap L_2} Z_{\rm F}(s^w x^f,q) \prod_{(w,f)\in L_1^\perp \cap L_2^\perp} Z_{\rm Ch}(s^w x^f,q) \times Z_{\rm boundary},
	\end{equation}
	where $(w,f)$ are gauge and flavor weights, $L_1\cap L_2$ denotes those weights that obey Dirichlet conditions on both boundaries, $L_1^\perp \cap L_2^\perp$ -- Neumann; and $Z_{\rm boundary}$ denotes contributions from the boundary chiral and Fermi multiplets.
	
	\item Dirichlet-Dirichlet. If we impose the $(2,2)$ Dirichlet boundary conditions on the gauge multiplets on both ends that involve no boundary vevs for matter, we get a bosonic zero mode from $\sigma+iA_y$. The corresponding partition function diverges. Turning on the appropriate boundary vevs for hypermultiplets can resolve this issue, and the minimal such vevs lead to the exceptional Dirichlet boundary conditions of \cite{Bullimore:2016nji}.
	
	Once the $\sigma$ zero mode issue is resolved by the boundary vevs, we can attempt to compute the partition function. It has a perturbative part, whose computation is straightforward and follows the above recipe: the gauge field is flat on the localization locus (and thus $A_\parallel\big|=s$ should be the same at both ends), and we simply quantify the interval zero modes and their one-loop determinants, whose functional form is again as in \eqref{oneloopdet}. In this case the gauge multipelt also contributes zero modes: an adjoint $(0,2)$ chiral $S$ and an adjoint Fermi $\Psi_\Phi$.
	
	Additionally, there are non-perturbative contributions corresponding to the possibility that $A_\parallel\big|_{y=0}$ and $A_\parallel\big|_{y=y_0}$ differ by a large gauge transformation (and so there is no global trivialization on $E_\tau\times I$, throughout which the gauge field remains flat). We do not analyze this, and leave it as an open question for the future. For this reason, we make sure that explicit computations in this paper do not rely on precises expressions for the Dirichlet-Dirichlet partition functions.

\end{itemize}

\section{Boundary conditions and boundary states}\label{app:bdry_states}
When we are in the Euclidean signature, we do not distinguish \emph{boundary} and \emph{initial} conditions: one can always treat the direction normal to the boundary as time, and view the boundary condition $\cB$ as the initial condition preparing some state $|\cB\rangle$. Local boundary conditions $\cB$ and their corresponding boundary states $|\cB\rangle$ play some role in this paper, so let us pause to discuss some of their properties.

The boundary conditions are generally required to be elliptic, meaning that the kinetic differential operators for various fields remain elliptic in their presence (or transversally elliptic in the gauged case). This allows to avoid various pathologies, such as infinite number of zero modes, that make perturbation theory ill defined. Such boundary conditions also correspond, in a certain sense, to ``good'' boundary states. Namely, while the state $|\cB\rangle$ itself is usually highly singular and does not belong to the conventional Hilbert space (being an unnormalizable infinite linear combination of states of infinitely high energy), the ``regularized state''
\begin{equation}
e^{-TH}|\cB\rangle,
\end{equation}
where $H$ is the Hamiltonian, sometimes lands in the physical Hilbert space. The unphysical (infinite-energy) part gets suppressed by the Euclidean evolution operator. In order for this to happen, roughly, two things must hold: the Hamiltonian must be sufficiently positive definite (so that $e^{-TH}$ is capable of killing the ``bad'' part of $|\cB\rangle$); and $|\cB\rangle$ must have some ``good'' part in it, so that $e^{-TH}|\cB\rangle$ is not just zero.  Let us briefly discuss this issue in a couple of examples, and make a semi-general proposal that elliptic boundary conditions lead to physically sensible regularized boundary states $e^{-TH}|\cB\rangle$.

\subsection{Bosonic fields}
We start with bosons, and in this case, the most important requirement is the positivity of $H$. The simplest case when it is \emph{not} positive is the topological quantum mechanics with $H=0$:
\begin{equation}
S=\int p\dd q.
\end{equation}
Consider the boundary conditions $q\big|=x$ in such a theory, and look at the interval partition function with $q=x_1$ and $q=x_2$ on the two ends. The Euclidean (or Lorentzian) interval partition function is $\langle x_1|e^{-TH}|x_2\rangle = \langle x_1|x_2\rangle=\delta(x_1-x_2)$. No matter what the length $T$ of the interval is, the wave function $\delta(x_1-x_2)$ is never $L^2$-normalizable. It is delta-function normalizable, of course, but this simple example illustrates what could go wrong in more general cases.

If we instead take $H=\frac12 p^2$, i.e., look at the free particle, then the boundary state $|\cB\rangle = |x\rangle$ is known to be better behaved. Indeed, the finite-time Euclidean evolution makes it $L^2$-normalizable:
\begin{equation}
\langle x| e^{-TH}|y\rangle \sim e^{-\frac{(x-y)^2}{2T}}.
\end{equation}
Notice that in the Lorentzian time, the exponential would be replaced by a pure phase, and would never become normalizable. Thus the contracting property of $e^{-TH}$ is indeed crucial here. However, if we took a momentum-$P$ ``eigenstate'' (a state from the continuous spectrum), that is a boundary state corresponding to the boundary condition 
\begin{equation}
\dot{q}\big| = P,
\end{equation}
then the corresponding boundary state $|P\rangle$ would never become $L^2$-normalizable because 
\begin{equation}
\langle x| e^{-TH}|P\rangle = e^{-T \frac{P^2}{2} + iPx}.
\end{equation}
The corresponding path integral on the interval, with $P$ fixed at both ends, has a bosonic zero mode rendering the answer infinite. In this theory, the Neumann boundary condition is not as well-behaved: Even with $e^{-TH}$, it produces a non-$L^2$ state from the continuous spectrum. 

Finally, let us look at the example of the harmonic oscillator. It is defined in terms of the creation-annihilation operators $[a,a^+]=1$, such that the position and the momentum are
\begin{equation}
p=\frac{i}{\sqrt 2}(a^+-a),\quad x=\frac1{\sqrt 2}(a^++a).
\end{equation}
In this case, $H$ has a positive discrete spectrum, which guarantees that essentially any conceivable boundary condition produces a state that becomes $L^2$ normalizable after a finite-time Euclidean evolution. In particular, we can express the ``position eigenstate'' as
\begin{equation}
|x\rangle = e^{-\frac12 (a^+)^2 +x\sqrt{2} a^+}|0\rangle,
\end{equation}
which is a non-$L^2$ state. After a time $T$ Euclidean evolution, the state
\begin{equation}
e^{-TH}|x\rangle = e^{-\frac12 e^{-2T} (a^+)^2 +x\sqrt{2} e^{-T} a^+}|0\rangle
\end{equation}
has a finite norm. For simplicity, we can just compute it at $x=0$:
\begin{equation}
\big| e^{-TH}|x=0\rangle\big|^2 = \frac1{\sqrt{1-e^{-4T}}},
\end{equation}
which is indeed finite for $T>0$, but diverges at $T=0$.

We could also work with the holomorphic quantization, in which case $a$ and $a^+$ are viewed as complex conjugate variables in the path integral, and the $a=0$ boundary condition, in fact, immediately produces the ground state $|0\rangle$, without the need to act with the contraction operators $e^{-TH}$.

The bottom line of this discussion is that for bosonic fields, the usual boundary conditions (such as Dirichlet or Neumann) produce a state (usually unphysical due to the excited infinite-energy modes), which after a finite-time Euclidean evolution, represented by $e^{-TH}$, can become a valid state from the Hilbert space of the theory. This requires the Hamiltonian to be positive definite, and might fail if there is a continuous spectrum. A way to detect whether this property holds is to consider the finite interval partition function: if it diverges, it indicates that the corresponding state is non-$L^2$. If the interval partition function is finite, it means the state, acted by $e^{-TH}$, is normalizable. We might also ask whether the state has a non-zero projection onto the ground state (because ultimately, in this paper, we use boundary conditions to mimic ground states). This can be tested by taking the $T\to \infty$ limit of the interval partition function: if it is non-vanishing (up to the zero point energy effect), it means that $e^{-TH}|\cB\rangle$ projects non-trivially onto the ground state.

In passing to quantum field theory, we assume that this property still holds, at least if the theory only has isolated massive vacua. More generally, if there are some non-isolated vacua with massless excitations, we have to study the interval partition function with the boundary conditions $\cB$ and ``$\cB^\dagger$'' at the endpoints. If it is finite (which usually happens for the elliptic boundary conditions), then the state $e^{-TH}|\cB\rangle$ is physical. Furthermore, if it does not vanish in the $T\to\infty$ limit (with the zero point energy effect subtracted), then the boundary state contains a ground state in its expansion.

\subsection{Fermionic fields}
Because we work with SUSY theories, fermions are always present, but the story we just sketched for bosons turns out to work quite differently for fermions. In quantum mechanics, we could just start with a complex fermion
\begin{equation}
S=\int\dd t\, \psi^* \dot\psi.
\end{equation}
Fields $\psi$ and $\psi^*$ are canonically conjugate, so we could just impose a boundary condition:
\begin{equation}
\psi\big|=0 \quad \Rightarrow \text{ a boundary state with the wave function } \delta(\psi).
\end{equation}
Unlike in the bosonic case, this state is perfectly physical. After all, when we quantize this system, we write
\begin{equation}
[\psi,\psi^*]_+=i,
\end{equation}
and either define a vacuum as $\psi|0\rangle=0$, which corresponds to the wave-function $\delta(\psi)$ above; or as $\psi^*|0\rangle=0$, which corresponds to $\delta(\psi^*)$. The norm is defined as
\begin{equation}
\int \dd\psi \dd\psi^*\, \delta(\psi^*)\delta(\psi)=1.
\end{equation}
We could also note that the Hilbert space is finite-dimensional, so there is barely any room for the infinite-energy states that plagued the bosonic case. 

That said, the trouble begins when we try to generalize this to fermions in $d\geq 2$ spacetime dimensions, because such boundary conditions usually do not produce the Dirac's vacuum (in extreme cases, one might end up with a completely empty Dirac's sea). Therefore, the boundary state $|\cB\rangle$ might include excitations that have infinite energy relative to the Dirac's state, and the regulator $e^{-TH}$ becomes necessary. It might also happen that $e^{-TH}|\cB\rangle$ is simply zero, -- this is the case when $|\cB\rangle$ only contains the infinite energy states, and is completely projected out. In the usual language, such boundary conditions are said to be non-elliptic, supporting infinitely many zero modes, and thus leading to zero answer for the path integral.

\paragraph{A sick example.} As a simple example, we start with the case where nothing works: a 2d chiral fermion with the action
\begin{equation}
S = \int\dd^2 x\, \bar\psi_- (\partial_1 + {\ii} \partial_2)\psi_-,
\end{equation}
which is written in the Euclidean signature. For boundary conditions at $x^2=0$, we think of $x^2$ as the Euclidean time, and $\bar\psi_-$, $\psi_-$ are the canonically conjugate variables. Naively, one could attempt to impose, for example,
\begin{equation}
\label{bad_fermi_bc}
\psi_-\big|=0.
\end{equation}
This results in the state with the wave functional
\begin{equation}
\delta[\psi_-],
\end{equation}
where square brackets signify that this is a ``functional'' delta, which imposes $\psi_-=0$ at every point of the boundary. One quantizes the Poisson bracket, resulting in:
\begin{equation}
[\psi_-(x^1), \bar\psi_-(y^1)]_+\ =\ {\ii}\delta(x^1-y^1).
\end{equation} 
Thinking of modes of $\bar\psi_-$ as the creation operators, and those of $\psi_-$ as the annihilation operators, the state, which obviously obeys
\begin{equation}
\psi_- \delta[\psi_-]=0,
\end{equation}
is identified as the ``wrong'' vacuum, the one with the empty Dirac's sea. The excitations due to $\bar\psi_-$ around this ``vacuum'' have both negative and positive energy, which is a well-known issue. The correct Dirac's vacuum has all the negative-energy modes filled, and the state $\delta[\psi_-]$, from this point of view, has the infinite energy. If we define the quantum Hamiltonian $H$ to be such that $H=0$ for the Dirac's vacuum (ignoring the zero-point energy, which would cancel in SUSY theories anyways), then
\begin{equation}
e^{-TH}\delta[\psi_-]=0.
\end{equation}
In other words, nothing is left in the physical Hilbert space. This signifies that any partition function with the boundary conditions \eqref{bad_fermi_bc} is automatically zero. This happens due to zero modes, and indicates that the boundary condition \eqref{bad_fermi_bc} is not elliptic. This is of course a known issue: chiral fermions do not have any good boundary conditions on their own.

\paragraph{A healthy example.} Now consider a Dirac fermion in two dimensions, the action being
\begin{equation}
S=\int\dd^2 x\,\left[\bar\psi_-(\partial_1 + {\ii}\partial_2)\psi_- - \bar\psi_+(\partial_1 - {\ii}\partial_2)\psi_+ \right].
\end{equation}
The boundary conditions $\psi_\pm\big|=0$, or $\bar\psi_\pm\big|=0$, or $\psi_-\big|=\bar\psi_+\big|=0$, or $\bar\psi_-\big|=\psi_+\big|=0$, all would lead to the same pathology as we have just observed, the corresponding states $|\cB\rangle$ being unphysical, and $e^{-TH}|\cB\rangle$ vanishing. But now we have more options, and we can choose instead:
\begin{equation}
(\psi_+ + \psi_-)\big| = (\bar\psi_+ - \bar\psi_-)\big|=0\qquad \text{(B type)}
\end{equation}
or
\begin{equation}
(\psi_+ + \bar\psi_-)\big| = (\bar\psi_+ - \psi_-)\big|=0\qquad \text{(A type)}.
\end{equation}
Both of these are known to be well-defined boundary conditions, so let us see how the boundary states work out. For simplicity, we will only consider the B case (the A type works the same way).

In Minkowski signature we have ${\ii}\partial_2=\partial_0$, and it is straightforward to analyze the Dirac equations in the momentum space, to find the following dispersion relations:
\begin{align}
&\text{For modes } \psi_-(p),\ \bar\psi_-(p):\qquad E=-p,\cr
&\text{For modes } \psi_+(p),\ \bar\psi_+(p):\qquad E=p.
\end{align}
The physical vacuum $|0\rangle$ is thus killed by the negative-energy modes:
\begin{align}
\psi_-(p)|0\rangle = \bar\psi_-(p)|0\rangle = \psi_+(-p)|0\rangle = \bar\psi_+(-p)|0\rangle=0,\quad \text{for }p>0,
\end{align}
where the anti-commutation relations are standard, $[\psi_+(p),\bar\psi_+(q)]\sim \delta(p+q)$, and  $[\psi_-(p),\bar\psi_-(q)]\sim \delta(p+q)$. Now the state $|B\rangle$ created by the B-type boundary conditions obeys:
\begin{align}
(\psi_+(p)+\psi_-(p))|B\rangle &= (\bar\psi_+(p)-\bar\psi_-(p))|B\rangle\cr 
= (\psi_+(-p)+\psi_-(-p))|B\rangle &= (\bar\psi_+(-p)-\bar\psi_-(-p))|B\rangle=0,\quad \text{for }p>0.
\end{align}
It is not too hard to find the Bogolyubov transformation that relates the two states:
\begin{equation}
|B\rangle=e^{\int_{p>0}\dd p\,\left[\psi_+(p)\bar\psi_-(-p) + \psi_-(-p)\bar\psi_+(p)\right]}|0\rangle.
\end{equation}
Thus we see that, predictably, $|B\rangle$ is an infinite linear combination of states with arbitrarily high energy. However, unlike the sick state $\delta[\psi_-]$, this linear combination involves the low energy states, including the Dirac's vacuum:
\begin{equation}
|B\rangle = |0\rangle + \dots
\end{equation}
Thus it makes sense to assume that $e^{-TH}|B\rangle$ is a proper physical state.

Ultimately, the test for a boundary state $|\cB\rangle$ to give a physically sensible state $e^{-TH}|\cB\rangle$ is that the interval partition function with the boundary conditions $\cB$ and $\cB^\dagger$ be finite and non-vanishing (in Euclidean signature). Here the $\cB^\dagger$ involves some proper notion of Hermitian conjugation, which we deliberately leave undefined. In this paper, we apply this to supersymmetric theories in three, two, and one dimensions, with the $\cN=(0,2)$ boundary conditions in 3d, and their reductions in 2d and 1d. All our boundary conditions have the nice properties describes above (in particular, are elliptic). Moreover, they do not depend on the interval length, as implied by the SUSY. This is consistent with the claim that such boundary conditions, at the level of Q-cohomology, mimic the SUSY vacua. We will work under the assumption that once elliptic, such boundary conditions, followed by the Euclidean evolution, prepare physically sensible states $e^{-TH}|\cB\rangle$.

\setlength{\unitlength}{1mm}

\newpage

\bibliographystyle{utphys}
\bibliography{refs}
\end{document}